\documentclass[]{aa}

\usepackage{amsmath}
\usepackage{graphicx}
\usepackage{graphics}
\usepackage{epsfig}
\usepackage[separate-uncertainty=true, multi-part-units=single, print-unity-mantissa=false]{siunitx} 
\usepackage{interval}
\usepackage[above]{placeins}
\usepackage{multirow}

\usepackage{tikz}
\usetikzlibrary{positioning}
\usepackage{subcaption}
\usepackage{orcidlink}

\usepackage{tablefootnote}
\usepackage{longtable}
\usepackage{booktabs}
\usepackage[flushleft]{threeparttable}
\usepackage[version=4]{mhchem}
\usepackage[above]{placeins}

\usepackage{txfonts}

\usepackage{url}
\usepackage{color}
\hypersetup{
  colorlinks,
  linkcolor=blue,
  citecolor=blue,
  urlcolor=blue
}

\usepackage{natbib,twoopt}
\bibpunct{\textcolor{black}{(}}{\textcolor{black}{)}}{\textcolor{black}{;}}{\textcolor{black}{a}}{\textcolor{black}{}}{\textcolor{black}{,}}

\usepackage{makecell}
\makeatletter
\newcommand{\bibnote}[2]{\global\@namedef{#1note}{#2}}
\newcommand{\biblink}[2]{\global\@namedef{#1link}{#2}}
\makeatother

\makeatletter
  \protected\def\stonyslink{%
     \def\hyper@linkstart##1##2{}\let\hyper@linkend\@empty}
  \newcommandtwoopt{\citeads}[3][][]{%
   \href{http://ui.adsabs.harvard.edu/abs/#3/abstract}%
        {\stonyslink \citealp[#1][#2]{#3}}%
   \biblink{#3}{\href{http://ui.adsabs.harvard.edu/abs/#3/abstract}{ADS}}}
 \newcommandtwoopt{\citepads}[3][][]{%
   \href{http://ui.adsabs.harvard.edu/abs/#3/abstract}%
        {\stonyslink \citep[#1][#2]{#3}}%
   \biblink{#3}{\href{http://ui.adsabs.harvard.edu/abs/#3/abstract}{ADS}}}
 \newcommandtwoopt{\citetads}[3][][]{%
   \href{http://ui.adsabs.harvard.edu/abs/#3/abstract}%
        {\stonyslink \citet[#1][#2]{#3}}%
  \biblink{#3}{\href{http://ui.adsabs.harvard.edu/abs/#3/abstract}{ADS}}}
 \newcommandtwoopt{\citeyearads}[3][][]{%
   \href{http://ui.adsabs.harvard.edu/abs/#3/abstract}%
        {\stonyslink \citeyear[#1][#2]{#3}}%
   \biblink{#3}{\href{http://ui.adsabs.harvard.edu/abs/#3/abstract}{ADS}}}
 \newcommandtwoopt{\citetaliasads}[3][][]{%
   \href{http://ui.adsabs.harvard.edu/abs/#3/abstract}%
        {\stonyslink \citetalias[#1][#2]{#3}}%
   \biblink{#3}{\href{http://ui.adsabs.harvard.edu/abs/#3/abstract}{ADS}}}
\makeatother

\def\linkadspage#1#2#3{\href{https://ui.adsabs.harvard.edu/link_gateway/#1/pub_pdf\#page=#2}{#3}}

\def\linkoldadspage#1#2#3{\href{https://ui.adsabs.harvard.edu/link_gateway/#1/ads_pdf\#page=#2}{#3}}

\defcitealias{2020A&A...634A.111F}{F20}
\defcitealias{2020MNRAS.492.4553R}{R20}

\def\MgII{\ion{Mg}{II}}
\def\SII{[\ion{S}{II}]}
\def\OIII{[\ion{O}{III}]}
\def\NII{[\ion{N}{II}]}

\def\HeII{\ion{He}{II}}
\def\CIV{\ion{C}{IV}}
\def\SiIV{\ion{Si}{IV}}
\def\CIII{\ion{C}{III}]}
\def\FeII{\ion{Fe}{II}}
\def\AlIII{\ion{Al}{III}}
\def\FeIII{\ion{Fe}{III}}
\def\AlII{\ion{Al}{II}}
\def\CII{\ion{C}{II}}
\def\NV{\ion{N}{V}}

\DeclareCaptionSubType*[Alph]{table}
\DeclareCaptionLabelFormat{mystyle}{Table~\bothIfFirst{#1}{ }#2}
\begin{document}

\title{Extreme outflow velocities and weak UV emission lines indicate quasars shedding their dust cocoons}\titlerunning{Quasars with extreme outflow velocities and weak UV emission lines}

\author{
Guozhen~Ma\inst{\ref{inst1},\ref{inst2},\ref{inst3}}\orcidlink{0000-0003-0825-7911}
\and
Stefan~J.~Geier\inst{\ref{inst4},\ref{inst5}}\orcidlink{0000-0003-3154-2120}
\and
Johan~P.~U.~Fynbo\inst{\ref{inst1},\ref{inst2}}\orcidlink{0000-0002-8149-8298}
\and
Lise~Christensen\inst{\ref{inst1},\ref{inst2}}\orcidlink{0000-0001-8415-7547}
\and
Andrei~Berdyugin\inst{\ref{inst6}}\orcidlink{0000-0002-9353-5164}
\and
Rasmus~Frederiksen\inst{\ref{inst1}, \ref{inst2}}
\and
Kasper~E.~Heintz\inst{\ref{inst1},\ref{inst2}}\orcidlink{0000-0002-9389-7413}
\and
Phillip~D.~Henriksen\inst{\ref{inst1},\ref{inst2}}
\and
Jens-Kristian~Krogager\inst{\ref{inst8},\ref{inst2}}\orcidlink{0000-0002-4912-9388}
\and
C\'edric~Ledoux\inst{\ref{inst3}}\orcidlink{0000-0002-7864-3327}
\and
Vilppu~Piirola\inst{\ref{inst6}}\orcidlink{0000-0003-0186-206X}
\and
Palle~M\o ller\inst{\ref{inst7},\ref{inst2}}\orcidlink{0000-0002-9994-505X}
\and
Simone~Vejlgaard\inst{\ref{inst1},\ref{inst2}}\orcidlink{0000-0003-0471-5647}
\and
Hyunseop~Choi\inst{\ref{inst9},\ref{inst10},\ref{inst11}}\orcidlink{0000-0002-3173-1098}
}
\institute{
Cosmic DAWN Center,
\email{guozhen.ma@nbi.ku.dk}\label{inst1}
\and
Niels Bohr Institute, University of Copenhagen, Jagtvej 128, DK-2200, Copenhagen N, Denmark\label{inst2}
\and
European Southern Observatory, Alonso de C\'ordova 3107, Vitacura, Casilla 19001, Santiago, Chile\label{inst3}
\and
Instituto de Astrof{\'i}sica de Canarias, V{\'i}a L{\'a}ctea, s/n, 38205, La Laguna, Tenerife, Spain\label{inst4}
\and
Gran Telescopio Canarias (GRANTECAN), 38205 San Crist{\'o}bal de La Laguna, Tenerife, Spain\label{inst5}
\and
Finnish Centre for Astronomy with ESO (FINCA), Quantum, FI-20014 University of Turku, Finland\label{inst6}
\and
European Southern Observatory, Karl-Schwarzschildstrasse 2, D-85748 Garching, Germany\label{inst7}
\and
Universit\'e Lyon1, ENS de Lyon, CNRS, Centre de Recherche Astrophysique de Lyon UMR5574, F-69230 Saint-Genis-Laval, France\label{inst8}
\and
Department of Astronomy, University of Michigan, 1085 S. University Ave., Ann Arbor, MI 48109, USA\label{inst9}
\and
D\'epartement de Physique, Universit\'e de Montr\'eal, C. P. 6128, Succ. Centre-Ville, Montr\'eal, QC, H3C 3J7, Canada\label{inst10}
\and
Mila - Quebec Artificial Intelligence Institute, Montr\'eal, QC, H2S 3H1, Canada\label{inst11}
}
\authorrunning{Ma et al.}

\date{Received xxx / Accepted xxx}

\abstract{
The recently discovered low-ionisation broad absorption line (LoBAL) quasar GQ\,1309$+$2904 is unusual due to its very broad, highly blueshifted absorption troughs and an absence of broad emission lines except for ${\mathrm{H} \alpha}$. In this paper, we present observations of six quasars that appear very similar to GQ\,1309$+$2904 in the rest-frame ultraviolet (UV). We measure the systemic redshifts of these quasars to be $z\approx$ 2.07--3.28 from detected ${\mathrm{H} \alpha}$ emission lines. We confirm that all targets are quasars with highly blueshifted BALs possessing high-speed outflows with velocities up to $\sim \num{0.16}\,c$, and five of them are confidently identified as LoBAL quasars. Based on ${\mathrm{H} \alpha}$ emission, black hole masses and Eddington ratios of these quasars are $M_{\mathrm{BH}} \approx 10^{8.7}$--$10^{9.4}\,M_{\odot}$ and $L_{\mathrm{bol}} / L_{\mathrm{Edd}} \approx$ 0.14--0.34, indicating that their central black holes are very massive and active. Every quasar in our sample exhibits a very flat or reddened continuum. The spectral shapes of three objects are well-fitted by a normal quasar composite reddened by a Small-Magellanic-Cloud-like (SMC-like) extinction curve, while the other three require a steeper extinction law. Broad-band ($BVR$) polarimetry for two of the latter group (plus GQ\,1309$+$2904) reveals their low polarisations, consistent with low inclination (more face-on) angles. We propose that these objects are weak emission-line quasars (WLQs) observed through the disc wind, caught emerging from their dust cocoons. As quasars shed their cocoons, dust grains in the disc wind are shattered into smaller particles, producing the UV-steeper extinction curve observed along the outflow. We present a schematic illustration of this shedding process that can account for the peculiar spectral features observed in our sample.}

\keywords{quasars: general -- quasars: absorption lines -- quasars: emission lines --
quasars: individual: GQ\,1309$+$2904}

\maketitle
\nolinenumbers

\section{Introduction}     
\label{sec:introduction}

Since the discovery of quasars more than half a century ago \citepads{1963Natur.197.1040S}, they have been understood to be a crucial stage in the evolution of massive galaxies back to $z\gtrsim7$ (\citeads{2021ApJ...907L...1W}; \citeads{2022Natur.604..261F}). In this paper, we explore a subset of the quasars that display powerful outflows, the so-called broad absorption line (BAL, \citeads{1980ApJ...238..488T}) quasars. \citetads{1993ApJ...413...95V} presented the first systematic study of BAL quasars exhibiting low-ionisation broad absorption lines (LoBALs). These quasars are characterised by {\it (i)} \ion{Al}{iii} absorption stronger than \ion{Al}{ii} absorption, {\it (ii)} very deep and broad \ion{C}{iv} absorption troughs, and {\it (iii)} low-ionisation absorption lines that tend to be narrower than the high-ionisation absorption lines and located towards the low-velocity end of the \ion{C}{iv} troughs. \citetads{1993ApJ...413...95V} found that about 1.5\% of all optically selected quasars are of this kind, but these quasars are much more common among infrared-selected quasars.

\citetads{2020A&A...634A.111F}, hereafter \citetaliasads{2020A&A...634A.111F}, present the discovery of what appears to be an extreme LoBAL quasar GQ\,1309$+$2904 in a {\it Gaia}-assisted selection for red quasars partly based on proper
motion measurements from the {\it Gaia} survey (\citeads{2016A&A...595A...1G}). This quasar resembles the quasars discussed by \citetads{1983PASP...95..341F} and \citetads{1993ApJ...413...95V}, in terms of the large velocity range of BAL troughs. GQ\,1309+2904 attracted special attention in \citetaliasads{2020A&A...634A.111F} as its observer-frame optical spectrum is enigmatic. In particular, the systemic redshift is hard to pin down as the optical spectrum does not display any of the strong and broad emission lines (BELs) normally seen in type 1 quasars (see e.g. \citeads{1990MNRAS.243..231B}; \citeads{2001ApJ...546..775B}; \citeads{2016A&A...585A..87S}). Instead, the optical spectrum is characterised by a red continuum and several very deep and broad absorption troughs. For such an object one would normally determine the redshift from the position of the red edges of the absorption troughs in the spectrum. However, in the case of GQ\,1309$+$2904, there is also a weak and spatially extended emission line that most naturally is interpreted as extended Lyman-$\alpha$ (Ly$\alpha$) emission from the host galaxy or its circumgalactic environment. Such extended Ly$\alpha$ emission is fairly ubiquitous around bright quasars (\citeads{1992MNRAS.258P..23B}; \citeads{1999MNRAS.305..849F}; \citeads{2006A&A...459..717C}; \citeads{2019MNRAS.482.3162A}). In \citetaliasads{2020A&A...634A.111F} it is established that the redshift of GQ\,1309$+$2904 is 2.66 and that the enigmatic character of the optical spectrum results from the fact that the BALs are highly blueshifted, with no obvious emission lines or absorption lines in the velocity range 0--{\SI{22000}{\km\per\s}} relative to the systemic redshift, while the BAL velocities lie in the range {\num{22000}}--{\SI{40000}{\km\per\s}}.

Following the work of \citetaliasads{2020A&A...634A.111F}, we have searched for other quasars to investigate whether there are similar quasars with extremely blueshifted and potentially detached LoBALs. If so, we plan to characterise this sub-class of LoBAL quasars and determine where they fit in the wider picture of the quasar phenomena in general and BALs in particular. Concerning the last objective, it is puzzling why these quasars with very blueshifted and possibly detached absorption troughs have weak BELs. If the BAL troughs are detached, they ought to leave the BELs unabsorbed. This, however, may provide interesting constraints on the relative positions of the gas causing the absorption troughs and the broad-line region (BLR; see also the discussion in \citeads{2002ApJS..141..267H}), or indicate that the BLR emission is intrinsically weak. 

Weak emission lines have been found in a class of quasars called weak emission-line quasars (WLQs; \citeads{2007ApJ...663..103L}; \citeads{2010MNRAS.404.2028H}; \citeads{2012ApJ...747...10W}; \citeads{2015ApJ...805..122L}), and they show extremely weak or even missing BELs in the rest-frame ultraviolet (UV). \citetads{2025ApJ...994..213C} define WLQs as quasars with {\CIV} rest-frame equivalent width (REW) $< \SI{8.9(0.2)}{\AA}$, based on a sample of \num{371091} quasars from the Sloan Digital Sky Survey Data Release 16 (SDSS-DR16) quasar catalogue (\citeads{2022ApJS..263...42W}). They also define quasars with {\CIV} REW of \num{8.9}--\SI{19.3}{\AA} as `bridge quasars', which denote a transitional stage between WLQs and normal quasars ({\CIV} REW $> \SI{19.3(0.3)}{\AA}$). Although there are normally no BALs observed in WLQs, it is a selection effect because BAL quasars are commonly excluded for WLQ selections (e.g. \citeads{2012ApJ...747...10W}; \citeads{2015ApJ...805..122L}). In fact, WLQs are found to have blueshifted BALs (\citeads{2013AJ....145..157J}; \citeads{2022ApJ...930....5Y}), and our objects could therefore be a specific sub-population of WLQs; we will investigate this further.

\citetads{1998A&A...340..371H} found that for LoBAL quasars there is an anti-correlation between the degree of polarisation and the detachment of the absorption troughs. For this reason, we also present polarisation data for a subset of our sample for which we have been able to secure observations. The polarisation measurements can also help to determine the inclination angle (the angle between the line of sight and the normal line to the disc plane) of the quasar (\citeads{2004A&A...427..107L}).

Given the absence of strong emission lines in these quasars, the extremely blueshifted nature of their BALs is easily overlooked as there is no obvious feature in observer-frame optical spectra on which to base a determination of the systemic redshift. For this reason, it is not possible to simply search existing compilations of BAL quasars such as in \citetads{2002ApJS..141..267H}. Instead, we had to primarily rely on visual inspection. 

In this paper, we report on a sample of seven quasars with highly blueshifted BALs and weak UV emission lines. In Sect.~\ref{sec:sample}, we describe the selection of our sample. We describe our observations, including polarimetry of three of the quasars in Sect.~\ref{sec:data}. In Sect.~\ref{sec:results}, we describe how we infer the systemic redshifts and other fundamental properties of the systems. Finally, we present our conclusions on the nature of this class of quasars and discuss how one may place them in the wider context of active galactic nuclei (AGNs) in Sect.~\ref{sec:discussion}. We assume a flat $\mathrm{\Lambda }$ cold dark matter cosmology with $H_0=70$\,km\,s$^{-1}$\,Mpc$^{-1}$, $\Omega_\mathrm{M}=0.3$, and $\Omega_\Lambda=0.7$ throughout this work.

\section{Sample}    \label{sec:sample}

We searched two places for other quasars similar to GQ\,1309$+$2904: {\it (i)} in the sample of red quasars that our team has observed over the years (e.g. \citeads{2013ApJS..204....6F}; \citeads{2016ApJ...832...49K}; \citeads{2020A&A...644A..17H}; \citeads{2024A&A...683A.157V}) and {\it (ii)} in the Sloan Digital Sky Survey (SDSS) BAL quasar sample \citepads{2020ApJS..250....8L}. Another similar quasar PSS\,J0141$+$3334 is added from \citetads{2003AJ....126...53B}.

The features that we searched for are:
{\it (i)} very broad BAL troughs; 
{\it (ii)} an absence of BELs;
{\it (iii)} the presence of \ion{Al}{iii} absorption.
Criterion {\it (iii)} is based on the fact that GQ\,1309+2904 displays \ion{Al}{iii} absorption, which is common for LoBAL quasars \citepads{2002ApJS..141..267H}.

For over a decade our team has carried out various surveys for red quasars (e.g. \citeads{2013ApJS..204....6F}; \citeads{2016ApJ...832...49K}; \citeads{2020A&A...644A..17H}). The objective has primarily been to search for quasars reddened by foreground dusty Damped Lyman-$\alpha$ Absorbers (DLAs) and several such systems have indeed been found (e.g. \citeads{2016MNRAS.455.2698K}; \citeads{2017A&A...606A..13F}; \citeads{2018A&A...615A..43H}; \citeads{2019A&A...625L...9G}). These quasars have been selected as point sources not classified as quasars in SDSS and with optical, near-infrared (near-IR), and mid-IR colours inconsistent with being stars. We also use astrometry from {\it Gaia} to reject point sources with nonzero parallax and/or proper motion (see \citeads{2020A&A...644A..17H}, and references therein). Normal blue quasars reddened by foreground dusty DLAs only constitute a few per cent of the 534 quasars we have observed in these studies (for a complete sample list see the eHAQ+GAIA23 sample in \citeads{2024A&A...683A.157V}) and the majority are BAL quasars or quasars that are red for other reasons than intervening absorbers. We note that the eHAQ+GAIA23 sample is biased towards quasars that are redder than normal quasars by design, and is not sensitive to quasars affected by severe dust extinction due to flux limit. In the eHAQ+GAIA23 sample, we found three additional quasars resembling GQ\,1309$+$2904, namely GQ\,1237$+$1233, GQ\,0109$-$0719, and GQ\,1353$+$2554. A spectrum of GQ\,1237$+$1233 has previously been published in \citetads{2016ApJ...832...49K}.

In \citetads{2003AJ....126...53B}, we identified PSS\,J0141$+$3334, which fulfils the criteria. The approach that led \citetads{2003AJ....126...53B} to discover PSS\,J0141$+$3334 along with two other peculiar quasars was to take spectra of point sources in poorly populated regions of the $g-r$ vs. $r-i$ colour space. In other words, they studied point sources with optical colours unlike both stars and normal (typically blue) quasars. We have not found other similar quasars in other literature including the comprehensive work of \citetads{2002ApJS..141..267H}.

Finally, we carried out a visual inspection of quasars in the SDSS quasar catalogue \citepads{2020ApJS..250....8L} with magnitude $r<20$ and probability of being a BAL (BALP) larger than 90\%. We also restricted the search to redshifts $1.8<z<3.5$ where the main BAL troughs are in the visible spectral range. Several candidates were found and in this paper, we present follow-up observations for one of them, SDSS\,J1012$+$0358. An additional SDSS quasar, SDSS\,J0155$+$2543, was also identified as a potential `transition' quasar, as the spectrum contains both a \ion{Mg}{ii} and what seems to be a Ly$\alpha$ emission line, and we will also include this object in the sample.

\begin{table}[tb]
\centering
\begin{threeparttable}
\caption{Our sample.}
\begin{tabular}{lllc}
\hline \hline \noalign{\smallskip}
Name & R.A.  & Dec. & Ref. \\
    &   J\,2000.0 & J\,2000.0 & \\
\hline
\noalign{\smallskip}
GQ\,1309$+$2904 & 13:09:23.91 & +29:04:51.79 & 1 \\
GQ\,0109$-$0719 & 01:09:30.72 & $-$07:19:31.17 & 2 \\
GQ\,1353$+$2554 & 13:53:05.76 & +25:54:29.28 & 2 \\
GQ\,1237$+$1233 & 12:37:26.46 & +12:33:50.11 & 2 \\
PSS\,J0141$+$3334 & 01:41:32.79 & +33:34:24.69 & 3 \\
SDSS\,J1012$+$0358 & 10:12:30.16 & +03:58:56.89 & 4 \\
SDSS\,J0155$+$2543 & 01:55:56.81 & +25:43:34.67 & 4 \\
\noalign{\smallskip}
\hline
\noalign{\smallskip}
\hline
\end{tabular}
\tablebib{
(1)~\citetaliasads{2020A&A...634A.111F}; (2) eHAQ$+$GAIA23 (\citeads{2024A&A...683A.157V}); (3) \citetads{2003AJ....126...53B}; (4) SDSS \citepads{2020ApJS..250....8L};
}
\label{tab:sample}
\end{threeparttable}
\end{table}

Our final sample is listed in Table~\ref{tab:sample}. Four of the targets,  PSS\,J0141$+$3334, GQ\,1309$+$2904, GQ\,0109$-$0719, and GQ\,1353$+$2554 are not part of the SDSS quasar sample even though they are in the SDSS footprint. In other words, they do not fulfil the quasar selection criteria of SDSS due to their unusual optical colours.\\

\section{Observations and data reduction}    \label{sec:data}

Spectroscopic observations were collected at the Nordic Optical Telescope (NOT) with the instrument ALFOSC and at the Gran Telescopio Canarias (GTC) with the instruments OSIRIS and EMIR. Table~\ref{tab:log} presents a log of the observations.

The spectra were reduced using a set of Python scripts for the reduction of long-slit spectra. The code is available on GitHub\footnote{\href{https://github.com/keheintz/PyReduc}{https://github.com/keheintz/PyReduc}}. We calibrated the ALFOSC spectra based on He and Ne arc maps and OSIRIS spectra with HgAr arc frames. These optical arc line lists are in air wavelength, so we converted the wavelengths from air into vacuum following \citetads{1991ApJS...77..119M}. The spectra were flux calibrated using observations of spectrophotometric standard stars observed on the same nights.

For near-IR EMIR spectra, we used the same Python scripts for reduction. The spectra were wavelength calibrated using Ne, Xe, and HgAr arc frames, with calibration lines listed in vacuum wavelength. To correct for telluric absorption we used observations of telluric standard stars obtained immediately after science observations, and the flux calibration was done based on the known magnitude and stellar type of the telluric standard star.

The calibrated flux was compared with photometric data from several public surveys and adjusted to obtain a better spectrophotometric accuracy if necessary. The photometric data are from the optical SDSS-DR16 (\citeads{2020ApJS..249....3A}; \citeads{2020ApJS..250....8L}), the near-infrared UKIRT Infrared Deep Sky Survey (UKIDSS, \citeads{2007MNRAS.379.1599L}; \citeads{2007MNRAS.375..213W}), and the mid-infrared All-Sky Wide-field Infrared Survey Explorer (AllWISE, \citeads{2014yCat.2328....0C}). For quasars not observed by UKIDSS, we instead used the near-infrared data from the Two Micron All Sky Survey (2MASS, \citeads{2003yCat.2246....0C}; \citeads{2006AJ....131.1163S}). We also utilised the optical Panoramic Survey Telescope and Rapid Response System Telescope \#1 Data Release 1 (Pan-STARRS1-DR1, \citeads{2016arXiv161205560C}) for crosschecking. 

We converted all measurements in Vega magnitudes from UKIDSS, 2MASS, and AllWISE to AB magnitudes to calculate their corresponding photometric flux. The SDSS $u$ band is corrected by $u_{\mathrm{AB}} = u_{\mathrm{SDSS}} - 0.04~\mathrm{mag}$ and the $z$ band zero point is revised according to $z_{\mathrm{AB}} = z_{\mathrm{SDSS}} + 0.02~\mathrm{mag}$, based on the official SDSS photometric flux calibration web page\footnote{\href{https://www.sdss4.org/dr16/algorithms/fluxcal/}{https://www.sdss4.org/dr16/algorithms/fluxcal/}}.

We corrected all spectra for the foreground (Milky Way) dust reddening using \texttt{dust\_extinction}\footnote{\href{https://dust-extinction.readthedocs.io/en/stable/}{https://dust-extinction.readthedocs.io/en/stable/}} by \citet{gordon_2024_11235336} and the Milky Way extinction curve from \citetads{2023ApJ...950...86G}. We also obtained broad-band ($BVR$) polarisation observations of GQ\,1309$+$2904, GQ\,1353$+$2554, and GQ\,1237$+$1233 during a single night run with DiPol-UF at the Nordic Optical Telescope. DIPol-UF collects polarimetry in three optical filters, $B$, $V$, and $R$, simultaneously. After standard CCD calibrations, we extracted the difference between the brightness of
ordinary and extraordinary rays from the quasar and calculated the Stokes parameters. Further information on DiPol-UF and how the data are acquired and reduced can be found in \citetads{2021AJ....161...20P} and \citetads{2022MNRAS.514.2479K}. 

\section{Analysis and results}    \label{sec:results}

\subsection{Systemic redshifts and general spectral features}
\label{sect:redshifts}
First of all we set out to determine the systemic redshifts of the quasars. We cannot rely on the red edges of their absorption troughs as we hypothesise that the BALs of these quasars may well be strongly blueshifted relative to their systemic redshifts if indeed these quasars are similar to GQ\,1309$+$2904. Following \citetaliasads{2020A&A...634A.111F}, we determine the systemic redshifts through the subsequent three methods in order of precedence. 
\medskip

{\it (i)} Fitting the ${\mathrm{H} \alpha}$ emission line with a double-Gaussian model. The ${\mathrm{H} \alpha}$ emission lines, which appear to be ubiquitous in our targets, are the only clear emission lines that we can utilise to measure systemic redshifts. Low-ionisation BELs are also good options for obtaining the systemic redshift (\citeads{2010MNRAS.405.2302H}; \citeads{2016ApJ...831....7S}) as they are observed to be closely tied to narrow emission lines originating in the narrow-line region (NLR).

\begin{figure*}[th]
    \centering
    \includegraphics[width=17cm]{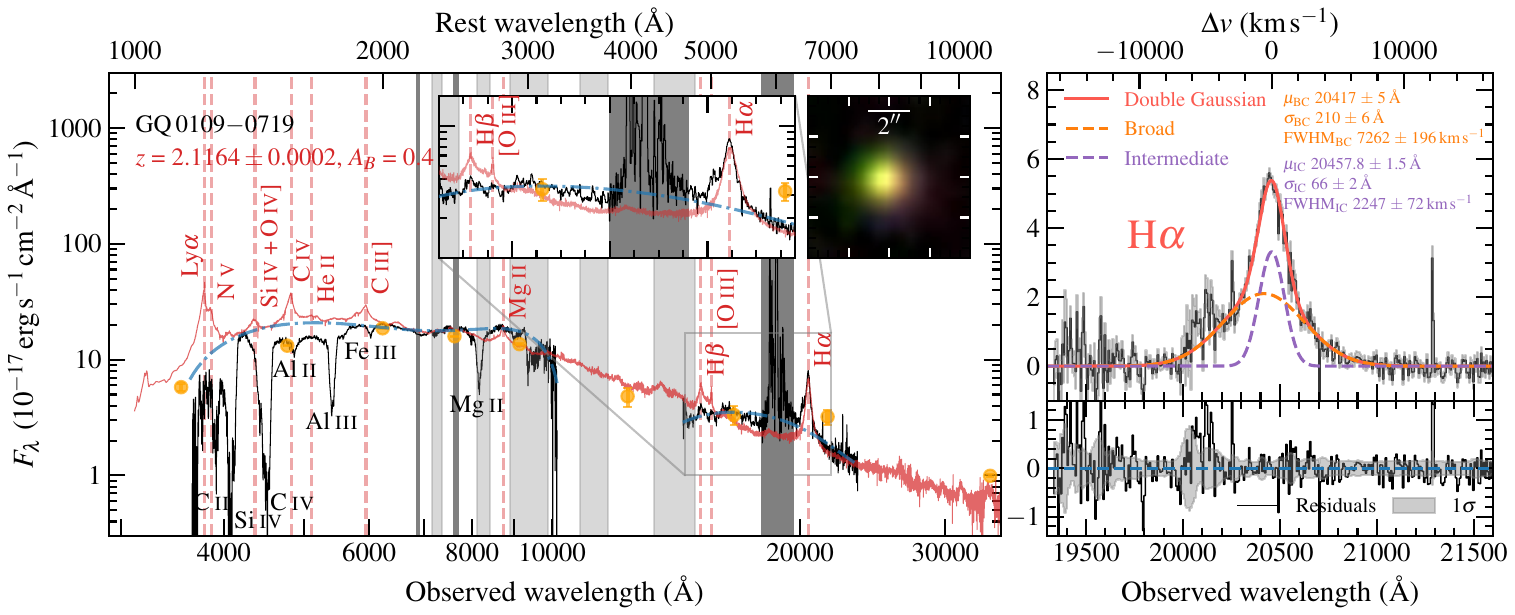}
    \caption{{\it Left:} Stitched optical and near-infrared spectra of GQ\,0109$-$0719 from NOT and GTC are shown in black, overplotted with $ugriz$, $JHK_{s}$, and $W1$ photometry from the SDSS, 2MASS, and AllWISE surveys as orange dots. The telluric absorption bands are displayed as shaded areas, with the dark grey areas indicating that the corresponding parts of the spectra are severely affected and not taken into account in the analysis, while the lightly shaded regions have negligible telluric absorption. The image inset in the upper right corner is the SDSS $8^{\prime\prime} \times 8^{\prime\prime}$ tricolour image ($irg$ to RGB) and a zoom in towards the spectral range covering ${\mathrm{H} \beta}$ to ${\mathrm{H} \alpha}$ is inserted. A quasar template spectrum from \citetads{2016A&A...585A..87S} is overplotted in red for comparison, scaled to match the $H$ band flux and we estimated the dust extinction $A_B = 0.4$\,mag assuming the \citetads{2015A&A...584A.100Z} extinction curve with $R_V = 2.21$. (for details see Sect.~\ref{sec:dust_extinction}). Our best estimate for the continuum is shown as the dash-dotted blue line.\\
   {\it Right:} Best-fit double-Gaussian model (in red) of the ${\mathrm{H} \alpha}$ line (in black, after continuum subtraction), including a broad component (BC; in orange), an intermediate component (IC; in purple), and 1$\sigma$ error shadings (in grey), together with the corresponding residual plot (bottom panel). The fitted mean ($\mu$), standard deviation ($\sigma$), and FWHM (corrected for the instrumental resolution) of each Gaussian component are listed in the upper right corner.
    }    
    \label{fig:GQ0109-0719_z_full}
\end{figure*}

\begin{figure*}[th]
\centering
\includegraphics[width=17cm]{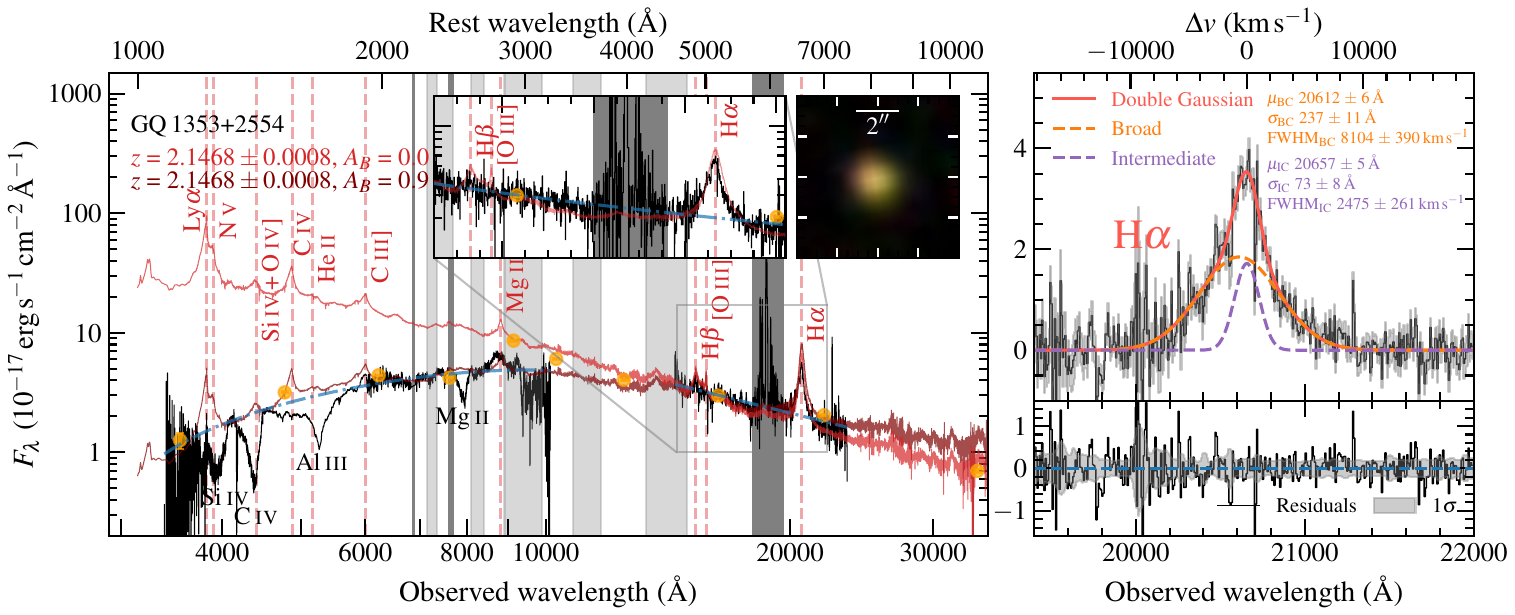}
\caption{Same as Fig.~\ref{fig:GQ0109-0719_z_full} but for GQ\,1353$+$2554, overplotted with $ugriz$ (SDSS), $YJHK$ (UKIDSS), and $W1$ (AllWISE) photometric bands. The \citetads{2016A&A...585A..87S} quasar template is overplotted in red ($A_B = 0.0$\,mag) and maroon ($A_B = 0.9$\,mag), assuming the Small-Magellanic-Cloud-like (SMC-like) extinction law. We can see that the observed optical data and the observed infrared data (especially the $W1$ measurement) cannot be fitted at the same time. We cannot find a proper fit with \citetads{2015A&A...584A.100Z} extinction curve as well. Hence, an extinction curve steeper than the \citetads{2015A&A...584A.100Z} extinction curve in the rest-frame UV is expected.
}
\label{fig:GQ1353+2554_z_full}
\end{figure*}

\begin{figure*}[th]
\centering
\includegraphics[width=17cm]{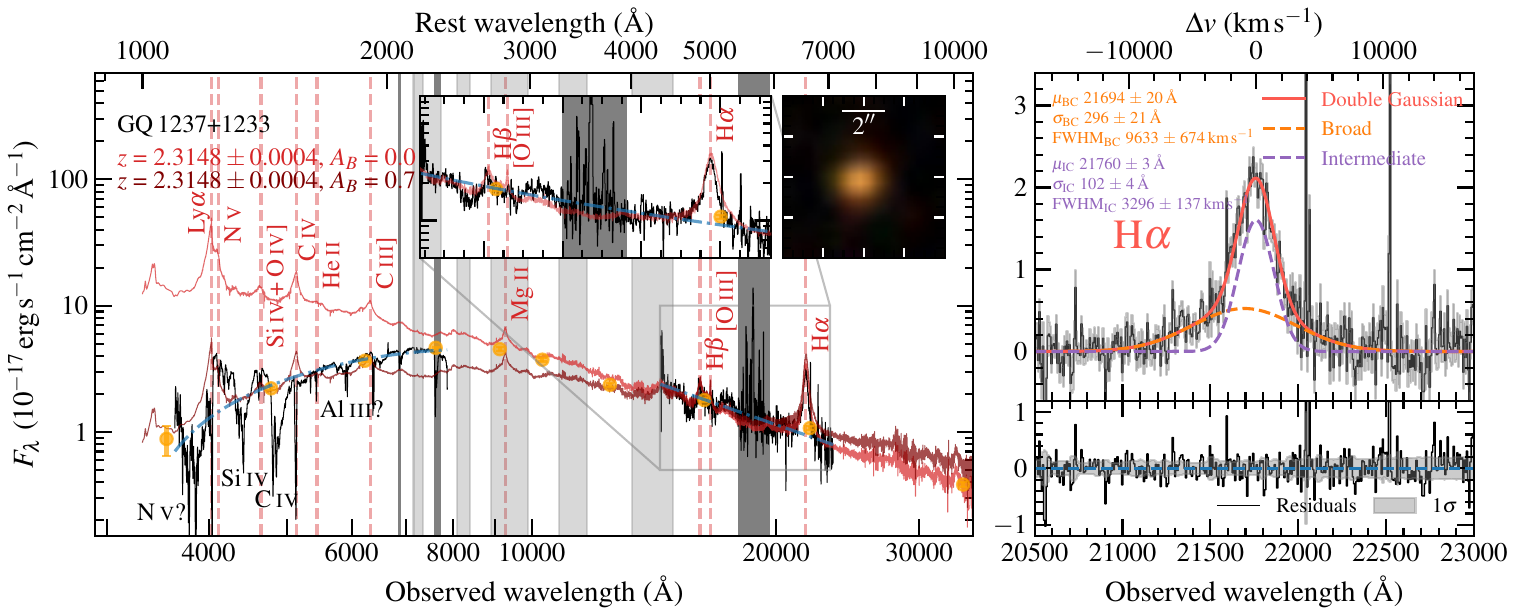}
\caption{Same as Fig.~\ref{fig:GQ0109-0719_z_full} but for GQ\,1237$+$1233. Photometric measurements from the $ugriz$ (SDSS), $YJHK$ (UKIDSS), and $W1$ (AllWISE) bands are shown as orange dots. Except for the $u$ band with an error of 0.34\,mag, the error bars are smaller than the size of the dots. Similar to GQ\,1353$+$2554, here we also expect a steeper extinction curve than the \citetads{2015A&A...584A.100Z} one to make the \citetads{2016A&A...585A..87S} template fit the full spectrum and photometric measurements.
}
\label{fig:GQ1237+1233_z_full}
\end{figure*}

\begin{figure*}[th]
\centering
\includegraphics[scale=0.8]{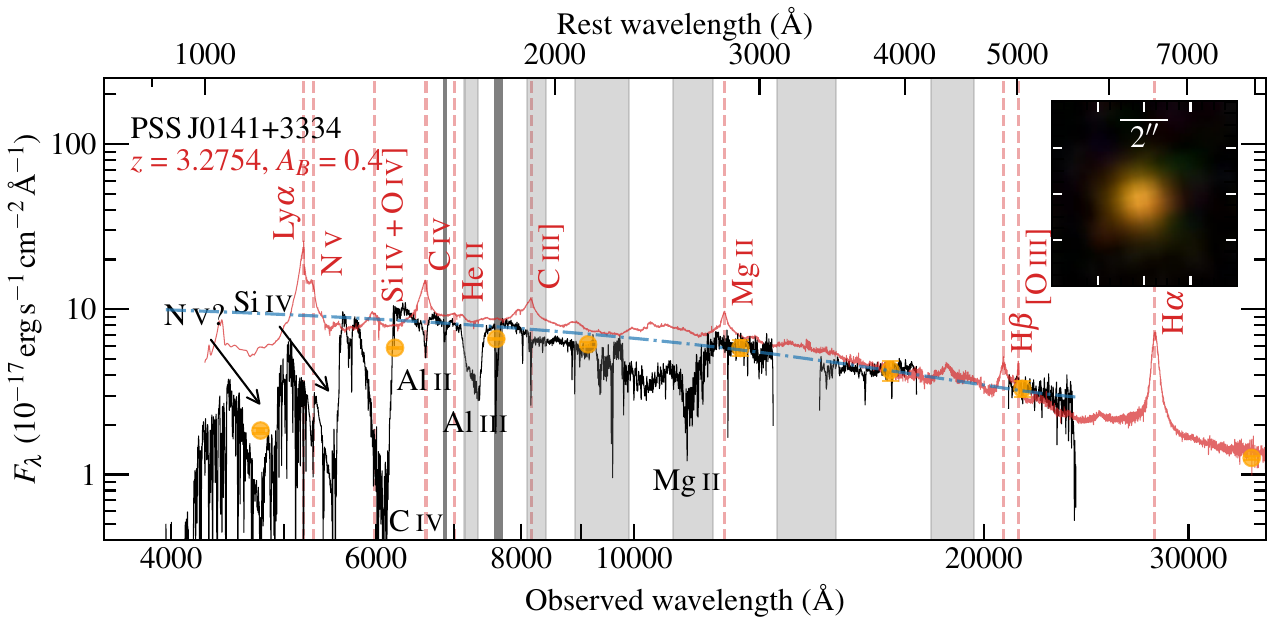}
\caption{Same as Fig.~\ref{fig:GQ0109-0719_z_full} but for PSS\,J0141$+$3334. ${\mathrm{H} \alpha}$ emission is not covered in the observed spectrum. The quasar template redshifted to $z=3.2754$ is plotted in red, based on our estimate for the lower limit of its systemic redshift. The $griz$, $JHK$, and $W1$ photometric results from the SDSS, 2MASS, and WISE surveys are displayed as orange points. The quasar composite is also set to be reddened by the SMC extinction curve with $A_B = 0.4$\,mag, which matches the overall continuum of the whole spectrum and most of the photometric data from blue to red.
}
\label{fig:PSSJ0141+3334_z_full}
\end{figure*}

\begin{figure*}[th]
\centering
\includegraphics[width=17cm]{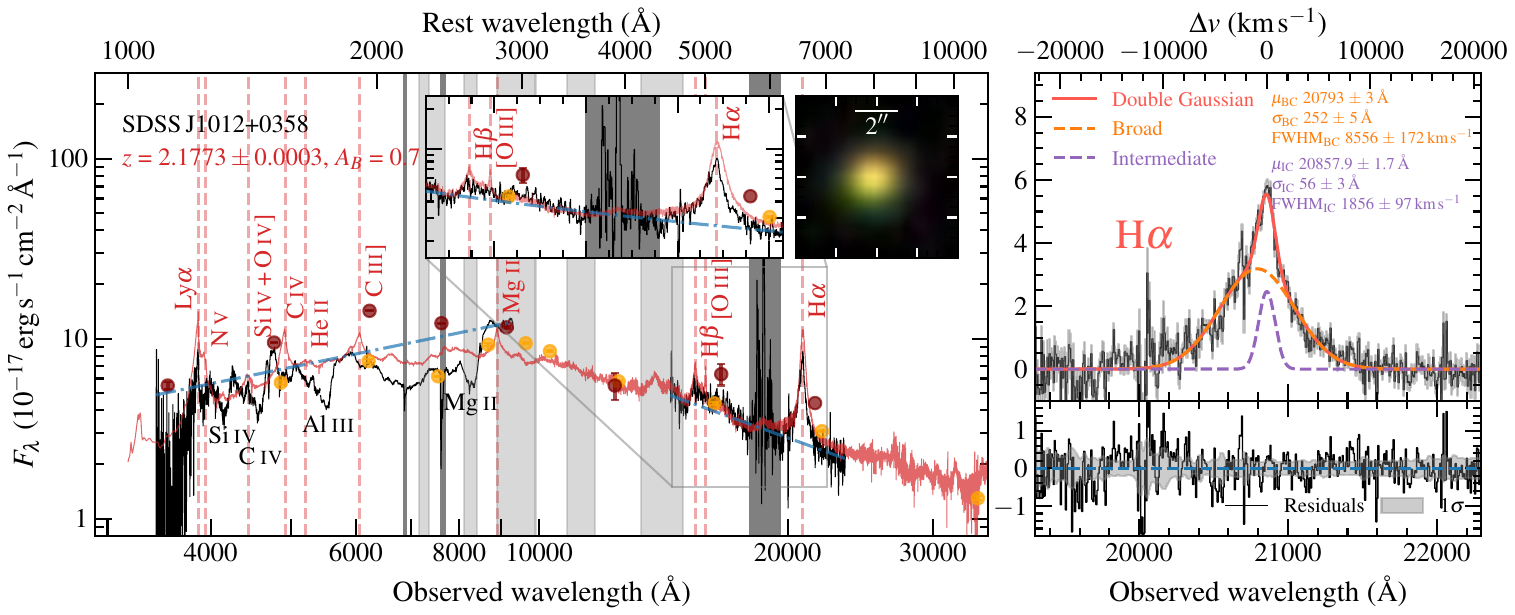}
\caption{Same as Fig.~\ref{fig:GQ0109-0719_z_full} but for SDSS\,J1012$+$0358, shown together with the more recent photometric data of $grizy_\mathrm{P1}$ (Pan-STARRS1), $YJHK$ (UKIDSS), and $W1$ (WISE) bands in orange, of which the observations were made in around 2010. The maroon dots represent the older $ugriz$ (SDSS) and $JHK$ (2MASS) photometric measurements from 2000. Note that there are no error bars for the $K$ band data from 2MASS since it is known to be a low-quality result. The photometric variability of this quasar is analysed in Sect.~\ref{sec:variability}. We estimated that this quasar is affected by a reddening of $A_B = 0.7$\,mag presuming the SMC-like extinction law to match the more recent photometry (in orange).
}
\label{fig:SDSSJ1012+0358_z_full}
\end{figure*}

\begin{figure*}[th]
\centering
\includegraphics[width=17cm]{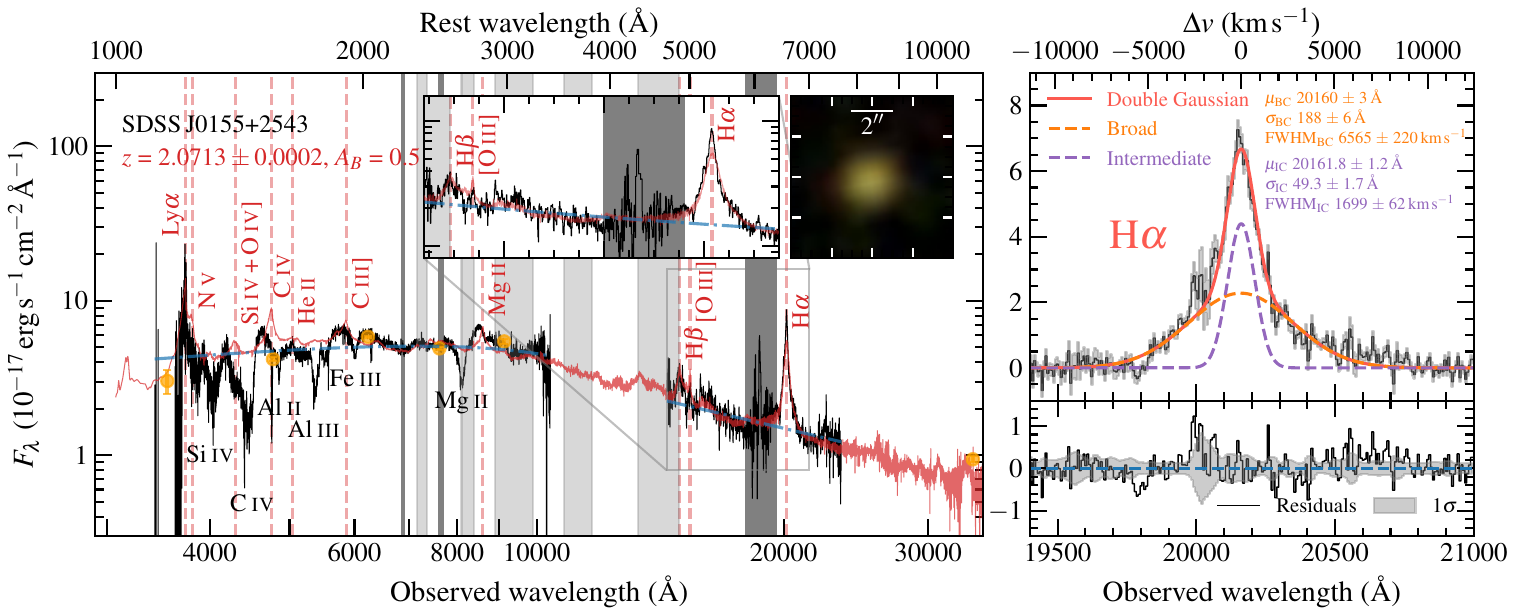}
\caption{Same as Fig.~\ref{fig:GQ0109-0719_z_full} but for SDSS\,J0155$+$2543, with $ugriz$ (SDSS) and $W1$ (AllWISE) magnitudes plotted on top as orange dots. This quasar is estimated to be reddened by $A_B = 0.5$\,mag applying an SMC extinction curve, fitting the quasar template to the continuum shape and the photometry.
}
\label{fig:SDSSJ0155+2543_z_full}
\end{figure*}

We fit ${\mathrm{H} \alpha}$ lines with multiple Gaussian components, following the line-based redshift estimation routine from previous works (e.g. \citeads{2010MNRAS.405.2302H}; \citeads{2011AJ....141..167R}; \citeads{2016ApJ...831....7S}). To perform an appropriate fit, as many Gaussian components as needed are included. The goodness of fit is illustrated by the residual plots and with the spectrum errors, shown in the bottom right panels of Figs.~\ref{fig:GQ0109-0719_z_full}--\ref{fig:SDSSJ0155+2543_z_full}.

For our sample, two broad Gaussian components (FWHM $\geqslant$ \SI{1000}{\km\per\s}) can fit all the ${\mathrm{H} \alpha}$ emission lines properly and hence there is no need to add an extra narrow component (FWHM < \SI{1000}{\km\per\s}). Besides, no narrow emission lines (such as {\OIII}\,$\lambda\lambda$4960.29, 5008.24, {\NII}\,$\lambda\lambda$6549.86, 6585.27, or {\SII}\,$\lambda\lambda$6718.29, 6732.67) can be found to constrain the potential narrow component of the ${\mathrm{H} \alpha}$ emission. Although narrow ${\mathrm{H} \alpha}$ components seem to exist in some selected quasars, such as SDSS\,J1012$+$0358 (see Fig.~\ref{fig:SDSSJ1012+0358_z_full}) and SDSS\,J0155$+$2543 (see Fig.~\ref{fig:SDSSJ0155+2543_z_full}), they are so weak that the results are virtually unaffected if we only consider the broad components. 

The peak of the potential narrow component is also located very closely to that of the fitted double-Gaussian model or the `intermediate component' (the narrower component of the fitted two), with a velocity shift around \SI{100}{\km\per\s} at most. \citetads{2016ApJ...831....7S} conclude that even though a strong, narrow component exists, the peak of the fitted model is still almost the same as that of the narrow component.

Either the peak of the multi-Gaussian model (\citeads{2016ApJ...831....7S}) or its centroid (\citeads{2010MNRAS.405.2302H}) is normally considered for redshift determination, while our decision was to compute the systemic redshifts for our targets using the peak of the intermediate component. This choice was made because the peak of the double-Gaussian model could potentially be influenced by the blueshifted `broad component' (the broader one between the two components), as seen in other lines (e.g. {\CIV}; \citeads{1992ApJS...79....1T}; \citeads{2002AJ....124....1R}), resulting in the ${\mathrm{H} \alpha}$ lines exhibiting blue wings (see Fig.~\ref{fig:SDSSJ1012+0358_z_full} for example). However, the velocity differences between the two peaks remain minor and they are less than \SI{50}{\km\per\s} for our sample.

{\it (ii)} Searching for spatially extended Ly${\alpha}$ emission.
The spatially extended Ly$\alpha$ emission is generally assumed to originate from the galactic or circumgalactic medium of quasars. It is used to determine the redshift of GQ\,1309$+$2904 by \citetaliasads{2020A&A...634A.111F}. Similarly, such emission is found in SDSS\,J1012$+$0358 among our new sample and can act as a validation of the systemic redshift estimated from the ${\mathrm{H} \alpha}$ line. For more details, see Sect.~\ref{sec:SDSSJ1012}.
\smallskip

{\it (iii)} Disentangling the onset of the Ly${\alpha}$ forest from the continuum and intrinsic absorption.
This has been done in \citetaliasads{2020A&A...634A.111F} to refute the redshift derived from the red edges of the absorption lines. Here we applied this method to PSS J0141+3334, where no obvious emission lines are detected, ${\mathrm{H} \alpha}$ is outside the spectral range, and no spatially extended Ly$\alpha$ emission is observed. By doing so, we can infer a lower limit of its systemic redshift and confirm the large blueshifts of the absorption troughs present in the spectrum, which is discussed in detail in Sect.~\ref{sec:PSSJ0141}.
\medskip

Figures~\ref{fig:GQ0109-0719_z_full}--\ref{fig:SDSSJ0155+2543_z_full} show the observed optical to near-infrared spectra of quasars in our sample, smoothed by a 3-pixel-wide boxcar filter. In the following, we describe how the systemic redshift of each quasar was determined and what spectral features were identified. The systemic redshifts are summarised in Table~\ref{tab:properties}.

\subsubsection{GQ\,1309$+$2904}

The redshift of GQ\,1309$+$2904 was determined in \citetaliasads{2020A&A...634A.111F} to be $z=2.66\pm0.01$ based on all three approaches mentioned above.  

\subsubsection{GQ\,0109$-$0719} \label{sec:GQ0109}

The full spectrum of GQ\,0109$-$0719 is shown in the left panel of Fig.~\ref{fig:GQ0109-0719_z_full}. The red edge of the \ion{C}{iv} absorption line has a wavelength of \SI{4587}{\AA} corresponding to a redshift of $z=1.96$. We determine the systemic redshift of this object to be $z=2.1164\pm0.0002$ via applying a double-Gaussian fit to the ${\mathrm{H} \alpha}$ emission line detected in the EMIR spectrum. 

The best-fit model is shown in the right panel of Fig.~\ref{fig:GQ0109-0719_z_full}. Some residuals of removed telluric absorption and sky emission plus bad pixels are present as narrow lines in the EMIR spectrum. The bump present in the blue wing of the ${\mathrm{H} \alpha}$ line is also an artefact, originating from the correction for the telluric $\ce{H2O}+\ce{CO2}$ band starting at \SI{2}{\um} (see \linkadspage{2015A&A...576A..77S}{4}{Fig.~1} of \citeads{2015A&A...576A..77S}). These artefacts are excluded in the fitting process but are not marked in Fig.~\ref{fig:GQ0109-0719_z_full} for visual clarity. 

One broad component with a full width at half maximum (FWHM) of \SI{7262(196)}{\km\per\s} at \SI{20417(5)}{\AA} and one intermediate component with a FWHM of \SI{2247(72)}{\km\per\s} at \SI{20457.8(1.5)}{\AA} are found to provide the best fit. The shown FWHM has been corrected from the fitted FWHM to remove the instrumental FWHM based on $\mathrm{FWHM} = \sqrt{{\mathrm{FWHM}_\mathrm{fit}}^2 - {\mathrm{FWHM}_\mathrm{inst}}^2}$. The $\mathrm{FWHM}_\mathrm{inst}$ was calculated from the EMIR/LR(HK) spectral resolution of $R = 987$ by $\mathrm{FWHM}_\mathrm{inst} = c/R = \SI{304}{\km\per\s}$, with $c$ being the speed of light. The correction was also applied to the rest of the sample.

With determining the systemic redshift from the intermediate component, the broad component is shifted by \SI{-594(82)}{\km\per\s}, where the negative sign indicates that it is blueshifted. Hereafter the velocity shift is calculated based on either the peak of the emission line or the trough of the absorption line compared with the corresponding line at the systemic redshift, unless specified.

The absorption feature bluewards of the expected {\MgII} emission is considered to be the {\MgII} BAL, although it is mostly covered by a telluric band. It is confirmed due to the common existence of {\MgII} BALs in our sample, especially because they are still present when they are not affected by the telluric bands, and the {\MgII} BAL in this quasar also spreads out of the telluric band.

All the BALs are greatly detached from the systemic redshift. Highly blueshifted {\SiIV}, {\CIV}, {\AlIII}, and {\MgII} BALs are identified, in line with our assumption, with the velocity offset being at least around \SI{-8000}{\km\per\s} (see the $v_{\mathrm{min}}$ in Table~\ref{tab:outflow_velocities}).

There are three additional absorption features at about \SI{3900}{\AA}, \SI{4870}{\AA}, and \SI{6026}{\AA} in the spectrum, which originate from \ion{C}{II}\,$\lambda$1335.31, \ion{Al}{II}\,$\lambda$1670.79, and \ion{Fe}{iii} UV 48\,$\lambda\lambda\lambda$2062.21, 2068.90, 2079.65. Similar features were found in objects of \citetads{2002ApJS..141..267H}. The three absorption lines are also quite broad and all have similar blueshifts to the other BALs in the spectrum (see Table~\ref{tab:outflow_velocities}). This object is an iron LoBAL quasar (FeLoBAL; \citeads{2002ApJS..141..267H}) based on the presence of excited iron absorption.

However, the ${\mathrm{H} \beta}$ line flux at the systemic redshift is unexpectedly low, with the REW of \SI{23(2)}{\AA} (see the zoom-in inset of Fig.~\ref{fig:GQ0109-0719_z_full} and Table~\ref{tab:properties}). The other quasars also have similar weak ${\mathrm{H} \beta}$ and a detailed discussion is given in Sect.~\ref{sec:smaller_dust}. {\HeII}\,$\lambda$1640.42 emission is missing in this quasar, while {\OIII} emission is marginally detected. Other common strong BELs, especially the high-ionisation ones, such as {\SiIV} and {\CIV}, remains elusive in this quasar and most of the other quasars in our sample. Considering the high-speed outflows indicated by the extremely blueshifted BALs, the high-ionisation emission lines may also be blueshifted and can be intrinsically weak (for a more in-depth discussion, see Sect.~\ref{sec:high-speed outflow}), though the possibility of them being absorbed by the BALs cannot be ruled out. We tentatively calculated a 3$\sigma$ upper limit of the {\CIV} REW to be $< \SI{0.6}{\AA}$ for this object, where we assume that the underlying {\CIV} emission line has the same width as that of the ${\mathrm{H} \alpha}$ broad component (\SI{7262}{\km\per\s}), integrated around the rest-frame wavelength of {\CIV}. According to \citetads{2025ApJ...994..213C}, GQ\,0109$-$0719 is a WLQ ($\mathrm{REW} < \SI{8.9(0.2)}{\AA}$).

The detection of these emission lines is complicated by the extensive amount of neighbouring broad and blueshifted troughs that can heavily intertwine with them. Therefore, it is difficult to determine whether the `peak' between the troughs signifies a genuine emission feature or simply a residual continuum. This also makes finding the optical continuum of most of our sample extremely hard. We cannot clearly identify any part in the shorter wavelength of the optical spectrum as the continuum, so we only show our best estimate for the continuum of each quasar as the dash-dotted blue line in Figs.~\ref{fig:GQ0109-0719_z_full}--\ref{fig:SDSSJ0155+2543_z_full}, with the assumption that the high-ionisation emission lines are intrinsically weak.

\subsubsection{GQ\,1353$+$2554} \label{sec:GQ1353}

\begin{figure}[th]
\centering
\resizebox{\hsize}{!}{\includegraphics[scale=0.43]{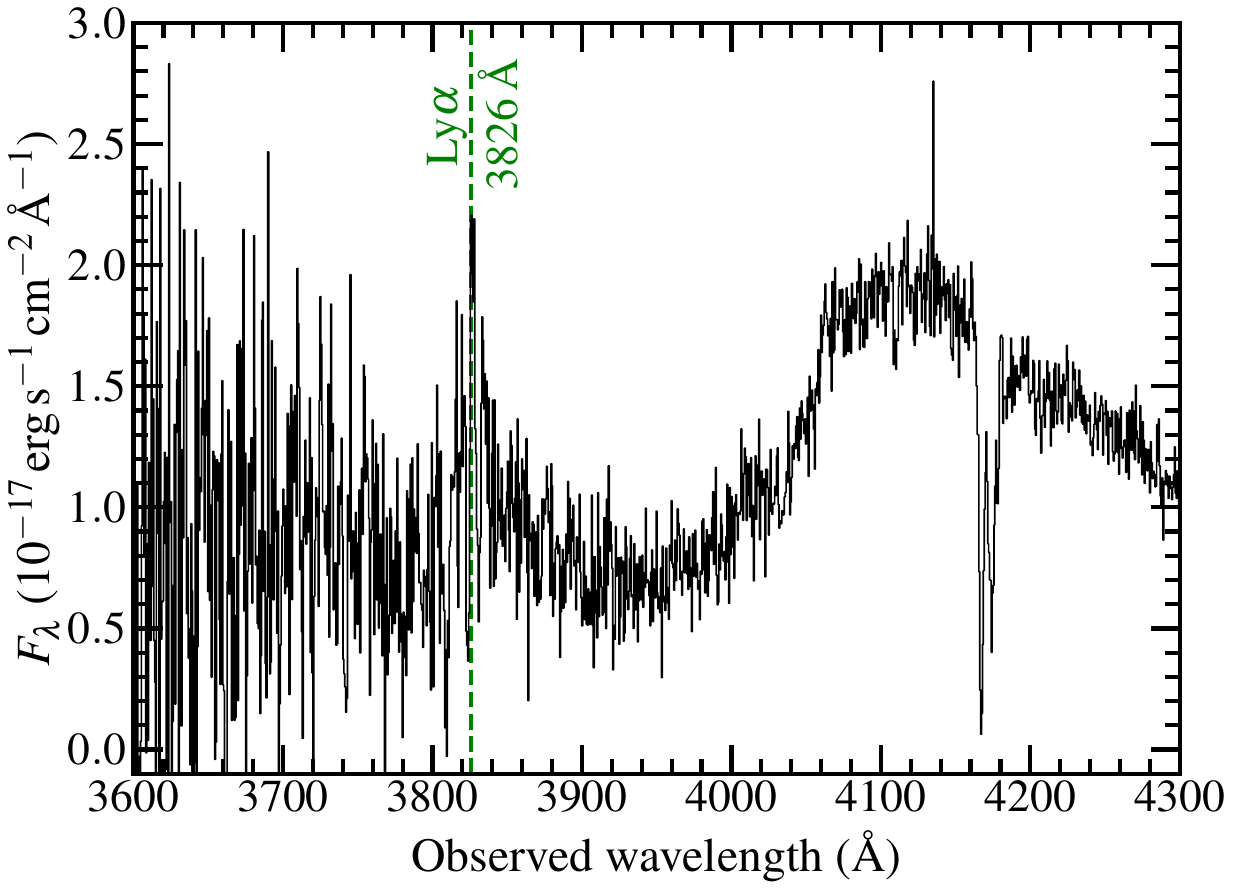}}
\caption{Identification of a weak BEL centred at \SI{3826}{\AA} as the Ly$\alpha$ emission line (marked with a vertical green line) for GQ\,1353$+$2554. It leads to a redshift of $z=2.1472$, further corroborating the redshift measurement $z=2.1468\pm0.0008$ based on ${\mathrm{H} \alpha}$ emission. The narrow absorption doublet around 4170 {\AA} is a \ion{C}{iv} absorber at $z=1.691$ for which we also detect \ion{Al}{ii}, \ion{Fe}{ii}, and \ion{Mg}{ii}. 
}
\label{fig:GQ1353+2554_z}
\end{figure}

The full spectrum of GQ\,1353$+$2554 is shown in the left panel of Fig.~\ref{fig:GQ1353+2554_z_full}. The red edge of the \ion{C}{iv} absorption trough has a wavelength of 4480 {\AA} corresponding to a redshift of $z=1.891$. In our spectra, we do not detect spatially extended Ly$\alpha$ emission. There are narrow absorption lines (NALs), which we can mostly match to an intervening absorber at $z=1.69$ (see Table~\ref{tab:abslines}). In the bluest spectrum there is an emission feature, which we interpret as Ly$\alpha$ emission corresponding to $z=2.1472$ (see Fig.~\ref{fig:GQ1353+2554_z}). The {\MgII} emission line seems to be present in the spectrum as well, but with its relatively weak nature and its proximity to the telluric absorption band, we cannot confirm its detection.

We subsequently managed to secure a near-IR spectrum in which we detect the ${\mathrm{H} \alpha}$ emission line (see Fig.~\ref{fig:GQ1353+2554_z_full}) at $z=2.1468\pm0.0008$, derived from a double-Gaussian model fit. We fit the ${\mathrm{H} \alpha}$ line profile with one broad component with a FWHM of \SI{8104(390)}{\km\per\s} at \SI{20612(6)}{\AA} and one intermediate component with a FWHM of \SI{2475(261)}{\km\per\s} at \SI{20657(5)}{\AA}. According to the systemic redshift derived from the intermediate component, the broad component is shifted by \SI{-665(113)}{\km\per\s}.

Extremely weak ${\mathrm{H} \beta}$ emission is found in this object ($\mathrm{REW} = \SI{13(4)}{\AA}$) and so is the {\HeII} emission. The expected {\CIV} emission line at the systemic redshift is also absent, and we estimated the 3$\sigma$ upper limit of the {\CIV} REW to be $< \SI{0.5}{\AA}$, assuming the FWHM of \SI{8104}{\km\per\s} from the ${\mathrm{H} \alpha}$ broad component. GQ\,1353$+$2554 is a WLQ based on the definition from \citetads{2025ApJ...994..213C}.

{\SiIV}, {\CIV}, {\AlIII}, and {\MgII} BALs are measured to be the most blueshifted absorption troughs in our sample (see Fig.~\ref{fig:velocities} and Table~\ref{tab:outflow_velocities}). The potential rest-frame UV emission lines are also weaker than those in other quasars, consistent with what we discuss in Sect.~\ref{sec:high-speed outflow}. We lack confirmation of the {\AlII} and {\FeIII} UV 48 absorption lines in this quasar, despite tentative indications of their presence in the spectrum.

\subsubsection{GQ\,1237$+$1233} \label{sec:GQ1237}

The full spectrum of GQ\,1237$+$1233 is shown in Fig.~\ref{fig:GQ1237+1233_z_full} and it displays no obvious rest-frame UV emission lines as well. There are several NALs: most of them are from an intervening absorber at $z=1.28$; the rest come from a system at $z = 2.30$ (see Table~\ref{tab:abslines}). We detect a strong ${\mathrm{H} \alpha}$ line in the near-IR. Based on the best-fit double-Gaussian model, we identified an intermediate component at \SI{21760(3)}{\AA} (see the right panel of Fig.~\ref{fig:GQ1237+1233_z_full}) and determined a systemic redshift of $z=2.3148\pm0.0004$.

The intermediate component of the ${\mathrm{H} \alpha}$ line is much stronger than the broad component, shown in the right panel of Fig.~\ref{fig:GQ1237+1233_z_full}. The intermediate component is likely to be the main component of GQ\,1237+1233, different from other quasars in our sample. The FWHM of the intermediate component is measured to be \SI{3296(137)}{\km\per\s} and the broad component is fitted to be at \SI{21694(20)}{\AA} with a FWHM of \SI{9633(674)}{\km\per\s}, offset by \SI{-918(280)}{\km\per\s} from the intermediate one. Weak ${\mathrm{H} \beta}$ emission, with REW of \SI{15(5)}{\AA}, is also detected and {\HeII} emission is missing. Non-detected {\CIV} emission is estimated to have a 3$\sigma$ upper limit of $< \SI{0.7}{\AA}$, assumed to have a width of \SI{9633}{\km\per\s} from the ${\mathrm{H} \alpha}$ broad component. This object is also a WLQ according to \citetads{2025ApJ...994..213C}.

The BAL troughs of this quasar are less blueshifted than other quasars in our sample and the {\CIV} BAL has the lowest ejection velocity $v_{\mathrm{min}}$ of \SI{-5000}{\km\per\s}, as shown in Table~\ref{tab:outflow_velocities} and Fig.~\ref{fig:velocities}, and the {\SiIV} BAL has the $v_{\mathrm{min}}$ of \SI{-10000}{\km\per\s}. The velocity profile of this quasar is also distinctive from others. It shows a narrower blue wing than the red wing, which is the inverse of BAL profiles of other quasars (see Sect.~\ref{sec:velocities}).

Only blueshifted {\SiIV} and {\CIV} BALs are certainly identified in this object, while the {\AlIII} BAL is relatively vague as seen in Fig.~\ref{fig:velocities}, which may indicate lower column densities of the BAL clouds, different from other quasars in our sample. It is uncertain whether the {\NV} BAL that is also observed in other quasars (e.g. \citeads{1991ApJ...373...23W}) exists. Although the tentative {\NV} BAL arises roughly at the same speed as the {\CIV} BAL, it extends to higher velocities and show a different profile. {\AlII} and {\FeIII} UV 48 absorption lines are not detected. {\MgII} is not covered in our observed spectrum because it has been redshifted out of the OSIRIS/R1000B spectral range. Thus, there is only weak evidence to conclude that GQ\,1237$+$1233 is a LoBAL quasar but not a high-ionisation BAL (HiBAL) quasar, based on the really weak {\AlIII} BAL.

\subsubsection{PSS\,J0141$+$3334} \label{sec:PSSJ0141}

\begin{figure*}[th]
\centering
\includegraphics[width=17cm]{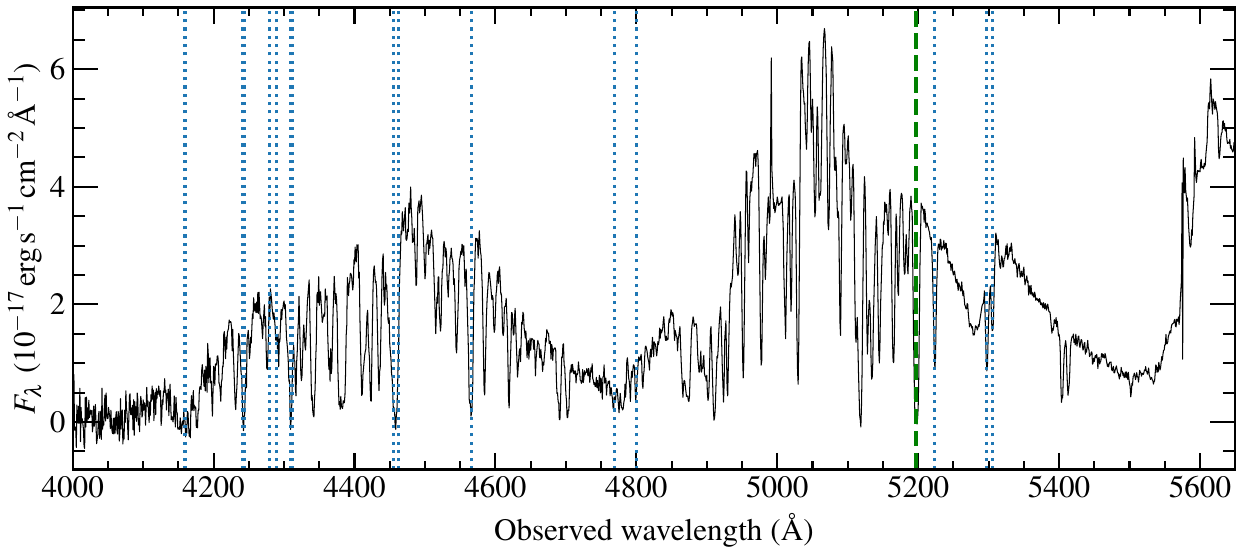}
\caption{Onset of the Ly$\alpha$ forest in the spectrum of PSS\,J0141+3334. The dotted blue lines are associated with an intervening DLA at $z=2.4215$. The dashed green line marks the wavelength from which we estimate the Ly$\alpha$ forest begins corresponding to $z=3.2754$, which we consider a rough estimate of the redshift of PSS\,J0141+3334, or at least a robust lower limit on its redshift.
}
\label{fig:PSSJ0141+3334_forest}
\end{figure*}

The full spectrum of PSS\,J0141$+$3334 is shown in Fig.~\ref{fig:PSSJ0141+3334_z_full}. \citetads{2003AJ....126...53B} determined the redshift of this source to be $z=3.005\pm0.005$ based on the red edges of BAL absorption troughs. At this redshift ${\mathrm{H} \alpha}$ is shifted out of the $K$ band. In our near-IR spectra covering the $J$, $H$, and $K$ bands we do not detect any emission lines. There is also no extended Ly$\alpha$ emission line in the spectrum covering the expected position of Ly$\alpha$. Instead, we can only estimate the redshift using the onset of the Ly$\alpha$ forest, as shown in Fig.~\ref{fig:PSSJ0141+3334_forest}. To do this we first identified all the metal line systems in the spectrum. Information about the identified lines can be found in Table~\ref{tab:abslines}. Most of the narrow metal lines are from an intervening DLA with a hydrogen column density of $\log_{10} ({N_{\ion{H}{i}}/\mathrm{cm^{-2}})} = 20.5$ at $z=2.4215$. We find that the Ly$\alpha$ forest most likely starts at \SI{5197.5}{\AA}, corresponding to $z \gtrsim 3.2754$.

We overplotted the \citetads{2016A&A...585A..87S} template spectrum redshifted to $z = 3.2754$ on the observed spectrum in Fig.~\ref{fig:PSSJ0141+3334_z_full}. With setting $z = 3.2754$ as a tentative redshift estimate or at least a lower limit of the systemic redshift, the BALs ({\NV}, {\SiIV}, {\CIV}, {\AlII}, and {\MgII}) in the spectrum are all highly blueshifted, although we cannot accurately determine the outflow velocity for this quasar and the detection of {\NV} BAL is a bit uncertain. {\FeIII} UV 48 absorption is not found in this object. The additional absorption around the blue wing of the {\MgII} trough could originate from a broad, blended, but shallow {\FeII} opacity as seen in overlapping-trough FeLoBALs (\citeads{2002ApJS..141..267H}; \citeads{2022ApJ...937...74C}).

\subsubsection{SDSS\,J1012$+$0358} \label{sec:SDSSJ1012}
\begin{figure}[th]
\centering
\resizebox{\hsize}{!}{\includegraphics[scale=0.44]{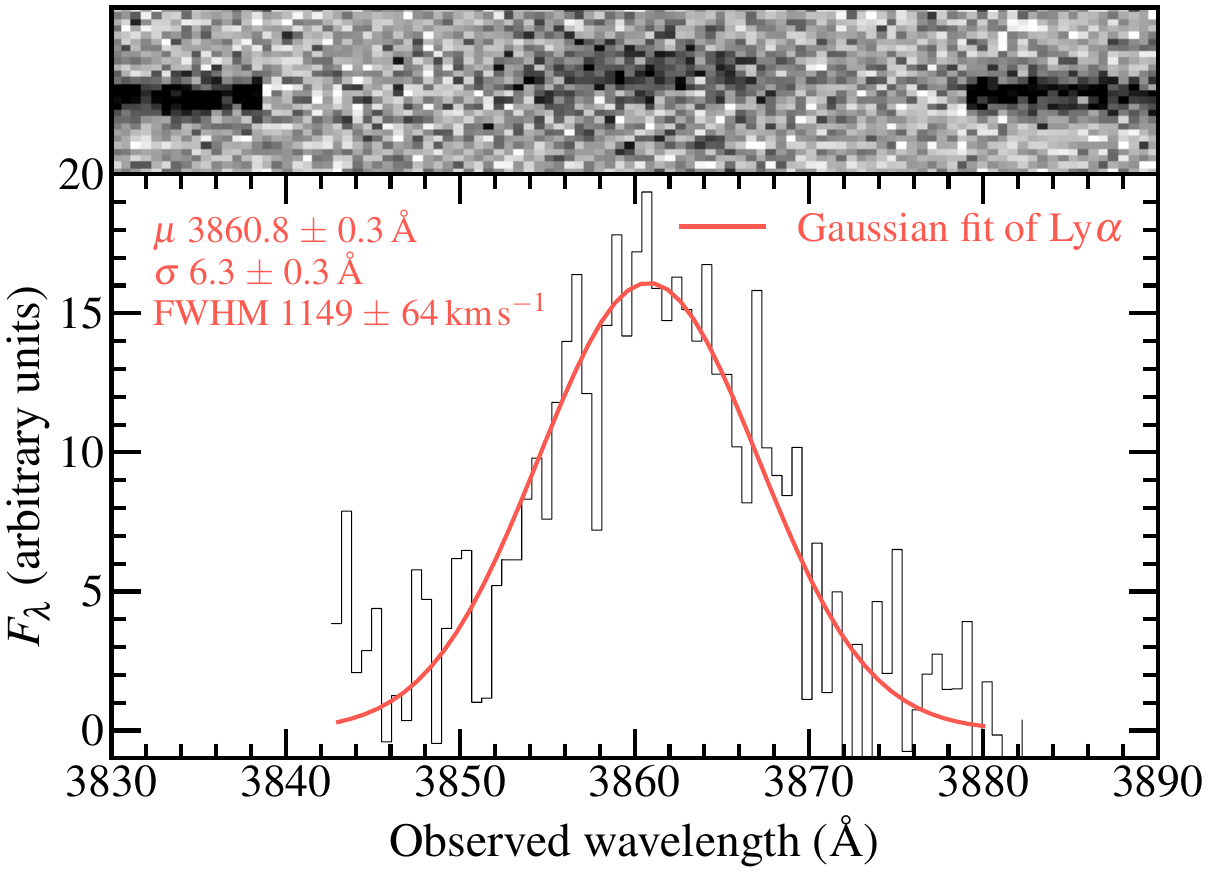}}
\caption{Extended Ly$\alpha$ emission line of SDSS\,J1012$+$0358 centred at \SI{3860.8(0.3)}{\AA}. \textit{Top}: 2D spectrum (after subtraction of the quasar spectral point-spread function). \textit{Bottom}: Extracted 1D spectrum. The central wavelength of the line corresponds to a redshift of $z=2.1759 \pm 0.0003$, indicating a velocity shift of about \SI{-100}{\km\per\s} based on the redshift from the ${\mathrm{H} \alpha}$ line. The width (FWHM) of the line is \SI{1149(64)}{\km\per\s}. The red curve in the bottom panel shows the Gaussian fit to the emission line. In the SDSS quasar catalogue, the redshift is given as $z=2.1000$ \citepads{2020ApJS..250....8L}. 
}
\label{fig:SDSS1012+0358_z}
\end{figure}

The full spectrum of SDSS\,J1012$+$0358 is shown in Fig.~\ref{fig:SDSSJ1012+0358_z_full}. For this quasar, we can determine the systemic redshift both from a broad ${\mathrm{H} \alpha}$ emission line and a spatially extended Ly$\alpha$ emission line. \citetads{2020ApJS..250....8L} estimated its systemic redshift to be $z=2.1000$ from visual inspection of the SDSS spectrum. Figure~\ref{fig:SDSS1012+0358_z} shows the spectrum of the extended Ly$\alpha$ line, which we extracted in the following way. First, we subtracted the spectrum of the quasar using the method described in \citetads{2000Msngr..99...31M}. Then we extracted the spectrum of the extended emission itself by using the spatial profile of the emission as weight following \citetads{1986PASP...98..609H}. 

The extended Ly$\alpha$ emission line shape is consistent with a Gaussian profile, and the best-fit Gaussian model shows that the Ly$\alpha$ line peaks at \SI{3860.8(0.3)}{\AA}, corresponding to a redshift of $z=2.1759 \pm 0.0003$, with a FWHM of \SI{1149(64)}{\km\per\s}. Velocity widths of \num{1000}--\SI{1500}{\km\per\s} are typical in spatially extended Ly$\alpha$ emitting halos around regular broad-emission line radio-loud quasars \citepads{1991ApJ...381..373H}, whereas extended halos around radio-quiet quasars were found to have smaller FWHM \citepads{2006A&A...459..717C}, although no discernible difference in the halo kinematics was seen in a larger sample of luminous quasars at $z\sim3$ \citepads{2019MNRAS.482.3162A}.

In radio-loud quasars, the Ly$\alpha$ emitting halos are highly asymmetric and aligned with their radio-jets \citepads{1991ApJ...381..373H}. The spatially extended emission illustrated in the 2D spectrum shown in Fig.~\ref{fig:SDSS1012+0358_z} appears mostly to one side of the quasar emission trace, and can be fitted well by a Gaussian profile in the spatial direction with a FWHM of $11\pm1$\,pixels. With a pixel size of 0.254\,arcsec/pixel in binned mode, the spatial FWHM is $2.8\pm0.3$\,arcsec, corresponding to $23\pm2$\,kpc at the redshift of the quasar. Rather than a Gaussian profile in the spatial direction, the radial surface brightness distribution is often fitted by an exponential profile \( \mathrm{SB}(r) \propto \exp(-r/r_h) \). The exponential scale-length, $r_h$, is in this case $14.2\pm2.3$\,kpc, similar to that seen in other bright Ly$\alpha$ emitting halos \citepads{2019MNRAS.482.3162A}.

We later secured a near-IR spectrum (see Fig.~\ref{fig:SDSSJ1012+0358_z_full}), in which we determined the systemic redshift to be $z=2.1773\pm0.0003$ based on the ${\mathrm{H} \alpha}$ emission line. The ${\mathrm{H} \alpha}$ systemic redshift is consistent with the measurement from the extended Ly$\alpha$ emission line, $z=2.1759 \pm 0.0003$, indicating a minor shift around \SI{-100}{\km\per\s} of the extended Ly$\alpha$ emission. 

In the best-fit double-Gaussian model, an intermediate component with a FWHM of \SI{1856(97)}{\km\per\s} is obtained, along with a broad component with a FWHM of \SI{8556(172)}{\km\per\s}, blueshifted by \SI{928(52)}{\km\per\s}. The large blueshifts of {\SiIV}, {\CIV}, {\AlIII}, and {\MgII} are again validated based on the systemic redshift. Besides, it is not clear whether the {\AlII} and {\FeIII} UV 48 absorption lines are present in this quasar. As stated in Sect.~\ref{sec:PSSJ0141}, the blue part of the {\MgII} trough of this quasar may also be affected by the blended {\FeII} troughs.

Weak {\CIV}, {\HeII}, and ${\mathrm{H} \beta}$ emission lines are observed in this object. Using the quasar redshift and estimated continuum, we measured the REW of the {\CIV} emission to be \SI{9.4(0.9)}{\AA} (hence $\log_{10} \mathrm{({\CIV}~REW/\AA)} = \num{0.97(0.04)}$) and the {\CIV} emission peak is blueshifted by \SI{10640(63)}{\km\per\s}. This quasar is classified as a bridge quasar based on the criterion ($\num{8.9} < \mathrm{{\CIV}\ REW}< \SI{19.3}{\AA}$) from \citetads{2025ApJ...994..213C}, representing the transition from WLQs to normal quasars. {\HeII} emission of this quasar is extremely weak with REW of \SI{0.3(0.2)}{\AA} and $\log_{10} \mathrm{({\HeII}~REW/\AA)}$ of \num{-0.6(0.3)}, calculated by fitting the local continuum. ${\mathrm{H} \beta}$ emission has a REW of \SI{24(3)}{\AA}. We give a summary of all the measured values in Table~\ref{tab:properties}.

\subsubsection{SDSS\,J0155$+$2543} \label{sect:SDSS0155}
\begin{figure}[t!]
\centering
\resizebox{\hsize}{!}{\includegraphics[scale=0.43]{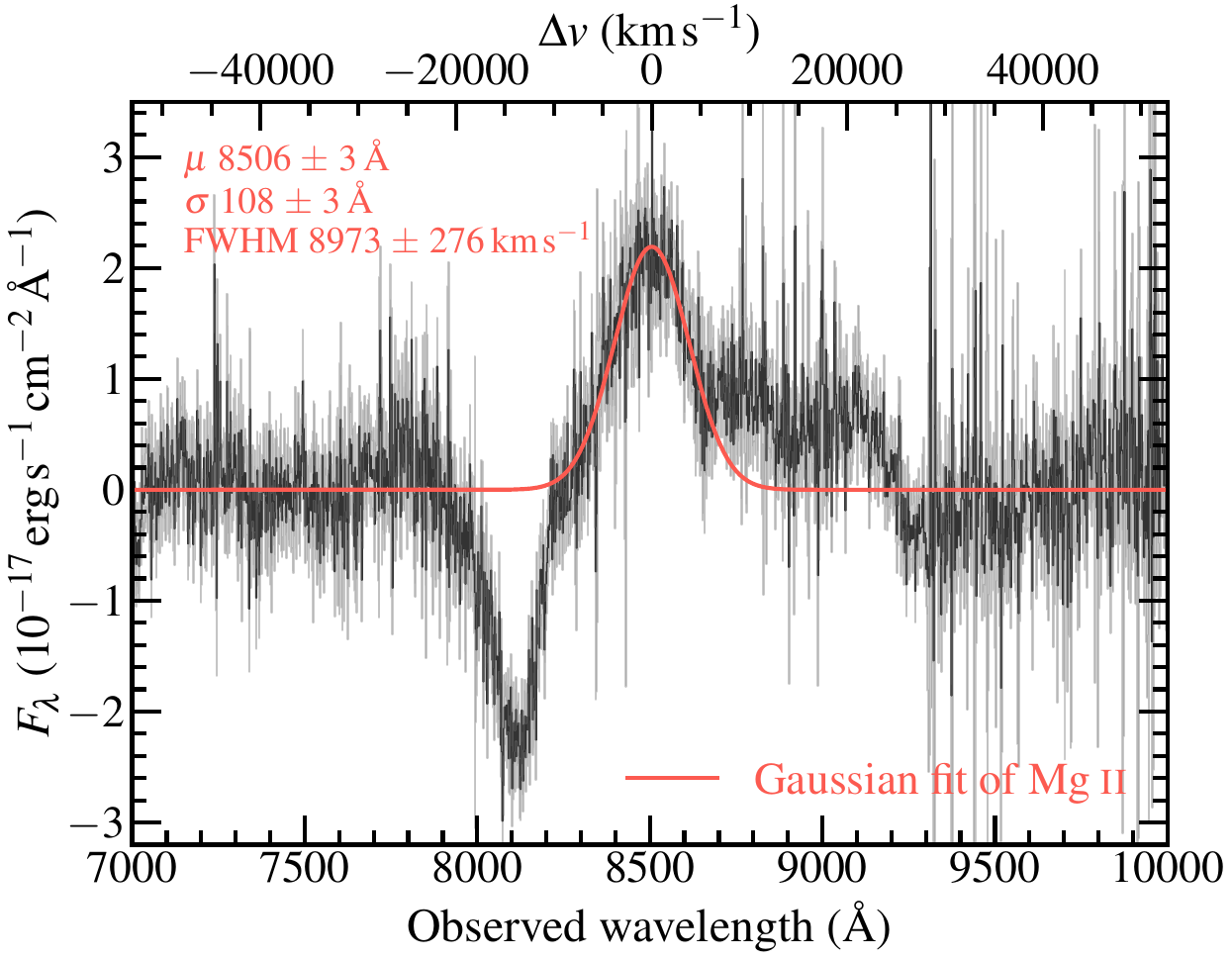}}
\caption{Potential {\MgII} emission line of SDSS\,J0155$+$2543 is centred at \SI{8506(3)}{\AA}, corresponding to $z=2.0387\pm0.0012$. The black line shows the SDSS spectrum around the {\MgII} line with contamination from {\FeII} emission. The red line is a Gaussian fit to the {\MgII} emission, with the shaded grey areas representing the 1$\sigma$ uncertainties. The width of the line is \SI{8973(276)}{\km\per\s}. In the SDSS quasar catalogue, the redshift is determined to be $z = 2.0300$. From the ${\mathrm{H} \alpha}$ line in the near-infrared spectrum, we infer a higher redshift of $z=2.0713\pm0.0002$ meaning that the {\MgII} emission line must be shifted by \SI{-3202(117)}{\km\per\s}.
}
\label{fig:SDSS0155+2543_z}
\end{figure}

The full spectrum of SDSS\,J0155$+$2543 is shown in Fig.~\ref{fig:SDSSJ0155+2543_z_full}. We first only relied on the optical spectrum from the SDSS database. It seems that there is a \ion{Mg}{ii} emission line detected in the SDSS spectrum corresponding to $z=2.0387\pm0.0012$ (see Fig.~\ref{fig:SDSS0155+2543_z}). We fitted a Gaussian model to the {\MgII} line for a rough inspection and the blended {\FeII} emission is not considered during the fit. 

To verify this redshift, we obtained a near-IR spectrum covering the position of ${\mathrm{H} \alpha}$ emission. We detect an ${\mathrm{H} \alpha}$ line with a FWHM of \SI{2310(85)}{\km\per\s}, which is narrower than that of the other quasars in the sample (see Table~\ref{tab:properties}). We inferred a higher redshift of $z=2.0713\pm0.0002$ from the best-fit double-Gaussian model and a fitted intermediate component at \SI{20161.8(1.2)}{\AA} with a FWHM of \SI{1699(62)}{\km\per\s}. A broad component is found at \SI{20160(3)}{\AA} with a FWHM of \SI{6565(220)}{\km\per\s}, offset by \SI{-28(54)}{\km\per\s}, which is the smallest velocity shift between the two components of ${\mathrm{H} \alpha}$ emission lines among our sample. 

Given that the systemic redshift from the ${\mathrm{H} \alpha}$ is $z=2.0713\pm0.0002$, the potential \ion{Mg}{ii} emission line is offset by \SI{-3202(117)}{\km\per\s}. This is not unheard of, as \citetads{2016MNRAS.459.2472R} report large {\MgII} blueshifts of over a thousand \si{\km\per\s} (up to \num{2800}--\SI{3800}{\km\per\s}) in FeLoBAL quasars and they attribute the large velocity offset between the {\MgII} line and the ${\mathrm{H} \beta}$ line to outflowing wind. {\MgII} blueshift relative to ${\mathrm{H} \beta}$ ($\sim \SI{2000}{\km\per\s}$) is also found in a changing WLQ by \citetads{2022ApJ...930....5Y}. 

Blueshifted {\SiIV}, {\CIV}, and {\CIII} emission lines are present in the spectrum as well, while the potential Ly$\alpha$ and {\OIII} emission features are located much closer to the systemic redshift. In more detail, the {\CIV} emission line is blueshifted by \SI{8546(171)}{\km\per\s} and has a $\log_{10} \mathrm{({\CIV}~REW/\AA)}$ of \num{1.0(0.2)} ($\mathrm{REW} = \SI{11(4)}{\AA}$). This object is also a bridge quasar according to the definition by \citetads{2025ApJ...994..213C}. A slightly noticeable {\HeII} emission line with REW of \SI{3(2)}{\AA} and $\log_{10} \mathrm{({\HeII}~REW/\AA)}$ of \num{0.5(0.2)} is also identified. The ${\mathrm{H} \beta}$ emission line is also found to be the most prominent in our sample, with REW of \SI{39(5)}{\AA}. Highly blueshifted {\SiIV}, {\CIV}, {\AlII}, {\AlIII}, {\FeIII}, and {\MgII} BALs are also confirmed in this object, making this object a FeLoBAL quasar.

\subsection{Black hole masses and Eddington ratios} \label{sec:blackhole_mass}
\citetads{2002ApJ...571..733V} was the first to show that the single-epoch AGN spectrum can be used to estimate the mass of the central black hole ($M_{\mathrm{BH}}$) in AGNs, based on the measurement of optical continuum strength and broad ${\mathrm{H} \beta}$ line width. \citetads{2005ApJ...630..122G} later extended the empirical relationship to use the broad ${\mathrm{H} \alpha}$ emission line for the estimation of $M_{\mathrm{BH}}$. With the detection of ${\mathrm{H} \alpha}$ lines in our quasars, $M_{\mathrm{BH}}$ of these targets can be determined from the following ${\mathrm{H} \alpha}$-based $M_{\mathrm{BH}}$ estimator:
\begin{equation}
\begin{aligned}[b]
M_{\mathrm{BH}}= &(2.67^{+0.5}_{-0.4}) \times 10^6 \\
& \times \left(\frac{L_{\mathrm{H} \alpha}}{10^{42} \mathrm{~erg} \mathrm{~s}^{-1}}\right)^{0.55 \pm 0.02}\left(\frac{\mathrm{FWHM}_{\mathrm{H} \alpha}}{10^3 \mathrm{~km} \mathrm{~s}^{-1}}\right)^{2.06 \pm 0.06} M_{\odot}\,,
\end{aligned}
\label{eq:black_hole_mass}
\end{equation}
which originates from \citetads{2005ApJ...630..122G} and is revised by \citetads{2022ApJS..261....5M} with substituting $f = 1$ for the virial factor ($f$). $f$ is used to convert the observed FWHM of emission lines into gas velocities in the BLR in three dimensions (\citeads{2000ApJ...533..631K}; \citeads{2003ApJ...583L...5N}), as shown by $M_{\mathrm{BH}}=G^{-1} R_{\mathrm{BLR}} v_{\mathrm{BLR}}^2 \propto f L_\lambda^\alpha \mathrm{FWHM}^2$ from \citetads{2022ApJS..261....5M}. 

The virial factor, $f$, is closely related to the geometry, kinematics, and orientation of the clouds in the BLR \citepads{2014ApJ...789...17H}. Instead of using $f = 0.75$ from \citetads{2005ApJ...630..122G}, which assumes a randomly oriented orbits of the BLR clouds and a spherical distribution of them (\citeads{2004MNRAS.352.1390M}; \citeads{2005ApJ...630..122G}), $f = 1$ is a more widely used virial factor (e.g. \citeads{2004MNRAS.352.1390M}; \citeads{2021ApJ...920....9L}; \citeads{2022ApJS..261....5M}). \citetads{2004MNRAS.352.1390M} concluded that it is a good choice no matter whether the BLR orbits are random (spherical) or flattened (disc-like). It is also consistent within the tolerance of the derived average virial factor ($f_{\mathrm{FWHM}} = 1.12^{+0.36}_{-0.27}$) from \citetads{2015ApJ...801...38W} based on the observational results.

We only fitted two broad Gaussian components to the ${\mathrm{H} \alpha}$ emission line, although the common approach is to deblend the contaminating narrow components from the entire profile because the virial $M_{\mathrm{BH}}$ estimation only relies on the information from the BLR clouds (\citeads{2005ApJ...630..122G}; \citeads{2022ApJS..261....5M}). However, the narrow components are mostly absent or very weak in our sample, having negligible impacts on the derived values of $L_{\mathrm{H} \alpha}$ and $\mathrm{FWHM}_{\mathrm{H} \alpha}$. There is no need to fit an extra narrow component apart from the two broad components, so we used the luminosity and FWHM of the entire fitted double-Gaussian model to determine the $M_{\mathrm{BH}}$ of our sample. This approach is similar to those taken by \citetads{2005ApJ...630..122G}, \citetads{2008ApJ...673..703M}, and \citetads{2020MNRAS.498.1469W}. Note that when converting the FWHM of the combined Gaussians to be in the unit of velocity (\si{\km\per\s}), we set the peak of the double-Gaussian model as the reference. According to the redshifts and fitted parameters of our sample from Sect.~\ref{sect:redshifts}, the intrinsic $\mathrm{FWHM}_{\mathrm{H} \alpha}$ that has been corrected for the instrumental FWHM (\SI{304}{\km\per\s}, see Sect.~\ref{sec:GQ0109}) and estimates of $M_{\mathrm{BH}}$ are shown in Table~\ref{tab:properties}.

The quasar bolometric luminosity ($L_{\mathrm{bol}}$) can be estimated from total continuum luminosity, $\lambda L_{\lambda}$, at \SI{5100}{\AA} (hereafter $L_{5100}$) since our spectra cover the continuum around \SI{5100}{\AA} in the rest frame. Following \citetads{2008ApJ...673..703M}, whose objects are much less luminous, we included all the luminosity acquired within the spectroscopic aperture in the calculation and did not remove the minor host galaxy contamination if it exits. After calculating Eddington luminosity ($L_{\mathrm{Edd}}$) from $M_{\mathrm{BH}}$, we can derive the Eddington ratios ($L_{\mathrm{bol}} / L_{\mathrm{Edd}}$) for our quasars, according to Eq.~\eqref{eq:eddington_ratio} from \citetads{2007ApJ...654..754N}:
\begin{equation}
L_{\mathrm{bol}} / L_{\mathrm{Edd}}=\frac{f_L L_{5100}}{1.5 \times 10^{38}\left[M_{\mathrm{BH}} / M_{\odot}\right]}\,.
\label{eq:eddington_ratio}
\end{equation}

To calculate $L_{5100}$, we followed the strategy in \citetads{2008ApJ...673..703M} starting from the mean flux between \SI{5050}{\AA} and \SI{5150}{\AA} in the rest frame. Besides, $M_{\mathrm{BH}}$ was set to be the BH mass determined from the ${\mathrm{H} \alpha}$ fit. $f_L$ in Eq.~\eqref{eq:eddington_ratio} is the bolometric correction factor ($L_{\mathrm{bol}} = {f_L L_{5100}}$) and we made the factor to be $f_L = 7$, adopted from \citetads{2007ApJ...654..754N}. The factor depends particularly on the luminosity of the intrinsic extreme UV continuum, which is thought to range from 5 to 10 and be larger in AGNs with a lower luminosity (\citeads{2007ApJ...654..754N}), and $f_L = 7$ is approximately the mean. The obtained Eddington ratios of our targets are listed in Table~\ref{tab:properties}.

We note that we chose not to correct for the intrinsic dust extinction so the $M_{\mathrm{BH}}$ and $L_{\mathrm{bol}} / L_{\mathrm{Edd}}$ can be underestimated. $L_{5100}$ is affected more by the dust extinction due to its shorter wavelength than that of ${\mathrm{H} \alpha}$ emission. The reason for the choice is that not all the quasars in our sample can be fitted well by commonly used extinction curves as we discuss in the following. Meanwhile, the dust extinction effect on the $M_{\mathrm{BH}}$ and $L_{\mathrm{bol}} / L_{\mathrm{Edd}}$ of our sample can be rather minor. For example, the underestimation is only $\sim 0.01$ on the values of $\log_{10} M_{\mathrm{BH}}$ and $L_{\mathrm{bol}} / L_{\mathrm{Edd}}$ for SDSS\,J1012$+$0358, which is the quasar that can be properly fitted by the SMC-like extinction law and possibly has the highest extinction in our sample. 

We also note that we typically do not assign any physical interpretation to the individual broad components of the fitted ${\mathrm{H} \alpha}$ model when estimating $M_{\mathrm{BH}}$ with the ${\mathrm{H} \alpha}$ emission (e.g. \citeads{2005ApJ...630..122G}; \citeads{2022ApJS..261....5M}). However, due to the extremely strong outflows indicated by the highly blueshifted BALs in our sample, their ${\mathrm{H} \alpha}$ emission may actually have non-virial contributions, similar to the {\CIV} emission researched by \citetads{2017MNRAS.465.2120C}. We might then overestimate the $M_{\mathrm{BH}}$ and $L_{\mathrm{bol}} / L_{\mathrm{Edd}}$, but the effect should not be large because of the small shift between the broad component and the intermediate component in the model ($< \SI{1000}{\km\per\s}$). No significant blueshifts are observed between the entire ${\mathrm{H} \alpha}$ profile and the systemic redshift as well, different from the {\CIV} emission. Nevertheless, the overestimation can only be quantified by comparing the single epoch ${\mathrm{H} \alpha}$ $M_{\mathrm{BH}}$ to the dynamical measurement of $M_{\mathrm{BH}}$ (e.g. \citeads{2024Natur.627..281A}).

\subsection{Dust extinction} \label{sec:dust_extinction}
The flat or red spectral slope of the quasars in our sample suggests a significant amount of extinction. It has been a normal procedure that the Small Magellanic Cloud (SMC) extinction curve can be used to describe the quasar dust extinction properly (\citeads{2003AJ....126.1131R}; \citeads{2004AJ....128.1112H}). This routine is still widely used in the analysis of quasar reddening nowadays (\citeads{2019MNRAS.486.4377K}; \citetaliasads{2020A&A...634A.111F}), including in studies of the eHAQ+GAIA23 sample, where the extinction observed in most quasars is well described by the SMC-like extinction curve (see \linkadspage{2016ApJ...832...49K}{12}{Table~4} in \citeads{2016ApJ...832...49K}).

To estimate the amount of dust extinction in our quasars, we applied the SMC-like extinction law, prescribed by \citetads{1992ApJ...395..130P}, to an X-Shooter composite spectrum from luminous blue quasars made by \citetads{2016A&A...585A..87S}. The X-Shooter composite spectrum is evaluated to be free of host galaxy contamination compared with the classic \citetads{2001AJ....122..549V} template. We tried to match the reddened quasar composite with the observed spectra and photometric data of our objects to get the best-fit dust extinction parameter. Table~\ref{tab:properties} lists the inferred amount of extinction at the rest-frame $B$ band ($A_B$) for our sample.

To illustrate the goodness of fit, we show the quasar templates that are reddened by the respective amount of extinction and are scaled according to their UKIDSS $H$ band magnitudes alongside the observed spectra in Figs.~\ref{fig:GQ0109-0719_z_full}--\ref{fig:SDSSJ0155+2543_z_full}. When the UKIDSS measurement is missing, 2MASS $H$ band magnitudes are used. The $H$ band magnitudes show the rest-frame red flux of quasars in our sample given their redshifts and hence the $H$ band measurements should be less affected by the dust extinction. It is therefore also better to scale the templates to the $H$ band regardless of whether these quasars are reddened. For three of the six quasars (PSS\,J0141$+$3334, SDSS\,J1012$+$0358, and SDSS\,J0155$+$2543), the template reddened by the SMC-like extinction curve can consistently fit the overall continuum of the spectrum and the photometric measurements from the observed optical SDSS $u$ band to the mid-IR AllWISE $W1$ band, showing the validity of assuming SMC-like extinction curve. 

However, the other three targets, GQ\,0109$-$0719, GQ\,1353$+$2554, and GQ\,1237$+$1233, appear to be affected by dust extinction laws that deviate from the standard SMC-like prescription. We can only match the SMC-like reddened quasar template with the observed spectra and photometric data in either the blue wavelengths or at the red end (especially $W1$ measurements, see Fig.~\ref{fig:GQ1353+2554_z_full} and \ref{fig:GQ1237+1233_z_full}). GQ\,1309$+$2904 from \citetaliasads{2020A&A...634A.111F} is also possibly affected by the anomalous dust extinction due to the discrepancy one can see between the $W1$ band measurement and the red template reddened by SMC-like extinction in their \linkadspage{2020A&A...634A.111F}{3}{Fig.~1}. It is similar to what was found in \citetads{2001ApJ...555..633C} and \citetads{2013ApJS..204....6F}. They suggested that in these situations, an extinction curve with steeper UV extinction than the SMC-like one is required. Our findings of anomalous extinction align with observations of WLQs hosting blueshifted BALs \citepads{2013AJ....145..157J, 2022ApJ...930....5Y}, suggesting this dust property can be a common characteristic of these sources with high-velocity outflows. They observed an extinction curve rising steeply at rest-frame wavelengths shorter than \SI{3000}{\AA}, while no significant reddening is found at longer wavelength. The mismatch between the reddened template ($A_B = 0.7$\,mag) and the spectrum of GQ\,1237$+$1233 around \SI{5400}{}--\SI{8000}{\AA} in the observed frame further corroborate that the SMC-like extinction law cannot be applied here.

We find that we indeed require an extinction curve with steeper UV extinction than the SMC-like one for these objects. As seen in Fig.~\ref{fig:GQ0109-0719_z_full}, GQ\,0109$-$0719 can be fitted properly with the extinction curve of \citetads{2015A&A...584A.100Z}, where they derived an average quasar extinction curve from quasars that cannot be fitted well by the SMC-like extinction curve. With the best-fit value of $R_V = 2.21 \pm 0.22$, the \citetads{2015A&A...584A.100Z} extinction curve has a steeper rise into the UV than the steepest Milky Way and SMC extinction curves. The smaller value of $R_V$ than that of SMC indicates the smaller mean grain size (\citeads{1989ApJ...345..245C}) of \citetads{2015A&A...584A.100Z} quasars and GQ\,0109$-$0719. 

We cannot get a proper fit for GQ\,1353$+$2554 and GQ\,1237$+$1233 with the extinction curve from either \citetads{2015A&A...584A.100Z} or \citetads{2013AJ....145..157J}. Probably an extinction curve steeper than the \citetads{2015A&A...584A.100Z} curve and flatter than the \citetads{2013AJ....145..157J} curve in the UV is needed. There is certainly some amount of dust extinction in the two quasars, but we can only offer the range of the dust extinction that these two objects could have in Fig.~\ref{fig:GQ1353+2554_z_full} and Fig.~\ref{fig:GQ1237+1233_z_full}, as well as in Table~\ref{tab:properties}.

It is also possible to fit the latter three quasars using a combination of a bluer intrinsic quasar continuum than that of \citetads{2016A&A...585A..87S}, $f_{\lambda} \propto \lambda^{-1.7}$, and the standard SMC-like extinction law. In more detail, we need a steeper quasar continuum of $f_{\lambda} \propto \lambda^{-2.7}$ and $A_B = 0.5$\,mag for GQ\,0109$-$0719, $f_{\lambda} \propto \lambda^{-2.4}$ and $A_B = 0.9$\,mag for GQ\,1353$+$2554, and $f_{\lambda} \propto \lambda^{-2.5}$ and $A_B = 0.7$\,mag for GQ\,1237$+$1233 to reach a reasonable fit. The required quasar continua are very steep and make the intrinsic quasars extremely blue, especially given that the \citetads{2016A&A...585A..87S} template is already a composite of luminous blue quasars, with the continuum steeper than other quasar composites (see their \linkadspage{2016A&A...585A..87S}{10}{Table~2}). Besides, it is not feasible for these quasars to have such extremely blue intrinsic continua, due to their lack of high-ionisation emission lines and the need for a very soft spectral energy distribution (SED) of these quasars (see Sect.~\ref{sec:high-speed outflow} for details).

The dust extinction can also be estimated by the Balmer decrement ${\mathrm{H} \alpha}$/${\mathrm{H} \beta}$. We observed very weak ${\mathrm{H} \beta}$ lines in most of our sources (see the zoom-in insets on the left panels of Figs.~\ref{fig:GQ0109-0719_z_full}--\ref{fig:SDSSJ0155+2543_z_full}) and it leads to high ${\mathrm{H} \alpha}$/${\mathrm{H} \beta}$ ratios. Higher ${\mathrm{H} \alpha}$/${\mathrm{H} \beta}$ ratios than that of the Case B recombination are normally considered to be an effect of dust extinction (e.g. \citeads{2010AJ....139..694L}). However, studies have found that the intrinsic Balmer decrement in the BLR of quasars can deviate a lot from the ${\mathrm{H} \alpha}$/${\mathrm{H} \beta}$ ratio of the Case B recombination, which is 2.86 for $T = 10^4$\,K and the electron density $n_e = $\SI{1e2}{\per\cubic\cm} (\citeads{1989agna.book.....O}). The ${\mathrm{H} \alpha}$/${\mathrm{H} \beta}$ ratios also show a broad distribution (e.g. 2.5--6.6 by \citeads{2016MNRAS.462.3570S}; 1.5--4 by \citeads{2016ApJ...832....8B}). The median Balmer decrement of quasars is found to be around 3 (\citeads{2006ApJS..166..128Z}; \citeads{2008MNRAS.383..581D}; \citeads{2012A&A...543A.142I}) and the Balmer decrement of the \citetads{2016A&A...585A..87S} template spectrum is about 4.54. The estimated amount of dust extinction is inversely proportional to the intrinsic Balmer decrement.

After subtracting the local continuum around ${\mathrm{H} \alpha}$ and ${\mathrm{H} \beta}$ emission of quasars in our sample, we get the true flux of ${\mathrm{H} \beta}$ by scaling the ${\mathrm{H} \alpha}$ model from previous analysis to fit the ${\mathrm{H} \beta}$ emission, and hence the ${\mathrm{H} \alpha}$/${\mathrm{H} \beta}$ ratios. We determined the colour excess $E(B-V)$ based on the Balmer decrement using \linkadspage{2010AJ....139..694L}{4}{Eq.~(1)} from \citetads{2010AJ....139..694L} and setting the intrinsic Balmer decrement to be 4.54. The calculated Balmer decrement and colour excess are presented in Table~\ref{tab:properties}.

The Balmer decrement of our objects leads to a higher amount of dust extinction than that estimated from the quasar continuum. For instance, SDSS\,J1012$+$0358 and SDSS\,J0155$+$2543, which are fitted well by the SMC-like extinction law, have ${\mathrm{H} \alpha}$/${\mathrm{H} \beta}$ ratios suggesting the extinction $A_B$ of \num{1.69(0.03)} and \num{1.19(0.04)}, assuming $R_V = 2.93$ for SMC according to \linkoldadspage{1992ApJ...395..130P}{3}{Table~2} of \citetads{1992ApJ...395..130P}. The observed Balmer decrement of GQ\,0109$-$0719 indicates $A_B$ of \num{1.61(0.02)} with $R_V = 2.21$ from \citetads{2015A&A...584A.100Z}.

\begin{table*}[t]
\centering
\caption{Determined spectral properties of our sample based on the observed emission lines.}
\addtolength{\tabcolsep}{-0.05cm}
\begin{tabular}{lccccc}
\hline \hline \noalign{\smallskip}
Spectral property & GQ\,0109$-$0719 & GQ\,1353$+$2554 & GQ\,1237$+$1233 & SDSS\,J1012$+$0358 & SDSS\,J0155$+$2554 \\
\hline
\noalign{\smallskip}
$z$ & $2.1164 \pm 0.0002$ & $2.1468 \pm 0.0008$ & $2.3148 \pm 0.0004$ & $2.1773 \pm 0.0003$ & $2.0713 \pm 0.0002$ \\
\noalign{\smallskip}
$\log_{10} L_{\mathrm{H} \alpha}$ ($\mathrm{erg\,s^{-1}}$) & $44.739 \pm 0.002$ & $44.685 \pm 0.003$ & $44.518 \pm 0.003$ & $44.922 \pm 0.001$ & $44.706 \pm 0.002$ \\
\noalign{\smallskip}
$\mathrm{FWHM}_{\mathrm{H} \alpha}$ (\si{\km\per\s}) & $3156 \pm 100$ & $4246 \pm 347$ & $3940 \pm 176$ & $4476 \pm 186$ & $2310 \pm 85$ \\
\noalign{\smallskip}
$\log_{10} M_{\mathrm{BH}}$ ($M_{\odot}$) & $8.96^{+0.11}_{-0.09}$ & $9.20^{+0.13}_{-0.12}$ & $9.04^{+0.11}_{-0.10}$ & $9.37^{+0.11}_{-0.10}$ & $8.66^{+0.11}_{-0.09}$ \\
\noalign{\smallskip}
$\log_{10} L_{\mathrm{bol}}$ ($\mathrm{erg\,s^{-1}}$) & $46.585 \pm 0.003$ & $46.549 \pm 0.008$ & $46.354 \pm 0.008$ & $46.755 \pm 0.005$ & $46.369 \pm 0.009$ \\
\noalign{\smallskip}
$L_{\mathrm{bol}} / L_{\mathrm{Edd}}$ & $0.28^{+0.06}_{-0.07}$ & $0.15^{+0.04}_{-0.04}$ & $0.14^{+0.03}_{-0.03}$ & $0.16^{+0.04}_{-0.04}$ & $0.34^{+0.07}_{-0.08}$ \\
\noalign{\smallskip}
$A_B$ (mag) & $0.4$\tablefootmark{\textcolor{blue}{a}} & \num{0}--\num{0.9}\tablefootmark{\textcolor{blue}{b}} & \num{0}--\num{0.7}\tablefootmark{\textcolor{blue}{b}} & $0.7$\tablefootmark{\textcolor{blue}{b}} & $0.5$\tablefootmark{\textcolor{blue}{b}} \\
\noalign{\smallskip}
{\CIV} REW (\si{\AA}) &  $< 0.6$  & $< 0.5$ & $< 0.7$ & \num{9.4(0.9)} & \num{11(4)} \\
\noalign{\smallskip}
$\log_{10} \mathrm{({\CIV}~REW/\AA)}$  & $< -0.2$ & $< -0.3$ & $< -0.1$ & \num{0.97(0.04)} & \num{1.0(0.2)} \\
\noalign{\smallskip}
{\CIV} blueshift (\si{\km\per\s}) & $\cdots$ & $\cdots$ & $\cdots$ & \num{10640(63)} & \num{8546(171)}\\
\noalign{\smallskip}
{\HeII} REW (\si{\AA}) & $\cdots$ & $\cdots$ & $\cdots$  & \num{0.3(0.2)} & \num{3(2)}\\
\noalign{\smallskip}
$\log_{10} \mathrm{({\HeII}~REW/\AA)}$ & $\cdots$ & $\cdots$ & $\cdots$ & \num{-0.6(0.3)} & \num{0.5(0.2)} \\
\noalign{\smallskip}
${\mathrm{H} \beta}$ REW (\si{\AA}) & \num{23(2)}  & \num{13(4)} & \num{15(5)} & \num{24(3)} & \num{39(5)}\\
\noalign{\smallskip}
${\mathrm{H} \alpha}$ REW (\si{\AA}) & \num{208(18)} & \num{227(34)} & \num{257(42)} & \num{275(17)} & \num{309(25)} \\
\noalign{\smallskip}
$F_{\mathrm{H} \beta}$ ($\mathrm{10^{-15}\,erg\,s^{-1}\,cm^{-2}}$) & \num{2.21(0.01)} & \num{1.29(0.03)} & \num{0.89(0.02)} & \num{3.39(0.03)} & \num{2.63(0.03)}\\
\noalign{\smallskip}
$F_{\mathrm{H} \alpha}$ ($\mathrm{10^{-15}\,erg\,s^{-1}\,cm^{-2}}$) & \num{16.52(0.07)} & \num{14.1(0.1)} & \num{7.99(0.05)} & \num{23.53(0.07)} & \num{16.14(0.08)}\\
\noalign{\smallskip}
${\mathrm{H} \alpha}$/${\mathrm{H} \beta}$  & \num{7.46(0.06)} & \num{10.9(0.2)} & \num{9.0(0.2)} & \num{6.95(0.06)} &  \num{6.13(0.07)}\\
\noalign{\smallskip}
$E(B-V)$~[${\mathrm{H} \alpha}$/${\mathrm{H} \beta}$]\tablefootmark{\textcolor{blue}{c}} (mag) & \num{0.503(0.008)} & \num{0.88(0.02)} & \num{0.69(0.02)} & \num{0.430(0.008)} & \num{0.30(0.01)} \\
\noalign{\smallskip}
$A_B$~[${\mathrm{H} \alpha}$/${\mathrm{H} \beta}$]\tablefootmark{\textcolor{blue}{d}} (mag) & \num{1.61(0.02)}\tablefootmark{\textcolor{blue}{a}}  & $\cdots$ & $\cdots$ & \num{1.69(0.03)}\tablefootmark{\textcolor{blue}{b}} & \num{1.19(0.04)}\tablefootmark{\textcolor{blue}{b}} \\
\noalign{\smallskip}
\hline
\noalign{\smallskip} \hline

\end{tabular}

\tablefoot{
PSS\,J0141$+$3334 is not included in the table because its redshift is not confirmed.
\tablefoottext{\textcolor{blue}{a}}{Rest-frame B-band extinction ($A_B$) assuming the \citetads{2015A&A...584A.100Z} extinction curve with $R_V = 2.21$.}
\tablefoottext{\textcolor{blue}{b}}{$A_B$ assuming the SMC-like extinction curve.}
\tablefoottext{\textcolor{blue}{c}}{Colour excess estimated from the Balmer decrement.}
\tablefoottext{\textcolor{blue}{d}}{$A_B$ estimated from the Balmer decrement.}
}

\centering
\label{tab:properties}
\end{table*}

\subsection{Narrow absorption lines and mini-BALs} \label{sec:NAL}
Several quasars show NALs at a range of redshifts. In Table~\ref{tab:abslines} we provide all identified NALs with their observed equivalent width ($\mathrm{EW}_{\mathrm{obs}}$), calculated after normalising with the local continuum, where the $\mathrm{EW}_{\mathrm{obs}}$ errors include both statistical noise and error from the normalisation (for more see \citeads{2009ApJS..185..526F}). The NAL systems are first identified through strong doublet absorption lines, such as {\SiIV}\,$\lambda\lambda$1393.75, 1402.77, {\CIV}\,$\lambda\lambda$1548.19, 1550.77, and {\MgII}\,$\lambda\lambda$2796.35, 2803.53. After determining the redshift of one NAL system, we search the spectrum for other NALs matching with commonly known atomic lines at the same redshift. All the identified NALs are marked on the normalised spectra as vertical solid lines in Figs.~\ref{fig:normalised_GQ0109}--\ref{fig:normalised_SDSSJ1012}.

An absorption profile without deblending can be classified as a mini-BAL if it has a FWHM of {\SI{500}{}}--{\SI{2000}{\km\per\s}}. For a single component of an absorption doublet (e.g. {\CIV}), the FWHM criterion of the mini-BAL can be extended to {\SI{70}{}}--{\SI{2000}{\km\per\s}} based on the {\CIV} mini-BAL study by \citetads{2021ApJ...907...84C}. There are several broader absorption lines that can be identified as {\SiIV} and {\CIV} mini-BALs in GQ\,1353$+$2554, GQ\,1237$+$1233, SDSS\,J1012$+$0358, and SDSS\,J0155$+$2543, as shown by the dash-dotted lines in Figs.~\ref{fig:normalised_GQ1353}, \ref{fig:normalised_GQ1237}, \ref{fig:normalised_SDSSJ1012}, and \ref{fig:normalised_SDSSJ0155}. We performed double-Gaussian fits on the potential mini-BAL doublets, as displayed in Figs.~\ref{fig:GQ1237_SiIV_blue}--\ref{fig:GQ1353_CIV}. We determined the number of {\CIV} doublet systems based on the identified {\SiIV} doublets, using the larger separation of the {\SiIV} components to resolve complex absorption profiles. In the absence of a corresponding {\SiIV} feature, we modelled the {\CIV} mini-BAL candidate with a single {\CIV} doublet, as accurate deblending of the trough is not possible. During the fitting process, the widths of the doublet components were tied, and their central wavelengths were scaled according to the redshift, while all other parameters were left free to vary. The calculated intrinsic FWHM has been corrected for the instrumental resolution, which is $R=2555$ for GQ\,1353$+$2554, $R = 1018$ for GQ\,1237$+$1233, $R = 2165$ for SDSS\,J1012$+$0358, and $R = 1850$ for SDSS\,J0155$+$2543.

We can confirm that these moderately broad absorption lines are indeed mini-BAL systems from the fitting results shown in Table~\ref{tab:mini-BALs}. All the mini-BALs are intrinsic to the quasars, due to their semi-broad line widths that are unlikely to be seen in intervening systems. The observed mini-BALs in these quasars display vastly different velocities. Some mini-BALs are close to the systemic redshift ($v \sim \SI{0}{\km\per\s}$; $z = 2.0719$ system in SDSS\,J0155$+$2543), while some are highly blueshifted ($v \sim \SI{-40000}{\km\per\s}$; $z = 1.6916$ system in GQ\,1353$+$2554). The series of discrete mini-BALs in GQ\,1237$+$1233 represent the clumpy feature of mini-BAL clouds in the outflows, consistent with the model by \citetads{2018MNRAS.479.4153Y}.

\begin{table}[!ht]
\centering
\begin{minipage}{0.45\textwidth}
\centering
\caption{Narrow metal absorption lines identified in the spectra.}
\begin{tabular}{cclc}
\hline \hline \noalign{\smallskip}
Wavelength & $\mathrm{EW}_{\mathrm{obs}}$  & ID & $z$ \\
({\AA})    & ({\AA})    &     &     \\
\hline
\noalign{\smallskip}
\multicolumn{4}{c}{GQ\,0109$-$0719}\\
\hline
\noalign{\smallskip}
4724.15  & \SI{0.63(0.06)}{} & \ion{C}{iv}\,$\lambda$1548.19 &  2.0514 \\                 
4732.02  & \SI{0.39(0.05)}{} & \ion{C}{iv}\,$\lambda$1550.77 &  2.0514 \\
4781.28  & \SI{0.97(0.05)}{} & \ion{C}{iv}\,$\lambda$1548.19 &  2.0883 \\
4789.24  & \SI{0.69(0.05)}{} & \ion{C}{iv}\,$\lambda$1550.77 &  2.0883 \\
4806.67  & \SI{1.03(0.05)}{} & \ion{C}{iv}\,$\lambda$1548.19 &  2.1047 \\
4814.68  & \SI{0.80(0.05)}{} & \ion{C}{iv}\,$\lambda$1550.77 &  2.1047 \\
\noalign{\smallskip}
\hline
\noalign{\smallskip}
\multicolumn{4}{c}{GQ\,1353$+$2554}\\
\hline
\noalign{\smallskip}
4485.71  & \SI{0.35(0.03)}{} & \ion{Al}{ii}\,$\lambda$1670.79 & 1.6858 \\
6983.54  & \SI{0.6(0.1)}{} & \ion{Fe}{ii}\,$\lambda$2600.17 & 1.6858 \\
7511.84  & \SI{1.8(0.2)}{} & \ion{Mg}{ii}\,$\lambda$2796.35 & 1.6858 \\
7531.12  & \SI{1.5(0.2)}{} & \ion{Mg}{ii}\,$\lambda$2803.53 & 1.6858 \\
\noalign{\smallskip}
\hline
\noalign{\smallskip}
\multicolumn{4}{c}{GQ\,1237$+$1233}\\
\hline
\noalign{\smallskip}
3807.23  & \SI{2.8(0.2)}{} & \ion{Al}{ii}\,$\lambda$1670.79 & 1.2787 \\
5340.92  & \SI{2.3(0.2)}{} & \ion{Fe}{ii}\,$\lambda$2344.21 & 1.2787 \\
5410.68  & \SI{1.3(0.2)}{} & \ion{Fe}{ii}\,$\lambda$2374.46 & 1.2787 \\
5429.62  & \SI{3.5(0.2)}{} & \ion{Fe}{ii}\,$\lambda$2382.77 & 1.2787 \\
5895.16  & \SI{2.7(0.2)}{} & \ion{Fe}{ii}\,$\lambda$2586.65 & 1.2787 \\
5927.26  & \SI{3.6(0.2)}{} & \ion{Fe}{ii}\,$\lambda$2600.17 & 1.2787 \\
6372.92  & \SI{6.1(0.1)}{} & \ion{Mg}{ii}\,$\lambda$2796.35 & 1.2787 \\
6389.45  & \SI{5.7(0.1)}{} & \ion{Mg}{ii}\,$\lambda$2803.53 & 1.2787 \\
\noalign{\smallskip}
\hline
\noalign{\smallskip}
\multicolumn{4}{c}{PSS\,J0141$+$3334}\\
\hline
\noalign{\smallskip}
5223.64  & \SI{3.52(0.08)}{} & \ion{Si}{ii}\,$\lambda$1526.71 &  2.4215 \\
5297.13  & \SI{2.71(0.06)}{} & \ion{C}{iv}\,$\lambda$1548.19 &  2.4215 \\
5305.96  & \SI{1.94(0.06)}{} & \ion{C}{iv}\,$\lambda$1550.77 &  2.4215 \\
5403.18  & \SI{4.11(0.07)}{} & \ion{C}{iv}\,$\lambda$1548.19  & 2.4905 \\ 
5412.19  & \SI{3.41(0.06)}{} & \ion{C}{iv}\,$\lambda$1550.77  & 2.4905 \\
5716.61  & \SI{3.5(0.7)}{} & \ion{Al}{ii}\,$\lambda$1670.79  & 2.4215 \\
8020.71  & \SI{4.3(0.3)}{} & \ion{Fe}{ii}\,$\lambda$2344.21 & 2.4215 \\
8124.21  & \SI{2.4(0.2)}{} & \ion{Fe}{ii}\,$\lambda$2374.46 & 2.4215 \\
8156.04  & \SI{5.9(0.2)}{} & \ion{Fe}{ii}\,$\lambda$2382.76 & 2.4215 \\
8850.22  & \SI{2.7(0.3)}{} & \ion{Fe}{ii}\,$\lambda$2586.65 & 2.4215 \\
8896.48  & \SI{5.3(0.4)}{} & \ion{Fe}{ii}\,$\lambda$2600.17 & 2.4215 \\
9567.71  & \SI{8.7(0.5)}{} & \ion{Mg}{ii}\,$\lambda$2796.35 & 2.4215 \\
9592.28  & \SI{10.1(0.3)}{} & \ion{Mg}{ii}\,$\lambda$2803.53 & 2.4215 \\
\noalign{\smallskip}
\hline
\noalign{\smallskip}
\multicolumn{4}{c}{SDSS\,J1012$+$0358}\\
\hline
\noalign{\smallskip}
4862.55 & \SI{0.8(0.1)}{} & \ion{C}{iv}\,$\lambda$1548.19 & 2.1408 \\
4870.65 & \SI{0.4(0.1)}{} & \ion{C}{iv}\,$\lambda$1550.77 & 2.1408 \\
\noalign{\smallskip}
\hline
\noalign{\smallskip} \hline
\end{tabular}

\centering
\label{tab:abslines}
\end{minipage}
\end{table}

\begin{table}[!ht]
\centering
\begin{minipage}{0.45\textwidth}
\centering
\caption{Mini-BALs identified in our sample.}
\label{tab:mini-BALs}
\begin{tabular}{c@{\hspace{0.0cm}}c@{\hspace{0.1cm}}c@{\hspace{0.2cm}}l@{\hspace{0.2cm}}c}
\hline \hline \noalign{\smallskip}
Wavelength & FWHM & $\mathrm{EW}_{\mathrm{obs}}$  & ID & $z$ \\
({\AA})    & \SI{}{\km\per\s} & ({\AA})    &     &     \\
\hline
\noalign{\smallskip}
\multicolumn{5}{c}{GQ\,1237$+$1233}\\
\hline
\noalign{\smallskip}
4325.92 & \multirow{2}{*}{\num{1817(312)}} & \num{14(3)} & {\SiIV}\,$\lambda$1393.75 & \multirow{2}{*}{2.1038} \\
4353.92 &  & \num{5(5)} & {\SiIV}\,$\lambda$1402.77 & \\

4363.18 & \multirow{2}{*}{\num{1073(357)}} & \num{8(3)} & {\SiIV}\,$\lambda$1393.75 & \multirow{2}{*}{2.1305} \\
4391.41 &  & \num{5(3)} & {\SiIV}\,$\lambda$1402.77 & \\

4384.80 & \multirow{2}{*}{\num{1363(139)}} & \num{2(4)} & {\SiIV}\,$\lambda$1393.75 & \multirow{2}{*}{2.1460} \\
4413.18 &  & \num{16(2)} & {\SiIV}\,$\lambda$1402.77 & \\

4596.76 & \multirow{2}{*}{\num{331(203)}} & \num{2(1)} & {\SiIV}\,$\lambda$1393.75 & \multirow{2}{*}{2.2981} \\
4626.51 &  & \num{0.9(0.6)} & {\SiIV}\,$\lambda$1402.77 & \\

4610.25 & \multirow{2}{*}{\num{592(118)}} & \num{5(1)} & {\SiIV}\,$\lambda$1393.75 & \multirow{2}{*}{2.3078} \\
4640.08 & & \num{3.3(0.8)} & {\SiIV}\,$\lambda$1402.77 & \\

4794.60 & \multirow{2}{*}{\num{1934(569)}} & \num{9(107)} & {\CIV}\,$\lambda$1548.19 & \multirow{2}{*}{2.0969} \\
4802.59 &  & \num{20(99)} & {\CIV}\,$\lambda$1550.77 & \\

4844.25 & \multirow{2}{*}{\num{1368(1415)}} & \num{14(72)} & {\CIV}\,$\lambda$1548.19 & \multirow{2}{*}{2.1290} \\
4852.32 & & \num{5(89)} & {\CIV}\,$\lambda$1550.77 & \\

4874.64 & \multirow{2}{*}{\num{1581(3845)}} & \num{4(207)} & {\CIV}\,$\lambda$1548.19 & \multirow{2}{*}{2.1486} \\
4882.77 & & \num{9(183)} & {\CIV}\,$\lambda$1550.77 & \\

5104.80 & \multirow{2}{*}{\num{383(47)}} & \num{4.7(0.6)} & {\CIV}\,$\lambda$1548.19 & \multirow{2}{*}{2.2973} \\
5113.31 & & \num{6.3(0.7)} & {\CIV}\,$\lambda$1550.77 & \\

5121.91 & \multirow{2}{*}{\num{429(39)}} & \num{8.1(0.7)} & {\CIV}\,$\lambda$1548.19 & \multirow{2}{*}{2.3083} \\
5130.45 & & \num{7.4(0.7)} & {\CIV}\,$\lambda$1550.77 & \\
\noalign{\smallskip}
\hline
\noalign{\smallskip}
\multicolumn{5}{c}{SDSS\,J1012$+$0358}\\
\hline
\noalign{\smallskip}
4238.59 & \multirow{2}{*}{\num{401(433)}} & \num{1(1)} & {\SiIV}\,$\lambda$1393.75 & \multirow{2}{*}{2.0411} \\
4266.03 &  & \num{0.7(1.2)} & {\SiIV}\,$\lambda$1402.77 & \\

4703.00 & \multirow{2}{*}{\num{736(290)}} & \num{4(3)} & {\CIV}\,$\lambda$1548.19 & \multirow{2}{*}{2.0377} \\
4710.84 &  & \num{3(4)} & {\CIV}\,$\lambda$1550.77 & \\
\noalign{\smallskip}
\hline
\noalign{\smallskip}
\multicolumn{5}{c}{SDSS\,J0155$+$2543}\\
\hline
\noalign{\smallskip}
4755.96 & \multirow{2}{*}{\num{428(187)}} & \num{6(3)} & {\CIV}\,$\lambda$1548.19 & \multirow{2}{*}{2.0719} \\
4763.89 &  & \num{5(3)} & {\CIV}\,$\lambda$1550.77 & \\
\noalign{\smallskip}
\hline
\noalign{\smallskip}
\multicolumn{5}{c}{GQ\,1353$+$2554}\\
\hline
\noalign{\smallskip}
4167.11 & \multirow{2}{*}{\num{296(15)}} & \num{4.0(0.3)} & {\CIV}\,$\lambda$1548.19 & \multirow{2}{*}{1.6916} \\
4174.05 &  & \num{3.4(0.3)} & {\CIV}\,$\lambda$1550.77 & \\
\noalign{\smallskip}
\hline
\noalign{\smallskip} \hline
\end{tabular}
\tablefoot{
The FWHM column lists the fitted widths of individual components that have been corrected for the instrumental resolution. During the fitting process, the widths of doublet members were constrained to be identical. Large errors in some fitted values arise from the degeneracy between central wavelength, amplitude, and FWHM in regions where troughs strongly overlap.}
\end{minipage}
\centering
\end{table}

\subsection{Outflow velocities} \label{sec:velocities}

\begin{figure*}[th]
\centering
\includegraphics[width=17cm]{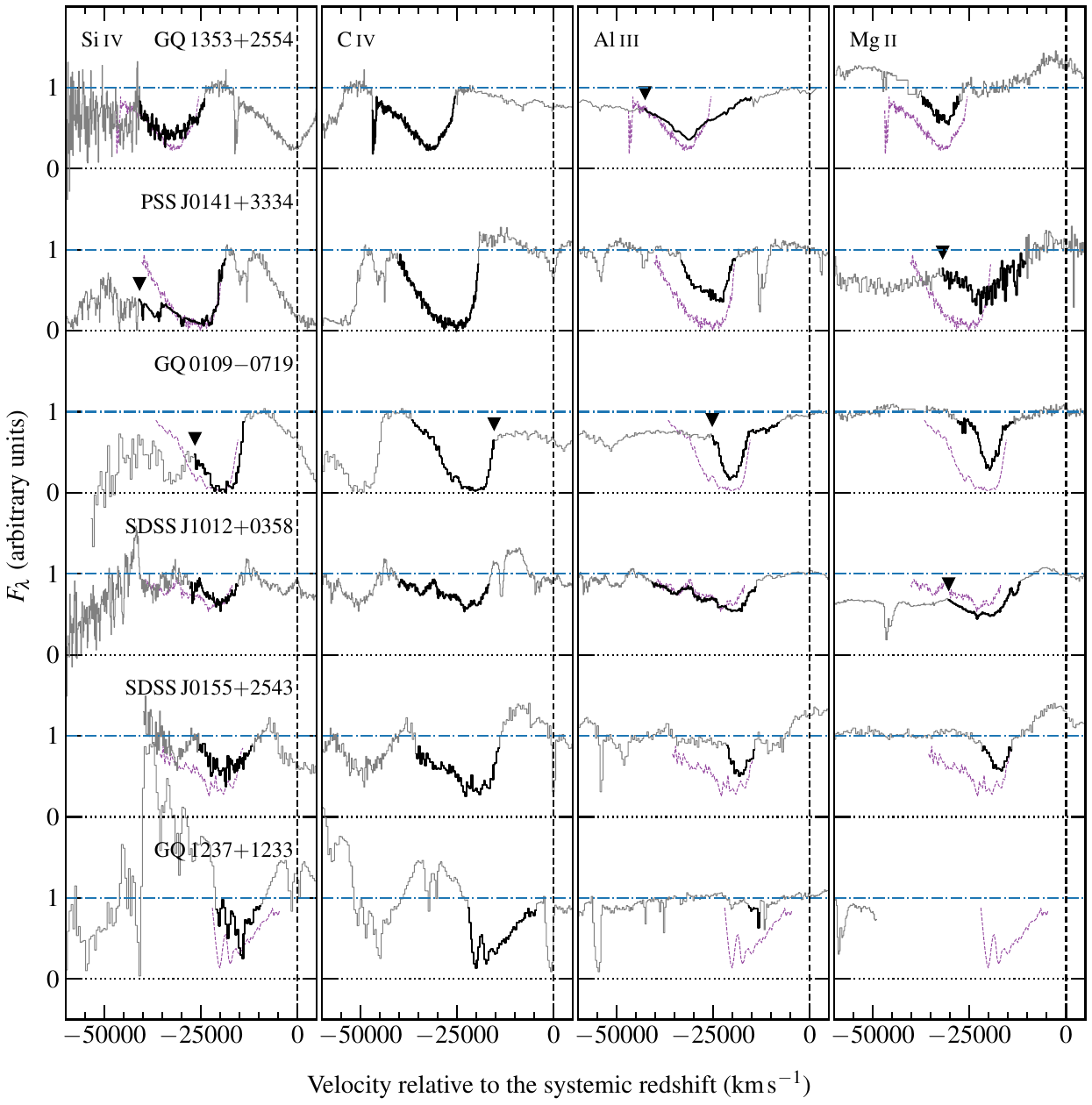}
\caption{Velocity profiles of {\SiIV}, {\CIV}, {\AlIII}, and {\MgII} (\textit{left to right}), ordered top to bottom by decreasing outflow velocity. They are plotted as solid black lines and defined as contiguous regions where the normalised flux is less than 0.9 \citepads{2009ApJ...692..758G}. The normalised quasar spectra (grey) are overplotted with their respective continua (dash-dotted blue). The vertical dashed line marks the $v = \SI{0}{\km\per\s}$ line defined from the systemic redshifts determined in Sect.~\ref{sect:redshifts}. The {\CIV} profile is overplotted in purple to directly compare with other BALs of the same quasar. The spectra are smoothed by a 3-pixel-wide boxcar filter (6-pixel for SDSS\,J0155$+$2543). The inverted triangles ($\blacktriangledown$) mark the manual determined endpoints of the velocity profiles. These are placed at points where the spectral slope changes drastically, and they are only used in cases where the flux does not recover above 0.9 within the displayed window.
}
\label{fig:velocities}
\end{figure*}

It is clear that all the BALs in our sample are highly blueshifted and well detached from the corresponding inferred wavelengths of emission lines based on the systematic redshift. Figure~\ref{fig:velocities} shows the velocity profiles of the {\SiIV}, \ion{C}{iv}, {\AlIII}, and {\MgII} lines for the quasars in our sample, ordered by their ejection velocities from top to bottom. Here we have defined $v=0\,\SI{}{\km\per\s}$ to be at the systemic redshift estimated in Sect.~\ref{sect:redshifts}. Velocities were calculated following \citetads{1986ApJ...307..504F}.

A complete list of ejection velocities and BAL $\mathrm{EW}_{\mathrm{obs}}$ is given in Table~\ref{tab:outflow_velocities}, deduced from the blueshifted BALs. In Fig.~\ref{fig:velocities}, the spectra (in grey) are normalised with the continua (the dash-dotted blue line), and the velocity profiles marked as black lines are defined to be BAL regions that fall below {90\%} of the continua \citepads{2009ApJ...692..758G}. The lowest ejection velocity ($v_{\mathrm{min}}$) and highest ($v_{\mathrm{max}}$) for a BAL trough were calculated based on the shortest and longest wavelengths of the corresponding velocity profile. BAL $\mathrm{EW}_{\mathrm{obs}}$ was also calculated based on the velocity profile. To prevent random noisy pixels from abruptly triggering our identification on the edges of the velocity profile, we smoothed the spectra using a 3-pixel–wide boxcar filter. For SDSS\,J0155$+$2543, which has a lower signal-to-noise ratio (S/N), we applied a wider 6-pixel boxcar smoothing. We also adopted a more conservative approach by adding an extra criterion that the start of the velocity profile is identified only when the first three consecutive smoothed pixels fall below 90\% of the continuum level, and the end is identified only when the last three consecutive smoothed pixels fall below 90\% of the continuum level. In a few cases, the absorbed flux does not recover above 90\% of the continuum within the velocity window shown in Fig.~\ref{fig:velocities} (\SI{-60000}{\km\per\s} to \SI{5000}{\km\per\s}). In these cases, we set the boundaries of velocity profiles at points where a significant change in the spectral slope occurs, and these locations are marked by inverted triangles ($\blacktriangledown$) in Fig.~\ref{fig:velocities}.

We can confirm that our objects have extremely high-speed outflows according to the $v_{\mathrm{max}}$ in Table~\ref{tab:outflow_velocities}, which is in agreement with \citetaliasads{2020A&A...634A.111F}. The BALs also show a wide outflow velocity range from \SI{-5000}{\km\per\s} to \SI{-47000}{\km\per\s} ($\sim \num{-0.16}\,c$).

GQ\,1353$+$2554 shows the highest outflow velocities among our sample, displayed by the {\SiIV}, \ion{C}{iv}, {\AlIII}, and {\MgII} BALs simultaneously. GQ\,0109$-$0719, SDSS\,J1012$+$0358, and SDSS\,J0155$+$2543 have lower outflow velocities, which are comparable to each other. GQ\,1237$+$1233 exhibits the lowest outflow speed. The BAL velocities of PSS\,J0141$+$3334 are estimated according to the lower limit of its systemic redshift from Sect.~\ref{sec:PSSJ0141}, and hence we chose to only display them in Fig.~\ref{fig:velocities} but not list them in Table~\ref{tab:outflow_velocities}.

All {\CIV} absorption lines are asymmetric. Most of them deepen rapidly from the red end and decay more slowly when approaching the highest-velocity blue end. The narrow red wings and broad blue wings of BALs are also observed in the literature (e.g. \citeads{1993ApJS...88..357K}; \citeads{2019MNRAS.483.1808H}). {\SiIV}, {\AlIII}, and {\MgII} BALs show similar profiles, but the resemblance is less obvious in the latter two species. It indicates that the deepest absorption trough is located at lower velocities of at least HiBALs, consistent with what is found by Rankine et al. (\citeyearads{2020MNRAS.492.4553R}, hereafter \citetaliasads{2020MNRAS.492.4553R}). The only exception is GQ\,1237$+$1233, which shows a broader red wing than the blue one.

Depending on different ionisation states, the BALs of different ions may represent different parts of the outflows, as seen from the difference between the highly ionised {\CIV} BALs and the lowly ionised {\AlIII} and {\MgII} BALs in Fig.~\ref{fig:velocities}. {\AlIII} and {\MgII} BALs are narrower than HiBALs in general (\citeads{2002ApJS..141..267H}) and our quasars show the same feature. The {\CIV} BALs reach higher $v_{\mathrm{max}}$ and span wider velocity ranges, whereas {\AlIII} and {\MgII} BALs require denser gas to form and therefore likely arise from clumpy clouds within the outflow.

\begin{table}[!t]
\centering
\begin{minipage}{0.47\textwidth}
\centering
\caption{Outflow velocities and $\mathrm{EW}_{\mathrm{obs}}$ of BALs in our sample.}
\begin{tabular}{lccc}
\hline \hline \noalign{\smallskip}
Outflows & $v_{\mathrm{max}}$ & $v_{\mathrm{min}}$  & $\mathrm{EW}_{\mathrm{obs}}$\tablefootmark{\textcolor{blue}{a}} \\
   & (\SI{}{\km\per\s})   &   (\SI{}{\km\per\s}) & ({\AA}) \\
\hline
\noalign{\smallskip}
\multicolumn{4}{c}{GQ\,0109$-$0719}\\
\hline
\noalign{\smallskip}
{\CII}   & \num{-23330}\tablefootmark{$\blacktriangledown$} & \num{-15268}\tablefootmark{$\blacktriangledown$} & \num{80(2)}\tablefootmark{$\blacktriangledown$} \\
{\SiIV}  & \num{-26552}\tablefootmark{$\blacktriangledown$} & \num{-13834} & \num{140(1)}\tablefootmark{$\blacktriangledown$} \\
{\CIV}   & \num{-36671} & \num{-15434}\tablefootmark{$\blacktriangledown$} & \num{205.2(0.9)}\tablefootmark{$\blacktriangledown$} \\
{\AlII}  & \num{-26854}\tablefootmark{$\blacktriangledown$} & \num{-15799}\tablefootmark{$\blacktriangledown$} & \num{67.3(0.5)}\tablefootmark{$\blacktriangledown$} \\
{\AlIII} & \num{-25276}\tablefootmark{$\blacktriangledown$} & \num{-8041}  & \num{136.4(0.5)}\tablefootmark{$\blacktriangledown$} \\
{\FeIII} & \num{-22630} & \num{-19555} & \num{9.8(0.2)} \\
{\MgII}  & \num{-28090} & \num{-13654} & \num{131.3(0.4)} \\
\noalign{\smallskip}
\hline
\noalign{\smallskip}
\multicolumn{4}{c}{GQ\,1353$+$2554}\\
\hline
\noalign{\smallskip}
{\SiIV}  & \num{-40763} & \num{-24095} & \num{104(1)} \\
{\CIV}   & \num{-46932} & \num{-25556} & \num{152.5(0.5)} \\
{\AlIII} & \num{-42648}\tablefootmark{$\blacktriangledown$} & \num{-15257} & \num{189.6(0.5)}\tablefootmark{$\blacktriangledown$} \\
{\MgII}  & \num{-37314} & \num{-27718} & \num{76.5(0.6)} \\
\noalign{\smallskip}
\hline
\noalign{\smallskip}
\multicolumn{4}{c}{GQ\,1237$+$1233}\\
\hline
\noalign{\smallskip}
{\NV}?    & \num{-21947} & \num{-10989} & \num{62(3)} \\
{\SiIV}  & \num{-21035} & \num{-9652}  & \num{50(1)} \\
{\CIV}   & \num{-22064} & \num{-4679}   & \num{136.5(0.7)} \\
{\AlIII}? & \num{-15828} & \num{-12982} & \num{9.9(0.3)} \\
\noalign{\smallskip}
\hline
\noalign{\smallskip}
\multicolumn{4}{c}{SDSS\,J1012$+$0358}\\
\hline
\noalign{\smallskip}
{\SiIV}  & \num{-27860} & \num{-15418} & \num{44.8(0.8)} \\
{\CIV}   & \num{-40161} & \num{-16781} & \num{87.6(0.7)} \\
{\AlIII} & \num{-40754} & \num{-14012} & \num{139.9(0.5)} \\
{\MgII}  & \num{-30395}\tablefootmark{$\blacktriangledown$} & \num{-11954} & \num{209.4(0.3)}\tablefootmark{$\blacktriangledown$} \\
\noalign{\smallskip}
\hline
\noalign{\smallskip}
\multicolumn{4}{c}{SDSS\,J0155$+$2543}\\
\hline
\noalign{\smallskip}
{\SiIV}  & \num{-25414} & \num{-11991} & \num{59(2)} \\
{\CIV}   & \num{-35585} & \num{-14190} & \num{147(2)} \\
{\AlII}  & \num{-18967} & \num{-15662} & \num{12.5(0.9)} \\
{\AlIII} & \num{-21511} & \num{-14491} & \num{45(1)} \\
{\FeIII} & \num{-17385} & \num{-16558} & \num{3.0(0.6)} \\
{\MgII}  & \num{-21648} & \num{-14215} & \num{61(1)} \\
\noalign{\smallskip}
\hline
\noalign{\smallskip} \hline
\end{tabular}

\tablefoot{
The velocities were determined based on the ${\mathrm{H} \alpha}$ systemic redshift. We marked the manually determined $v_{\mathrm{max}}$, $v_{\mathrm{min}}$, and affected $\mathrm{EW}_{\mathrm{obs}}$ with inverted triangles ($\blacktriangledown$).}

\centering
\label{tab:outflow_velocities}
\end{minipage}
\end{table}

\subsection{Polarimetric measurements} \label{sec:polarisation}
We obtained three-colour ($BVR$) polarimetric measurements for three quasars in the sample: GQ\,1309$+$2904 (\citetaliasads{2020A&A...634A.111F}), GQ\,1353$+$2554, and GQ\,1237$+$1233 using DiPol-UF. We note that weather conditions at the time of the polarimetric observation were not optimal (strong wind and bad seeing of $\sim5$\,arcsec), resulting in reduced S/N and larger measuring errors. The derived degree of linear polarisation is listed in Table~\ref{tab:polarimetry}. According to \citetads{1998A&A...340..371H}, the polarisation degree is very biased at low S/N but an unbiased estimator of the true polarisation degree ($P_0$) can be calculated as $P_0 = \sqrt{P^2 - 2\sigma^2}$ (\citeads{1985A&A...142..100S}; \citeads{2022MNRAS.514.2479K}), where $\sigma$ is the mean error on the observed degree of linear polarisation. Corresponding estimates of the true polarisation degree are also displayed in Table~\ref{tab:polarimetry}.

\begin{table}[!t]
\centering
\begin{minipage}{0.47\textwidth}
\centering
\caption{$BVR$ polarimetric measurements of GQ\,1309$+$2904, GQ\,1353$+$2554, and GQ\,1237$+$1233.}
\begin{tabular}{lccc}
\hline \hline \noalign{\smallskip}
Measurement & $B$ & $V$  & $R$ \\
\hline
\noalign{\smallskip}
\multicolumn{4}{c}{GQ\,1309$+$2904}\\
\hline
\noalign{\smallskip}
$P$ ($\%$)& $\cdots$ & \num{1.8(0.9)} & \num{0.9(0.5)}\\
$P_0$ ($\%$) & $\cdots$ & \num{1.3(0.9)} & \num{0.6(0.5)} \\
\noalign{\smallskip}
\hline
\noalign{\smallskip}
\multicolumn{4}{c}{GQ\,1353$+$2554}\\
\hline
\noalign{\smallskip}
$P$ ($\%$)& \num{1.9(0.9)} & \num{1.9(0.7)} & \num{1.6(0.2)}\\
$P_0$ ($\%$) & \num{1.4(0.9)} & \num{1.6(0.7)} & \num{1.6(0.2)} \\
\noalign{\smallskip}
\hline
\noalign{\smallskip}
\multicolumn{4}{c}{GQ\,1237$+$1233}\\
\hline
\noalign{\smallskip}
$P$ ($\%$)& $\cdots$ & $\cdots$ & \num{1.2(0.8)}\\
$P_0$ ($\%$) & $\cdots$ & $\cdots$ & \num{0.4(0.8)} \\
\noalign{\smallskip}
\hline
\noalign{\smallskip} \hline
\end{tabular}

\centering
\label{tab:polarimetry}
\end{minipage}
\end{table}

For GQ\,1309$+$2904, the source was too faint to secure a measurement in the $B$ band, and the estimated true polarisation degrees are both consistent with zero given their large errors. GQ\,1353$+$2554 has two of the measurements consistent with zero but in the $R$ band there is evidence of some polarisation. Due to the high Galactic latitude of this quasar ($\sim 76^{\circ}$), its interstellar polarisation component is expected to be of the order of or less than $\sim$ 0.1\%. We secured a single measurement of the degree of linear polarisation consistent with zero in the $R$ band for GQ\,1237$+$1233. In the $B$ and $V$ bands, the source is too faint to let us obtain a measurement.

We cannot evaluate whether the anti-correlation between the degree of polarisation and the BAL detachment of LoBAL quasars (\citeads{1998A&A...340..371H}) still holds for our sample due to the low quality of our polarimetric measurements in the $V$ band. If we tentatively use estimated values of the intrinsic $V$-band polarisation $P_{V,0}$, it seems that P-Cygni profiles may be present at least in GQ\,1309$+$2904 and GQ\,1353$+$2554 (see their \linkoldadspage{1998A&A...340..371H}{9}{Fig.~5}), hinting at blueshifted {\CIV} emission lines.

All the quasars with polarisation measurements in our sample have low polarisations ($P_0 < 4\%$, \citeads{2017A&A...607A..40M}), even if we consider the large errors of the measurements. According to the correlation between the inclination angle and the intrinsic polarisation degree in \linkadspage{2017A&A...607A..40M}{3}{Fig.~2} of \citetads{2017A&A...607A..40M}, we are observing these quasars with inclination angles $< 45^{\circ}$, at least above the torus horizon. It indicates that the anomalous dust extinction observed in GQ\,1353$+$2554, GQ\,1237$+$1233, and possibly in GQ\,1309$+$2904 is not from the torus.

\subsection{Photometric variability} \label{sec:variability}
We noticed a flux change in SDSS\,J1012$+$0358 between the older (in maroon) and more recent (in orange) photometric measurements displayed in Fig.~\ref{fig:SDSSJ1012+0358_z_full} and Table~\ref{tab:flux_change}. 
The dimming is around 0.5\,mag, 0.7\,mag, 0.8\,mag, and 0.4\,mag for the $griz$ band, respectively, within a timespan of around 10 years. Meanwhile, there is also a mild decrease in the brightness of the observed infrared $H$ band of about $0.4 \pm 0.2$\,mag. The decrease in the $K$ band flux is quite uncertain because the $K$ band magnitude from 2MASS is flagged as a measurement with a very poor quality or only indicating an upper limit. The change in the $J$ band measurements is negligible. We also checked the AllWISE and NEOWISE \citepads{2011ApJ...731...53M} $W1$ and $W2$ measurements of this quasar. A mild flux decrease between MJD \num{55334.8} to MJD \num{60274} is found.

\begin{table}[!t]
\centering
\begin{minipage}{0.47\textwidth}
\centering
\caption{Photometric measurements of SDSS\,J1012$+$0358 from different instruments and observation time.}
\begin{tabular}{lccc}
\hline \hline \noalign{\smallskip}
Filter & AB magnitude & Instrument & MJD \\
\hline
\noalign{\smallskip} 
\multirow{2}{*}{$g$} & \num{19.33(0.01)} & SDSS & \num{51960.2} \\
 & \num{19.84(0.03)} & Pan-STARRS1 & \num{56172.2} \\
\noalign{\smallskip}
\hline
\noalign{\smallskip}
\multirow{2}{*}{$r$} & \num{18.286(0.007)} & SDSS & \num{51960.2} \\
 & \num{18.99(0.03)} & Pan-STARRS1 & \num{56172.2} \\
\noalign{\smallskip}
\hline
\noalign{\smallskip}
\multirow{2}{*}{$i$} & \num{18.009(0.008)} & SDSS & \num{51960.2} \\
 & \num{18.768(0.008)} & Pan-STARRS1 & \num{56172.2} \\
\noalign{\smallskip}
\hline
\noalign{\smallskip}
\multirow{2}{*}{$z$} & \num{17.66(0.02)} & SDSS & \num{51960.2} \\
 & \num{18.02(0.02)} & Pan-STARRS1 & \num{56172.2} \\
\noalign{\smallskip}
\hline
\noalign{\smallskip}
\multirow{2}{*}{$J$} & \num{17.8(0.2)} & 2MASS & \num{51960.2} \\
 & \num{17.72(0.01)} & UKIDSS & \num{56172.2} \\
 \noalign{\smallskip}
\hline
\noalign{\smallskip}
\multirow{2}{*}{$H$} & \num{17.0(0.2)} & 2MASS & \num{51960.2} \\
 & \num{17.44(0.02)} & UKIDSS & \num{56172.2} \\
\noalign{\smallskip}
\hline
\noalign{\smallskip}
\multirow{2}{*}{$K$} & \num{16.82} & 2MASS & \num{51960.2} \\
 & \num{17.17(0.01)} & UKIDSS & \num{56172.2} \\
\noalign{\smallskip}
\hline
\noalign{\smallskip}
\multirow{2}{*}{$W1$} & \num{17.14(0.07)} & AllWISE & \num{55334.8} \\
 & \num{17.28(0.09)} & NEOWISE & \num{60274} \\
\noalign{\smallskip}
\hline
\noalign{\smallskip} \hline
\end{tabular}

\tablefoot{
The observation time is listed in the modified Julian date (MJD), and here we show the mean values of the measurements during the epoch of photometry.}

\centering
\label{tab:flux_change}
\end{minipage}
\end{table}

Other targets in our sample do not exhibit such variability, and the differences between their SDSS and Pan-STARRS1 photometric data are within 0.1\,mag. GQ\,1309$+$2904 is found to show minor variability of quasars by \citetaliasads{2020A&A...634A.111F}, but there is no evidence of a significant flux change similar to that of SDSS\,J1012$+$0358.

\subsection{Radio detection} \label{sec:radio_detection}
Most quasars in our sample are covered but not detected in the Faint Images of the Radio Sky at Twenty Centimeters (FIRST; \citeads{1995ApJ...450..559B}) survey. SDSS\,J1012$+$0358 also stands out as the sole quasar in our sample that is detected in the FIRST survey as a compact point source, with a peak flux density of $1.15\pm0.14$\,mJy/beam and an integrated flux density of 1.07\,mJy at an average epoch of observation on 16/02/1999 ($\mathrm{MJD} =$ \num{51225.2}). The notable variability and radio detection by FIRST may hint at the existence of a jet in this object, which is also consistent with the detected spatially asymmetric extended Ly$\alpha$ emission.

However, the quasars in our sample are all detected in the more sensitive Low-Frequency Array (LOFAR; \citeads{2017A&A...598A.104S}) survey when covered (GQ\,0109$-$0719 and GQ\,1237$+$1233 are not covered). The radio detection rate of our sample is therefore at least $5/7 = 71\%$. This is in agreement with the findings from \citetads{2021MNRAS.502.4154R} that the radio-detection fractions of quasars increase with the increasing {\CIV} emission blueshifts, while the radio-loud fractions decrease (according to their \linkadspage{2021MNRAS.502.4154R}{6}{Fig.~6}, almost all the radio-loud sources in LOFAR are also detected in FIRST).

\section{Discussion} 
\label{sec:discussion}

\subsection{High-speed outflows and weak UV emission lines} \label{sec:high-speed outflow}
We can certainly conclude that high-speed outflows exist ubiquitously in our sample, given $v_{\mathrm{max}}$ of the sample spanning from \SI{-16000}{\km\per\s} to \SI{-47000}{\km\per\s} (\num{-0.05}\,$c$ to \num{-0.16}\,$c$). The outflow velocities themselves are not extreme compared with Extremely High Velocity Outflows (EHVOs) that are investigated in \citetads{2020ApJ...896..151R}, but the outflows in our sample have much broader velocity profiles, showing much larger column densities of BAL clouds around these selected quasars. The existence of high-speed outflows is also supported by the high radio-detection fraction (\citeads{2021MNRAS.502.4154R}; \citeads{2022MNRAS.515.5159P}) and the detected weak radio emission can come from the interaction between high-speed outflows and the interstellar medium (ISM; \citeads{2022MNRAS.515.5159P}; \citeads{2024ApJS..271...61Y}).

\begin{figure}[t!]
\centering
\resizebox{\hsize}{!}{\includegraphics[scale=0.44]{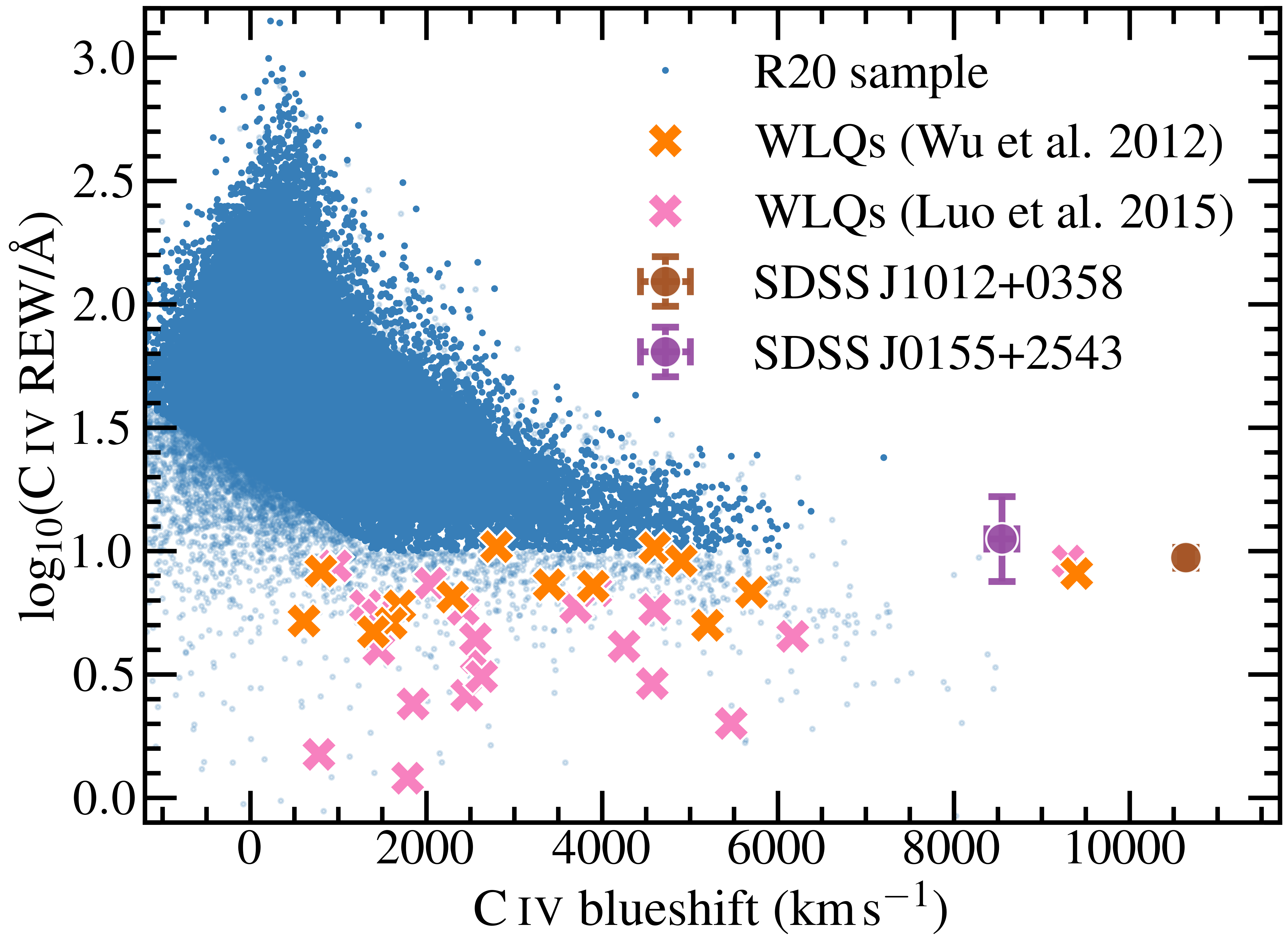}}
\caption{Positions of SDSS\,J1012$+$0358 (brown) and SDSS\,J0155$+$2543 (purple) in the {\CIV} emission space. For comparison, the non-BAL plus HiBAL quasars from \citetaliasads{2020MNRAS.492.4553R} are plotted in blue, with the `good' targets indicated by solid points). WLQs from \citetads{2012ApJ...747...10W} and \citetads{2015ApJ...805..122L} are shown in orange and pink, respectively. The remaining quasars in our sample are expected to lie below the two highlighted objects and to have at least comparably large {\CIV} blueshifts. Overall, our sample is distinctly separated from the normal quasar population (non-BAL and HiBAL) and is consistent with the WLQ population, suggesting a shared physical nature.
}
\label{fig:emission space}
\end{figure}

In addition to the highly blueshifted BALs in these targets, another noticeable common feature of the quasars is the absence or extreme weakness of the UV emission lines in their observed spectra. UV emission lines are predominant in type 1 quasars and HiBAL quasars. They are also obvious in most of the LoBAL quasars as shown by \citetads{2017ApJ...848..104S}. On the contrary, our quasars show very weak high-ionisation emission but rather normal ${\mathrm{H} \alpha}$ emission, and noticeable {\CIV} emission blueshift. These features have been seen in other WLQs (\citeads{2012ApJ...747...10W}; \citeads{2015ApJ...805..123P}; \citeads{2015ApJ...805..122L}; \citeads{2024ApJ...972..191C}), as one can also see in Fig.~\ref{fig:emission space}.

The weak high-ionisation emission lines and the existence of high-speed outflows are in fact closely related. \citetads{2011AJ....141..167R} discovered that the {\CIV} equivalent width (EW) is anti-correlated with the {\CIV} blueshift using $\sim\num{30000}$ quasars. The correlation between the high-speed outflow and weak emission line is further verified by \citetaliasads{2020MNRAS.492.4553R} from investigating about \num{144000} non-BAL and HiBAL quasars. They found an obvious resemblance in the distributions of non-BAL and HiBAL quasars in the {\CIV} emission blueshift versus EW space (hereafter {\CIV} emission space). They consolidate that the larger the {\CIV} intrinsic blueshift, the weaker the intrinsic {\CIV} emission, as shown in their \linkadspage{2020MNRAS.492.4553R}{9}{Fig.~8} for both kinds of quasars. This trend in the {\CIV} emission space is further extended to higher {\CIV} blueshifts and lower {\CIV} EW by \citetads{2022ApJ...939L..24R} researching quasars with EHVOs. A similar correlation is observed in other WLQs and the {\CIV} emission lines of WLQs are generally weaker and more blueshifted (see \linkadspage{2012ApJ...747...10W}{14}{Fig.~8} of \citeads{2012ApJ...747...10W}). This correlation and the trend in the {\CIV} emission space seem to be a general feature of quasars, regardless of their other spectral properties.

The trend observed in the {\CIV} emission space is governed by common physical mechanisms, as discussed by \citetads{2013MNRAS.432.1525B} and \citetaliasads{2020MNRAS.492.4553R}. In particular, \citetaliasads{2020MNRAS.492.4553R} find that a higher {\CIV} emission blueshift correlates with lower {\HeII} EW, higher quasar $L_{\mathrm{bol}}$, and a higher Eddington ratio. The {\HeII}\,$\lambda$1640.42 is one of the strongest {\HeII} recombination lines (\citeads{2001ApJ...553...73O}) and serves as a tracer of the amount of highly ionising photons with energy above 54\,eV. A lower {\HeII} EW thus indicates a weaker soft X-ray and softer UV continuum, which can produce less ionisation and become more effective in radiative acceleration (\citeads{2004ApJ...611..125L}; \citeads{2019MNRAS.483.1808H}). The disc wind can only be accelerated to high speed by line-driven radiation when the electrons are coupled to the nuclei, which requires that the SED of the quasar is soft enough and the high-energy ionising photons are few enough (\citetaliasads{2020MNRAS.492.4553R}). Therefore, BALs with deep troughs and large velocities are produced when the ionising continuum is soft (\citeads{2019MNRAS.483.1808H}), which also leads to weak {\CIV} but normal ${\mathrm{H} \alpha}$ emission. The anti-correlation between the blueshift of the {\CIV} emission line and the EW of the {\CIV} or {\HeII} emission arises naturally in the disc wind scenario.

We observe that {\CIV} emission lines and other high-ionisation emission lines are both weak and strongly blueshifted in SDSS\,J1012$+$0358 and SDSS\,J0155$+$2543. Their {\CIV} emission REW and blueshift are estimated in Sects.~\ref{sec:SDSSJ1012} and \ref{sect:SDSS0155}. With having stronger {\CIV} emission lines than those of other objects in our sample, we also identify weak {\HeII} lines in these two quasars (for REW see Table~\ref{tab:properties}). Due to the peculiar shapes of their optical spectra and the presence of strong, blueshifted BALs, we were unable to make the composite reconstruction as \citetaliasads{2020MNRAS.492.4553R} did to remove BALs. Therefore, our measured {\CIV} REWs and blueshifts are likely underestimates of their intrinsic values. Based on \linkadspage{2020MNRAS.492.4553R}{10}{Fig.~9} of \citetaliasads{2020MNRAS.492.4553R}, the reconstructed intrinsic {\CIV} REWs are only slightly larger than the observed values, and the blueshifts remain largely unchanged. This enables a reasonable comparison between these two quasars and the non-BAL and HiBAL quasars from \citetaliasads{2020MNRAS.492.4553R}, as well as the WLQs from \citetads{2012ApJ...747...10W} and \citetads{2015ApJ...805..122L} in Fig.~\ref{fig:emission space}.

For the rest of our sample, we cannot place them on the {\CIV} emission space diagram because their spectra show no clear {\CIV} emission, making it impossible to determine the {\CIV} blueshifts. We propose that their {\CIV} emission lines are intrinsically weak due to the high-speed outflow in these quasars, instead of being absorbed by the BALs. Previous studies (\citeads{2011AJ....141..167R}; \citetaliasads{2020MNRAS.492.4553R}; \citeads{2022ApJ...939L..24R}) find that the weaker and more blueshifted {\CIV} emission is often associated with stronger, faster {\CIV} BALs. This is consistent with the \citetads{2000ApJ...545...63E} disc wind model, which shows a continuous outflow where the BELs and BAL troughs originate from the same accelerating wind, and emission lines arise from the base of the wind. Based on the exceptionally strong and blueshifted BALs in our sample, our objects should also host high-speed BEL clouds and have intrinsically weak {\CIV} emission lines. If it is true that {\CIV} emission is not absorbed by BALs in our sample, GQ\,0109$-$0719, GQ\,1353$+$2554, and GQ\,1237$+$1233 are classified as WLQs, based on the upper limit of {\CIV} REW we estimated in Sects.~\ref{sec:GQ0109}--\ref{sec:GQ1237} and the criterion from \citetads{2025ApJ...994..213C}.

Another way to look into the origin of the potential trend in the {\CIV} emission space is to investigate the $M_{\mathrm{BH}}$ and $L_{\mathrm{bol}} / L_{\mathrm{Edd}}$ of quasars. \citetads{2023MNRAS.523..646T} examine the UV emission line features in $\sim$\num{190000} type 1 quasars at $z \sim 2$ based on their $M_{\mathrm{BH}}$ and $L_{\mathrm{bol}} / L_{\mathrm{Edd}}$. They find that only quasars with both high $M_{\mathrm{BH}}$ and high $L_{\mathrm{bol}} / L_{\mathrm{Edd}}$ show large {\CIV} blueshifts, low {\CIV} and {\HeII} EWs, as well as weak X-ray emission. Very strong correlations between all these features are only found in quasars with $L_{\mathrm{bol}} / L_{\mathrm{Edd}} \gtrsim 0.1$, and all our objects fulfil this condition. 

Different from what \citetads{2023MNRAS.523..646T} find, quasars in our sample have lower-than-expected $M_{\mathrm{BH}}$ and Eddington ratios when having very strong outflows. According to \linkadspage{2023MNRAS.523..646T}{7}{Fig.~3} of \citeads{2023MNRAS.523..646T} and \linkadspage{2020MNRAS.492.4553R}{15}{Fig.~14} of \citetaliasads{2020MNRAS.492.4553R}, quasars with $M_{\mathrm{BH}}$ of $10^{9}\,M_{\odot}$ and $L_{\mathrm{bol}} / L_{\mathrm{Edd}}$ of 0.2 have the {\CIV} emission blueshift around \SI{1000}{\km\per\s}, but our objects show much higher {\CIV} emission blueshifts with similar $M_{\mathrm{BH}}$ and $L_{\mathrm{bol}} / L_{\mathrm{Edd}}$. The discrepancy can indicate the fundamental differences between our sample and the non-BAL and HiBAL quasars.

\subsection{Anomalous dust extinction along high-speed outflows} \label{sec:smaller_dust}
Another interesting feature of our sample is that they cannot all be fitted by the SMC-like extinction curve, which is normally used to describe the dust extinction in quasars. It seems that the dust extinction of four objects in our sample, GQ\,0109$-$0719, GQ\,1353$+$2554, GQ\,1237$+$1233, and GQ\,1309$+$2904 from \citetaliasads{2020A&A...634A.111F}, is anomalous and extinction curves with steeper UV rise are needed to properly account for such reddening (\citeads{2001ApJ...555..633C}; \citeads{2013ApJS..204....6F}; \citeads{2020ApJ...891...53C}, \citeyearads{2022ApJ...937...74C}). 

The steeper rise in UV wavelengths of the extinction curve can be attributed to a preponderance of smaller dust grains along our lines of sight to the investigated quasars (\citeads{1989ApJ...345..245C}). We found that GQ\,0109$-$0719 can be fitted well by the extinction curve from \citetads{2015A&A...584A.100Z} that is steeper in the UV than the SMC-like curve, suggesting a significant amount of small dust grains existing around GQ\,0109$-$0719. The fact that GQ\,1353$+$2554 and GQ\,1237$+$1233 cannot be fitted by the \citetads{2015A&A...584A.100Z} extinction law indicates the need for a much UV-steeper extinction curve. Smaller dust grains can dominate the environments around these two quasars. 

Anomalous dust extinction has also been observed in other WLQs (e.g. \citeads{2013AJ....145..157J}; \citeads{2022ApJ...930....5Y}). They found that the SED of the WLQ can also be fitted by the unreddened quasar composite spectrum very well at rest-frame wavelength higher than \SI{4000}{\AA}, but deviates significantly at wavelengths lower than \SI{3000}{\AA}. This is very similar to what we see in Fig.~\ref{fig:GQ1353+2554_z_full} and Fig.~\ref{fig:GQ1237+1233_z_full}. 

What is more interesting is that we are probably observing these small dust grains along the high-speed outflows. The highly blueshifted BALs with very wide velocity profiles show that our line of sights can be aligned with the launched outflows. Low intrinsic polarisation degrees are found in three out of the four targets affected by the anomalous dust extinction (GQ\,1309$+$2904, GQ\,1353$+$2554, and GQ\,1237$+$1233). It indicates the low inclination angles of them, based on the conclusions from \citetads{2017A&A...607A..40M}. Dust grains with smaller mean size than that of the SMC-like dust may exist in the disc wind or on the outskirts of them. The idea of the dusty wind in quasars has been explored in numerous studies (e.g. \citeads{2006ApJ...648L.101E}; \citeads{2017ApJ...838L..20H}; \citeads{2023A&A...675A..43N}) and our objects seem to support this model.

Besides, our sample shows a high Balmer decrement, which indicates higher-than-expected dust extinction. The dust extinction inferred from the Balmer decrement is all larger than the amount of extinction estimated from the continuum, as shown by $A_B$ and $A_B$~[${\mathrm{H} \alpha}$/${\mathrm{H} \beta}$] in Table~\ref{tab:properties}. The differences may suggest that our selected LoBAL quasars have the higher intrinsic Balmer decrement and their physical conditions are very different from nebulous gas clouds and most of the normal quasars.

In summary, dust grains with smaller mean sizes, probably existing in the high-speed disc wind, are creating the anomalous extinction in four of our quasars. The rest of the sample are affected by SMC-like dust extinction. Our objects have larger observed ${\mathrm{H} \alpha}$/${\mathrm{H} \beta}$ ratios than those expected from the continuum estimated amount of dust extinction. It indicates that these quasars may have physical conditions that are distinct from the normal quasars, which leads to the larger intrinsic Balmer decrement of them.

\subsection{WLQs observed through the disc wind}
Our sample obviously differs from non-BAL and HiBAL quasars. Three of them (GQ\,0109$-$0719, GQ\,1353$+$2554, and GQ\,1237$+$1233) can be classified as WLQs and two of them (SDSS\,J1012$+$0358 and SDSS\,J0155$+$2543) to be bridge quasars, transitioning from WLQs to normal quasars, based on the definition from \citetads{2025ApJ...994..213C}, and assuming that our {\CIV} emission lines are indeed intrinsically weak and not absorbed by neighbouring BALs. Our sample and normal WLQs could be possibly observed at different inclinations. The search for WLQs normally excludes BAL quasars in general (e.g. \citeads{2012ApJ...747...10W}; \citeads{2015ApJ...805..122L}), but \citetads{2015ApJ...805..122L} mention that BALs can be observed in WLQs when the line of sight intercepts the disc wind, which could be exactly the case for our sample.

The observations also support the idea that our objects are WLQs but observed at different inclinations from normal WLQs. \citetads{2015ApJ...805..123P} detect broad ${\mathrm{H} \alpha}$ and ${\mathrm{H} \beta}$ emission in WLQs, and they are towards the weaker end of the expected EW distribution of typical quasars, but not as weak as the high-ionisation lines. \citetads{2024ApJ...972..191C} show that WLQs still have broad, strong ${\mathrm{H} \alpha}$ emission and weak or absent {\OIII} emission, same as our sample, although they observed comparable ${\mathrm{H} \beta}$ emission to normal quasars. Blueshifted {\MgII} emission lines are also found in WLQs (\citeads{2022ApJ...930....5Y}), as observed in SDSS\,J0155$+$2543. 

It is suggested that WLQs may generally host high-velocity outflows (\citeads{2015ApJ...805..122L}). This is plausible given their soft SEDs, as indicated by the absence of high-ionisation emission lines in their spectra. Under the disc wind scenario, the presence or absence of BALs is often attributed to different inclinations (e.g. \citeads{2000ApJ...545...63E}; \citeads{2022MNRAS.511.4946N}). If disc wind is present ubiquitously in WLQs, different inclinations can account for the observed BALs in our sample and their absence in normal WLQs, as suggested by \citetads{2015ApJ...805..122L}. \citetads{2013AJ....145..157J} and \citetads{2022ApJ...930....5Y} investigated WLQs with blueshifted BALs, in which they found anomalous dust extinction, same as some of our sample. If the combination of weak emission lines, blueshifted BALs, and anomalous dust extinction proves to be a general phenomenon that cannot be explained by the quasar unified model (e.g. \citeads{2012agn..book.....B}), extending observations to inclinations typically overlooked in WLQ research may help to constrain the physical origin of WLQs.

\subsection{Quasars emerging from the dust cocoon} \label{sec:transition phase}

The quasars studied in this paper share a set of unusual properties, including weak emission lines, powerful blueshifted BAL outflows, and in several cases anomalous dust extinction. We propose that these objects, some of which also exhibit FeLoBAL features (GQ\,0109$-$0719 and SDSS\,J0155$+$2543), represent quasars emerging from their surrounding dust cocoons, transitioning from the heavily dust-shrouded early-type quasars to normal quasars.

Generally, LoBAL quasars are considered a distinct group from non-BAL and HiBAL quasars, with properties indicating an early stage of quasar evolution (e.g. \citeads{1993ApJ...413...95V}; \citeads{2009ApJ...698.1095U}; \citeads{2011MNRAS.410..860A}; \citeads{2012ApJ...757...51G}). Within this population, the FeLoBAL subset has been frequently discussed as systems in which quasar feedback strongly interacts with a dusty environment, and proposed as candidates associated with transitional phases between ultra luminous infrared galaxies (ULIRGs) and normal quasars (\citeads{2007ApJ...662L..59F}; \citeads{2009ApJ...698.1095U}). \citetads{2011MNRAS.410..860A} argue that the FeLoBAL features represent the final stage of the process where the quasar shakes off its dust cocoon.

\citetads{2022ApJ...930....5Y} report a changing WLQ at redshift 2 experiencing the blowout phase and shedding its dust cocoon, based on the observed LoBAL to HiBAL transformation (\citeads{2019ApJ...870L..25Y}), high-speed BALs ($\sim \num{-0.1}\,c$), and a decrease in dust extinction. They found anomalous dust extinction for this WLQ, which cannot be explained by the SMC-like extinction law, indicating that large grains are destroyed or small grains are formed in the outflows (\citeads{2022ApJ...930....5Y}). The extremely powerful outflows in our sample are hence very likely to be capable of blowing out the dust from the quasars and removing their dusty cocoons, because they are much broader, stronger, and faster than those of the WLQ in \citetads{2022ApJ...930....5Y}. 

\begin{figure}[t!]
\centering
\resizebox{\hsize}{!}{\includegraphics[scale=0.19]{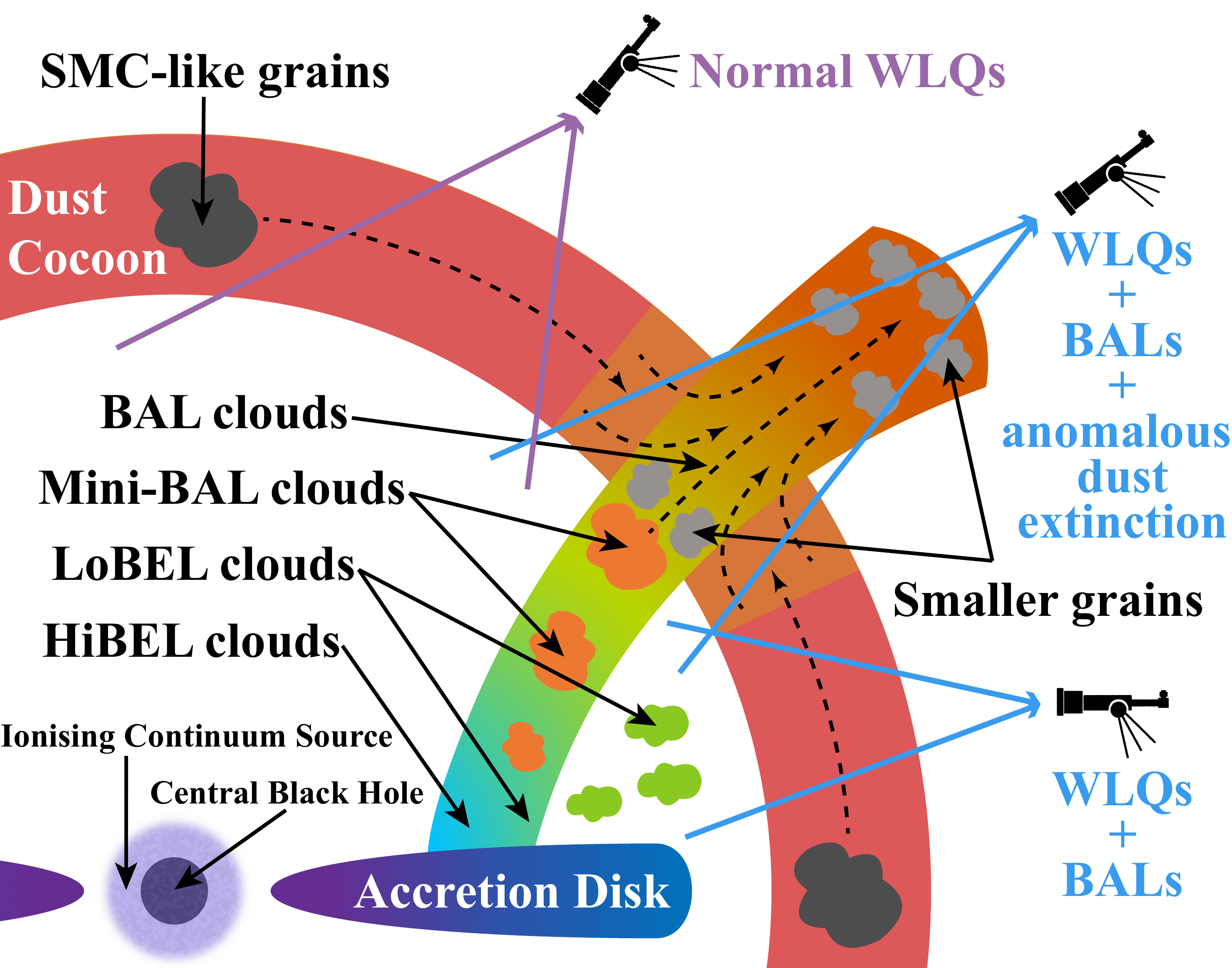}}
\caption{Schematic illustration (not to scale) of the geometry of the quasar blowout phase proposed to explain the peculiar spectral features observed in our objects. The dashed black arrows indicate the flow of gas and dust as the disc wind expels material from the cocoon. In this scenario, shocks shatter SMC-like grains into smaller dust grains within the outflow, where clumpy mini-BAL clouds can also form. The solid blue and purple arrows represent our lines of sight at different inclinations. Our objects, showing weak emission lines together with BALs or BALs plus anomalous dust extinction features, can be observed with the inclinations denoted by the blue arrows. Normal WLQs can be seen with the inclinations shown by the purple arrows.
}
\label{fig:evolution_model}
\end{figure}

To understand the peculiar spectral features of our sample and illustrate how the geometry of the blowout phase gives rise to these observations, we present a schematic illustration in Fig.~\ref{fig:evolution_model}. It shows the quasar blowout phase when the disc wind is piercing through the dust cocoon and expelling dust grains. As a result of shocks from the collision between high-speed outflows and the cocoon, SMC-like dust grains in the regions around outflows are destroyed into smaller grains via shattering (\citeads{1996ApJ...469..740J}). The outflow geometry is adapted from the disc‑wind model of \citetads{2000ApJ...545...63E}, in which BAL clouds, low‑ionisation broad emission line (LoBEL) clouds, and high‑ionisation broad emission line (HiBEL) clouds occupy different regions of the disc wind due to ionisation stratification and shielding effects. In quasars with soft SEDs, such as WLQs, the production of high‑ionisation emission lines can be suppressed, while LoBEL and BAL features remain prominent. This is why we observed normal ${\mathrm{H} \alpha}$ emission but weak HiBEL in normal WLQs (\citeads{2015ApJ...805..123P, 2024ApJ...972..191C}) and our sample.

The presence of anomalous dust extinction in some objects (GQ\,1309$+$2904, GQ\,0109$-$0719, GQ\,1353$+$2554, and GQ\,1237$+$1233) in our sample is naturally explained if low‑inclination sightlines intersect the disc wind containing small dust grains. This geometry is consistent with the low polarisation observed in three out of the four sources (GQ\,1309$+$2904, GQ\,1353$+$2554, and GQ\,1237$+$1233), and a similar viewing angle may apply to SDSS\,J0827$+$4252 (\citeads{2022ApJ...930....5Y}) and the WLQ IRAS\,14026$+$4341 (\citeads{2013AJ....145..157J}) with anomalous dust extinction. At even lower inclinations (almost face-on), without seeing through the disc wind, we can see the normal WLQs, as suggested by \citetads{2015ApJ...805..122L}. This aligns with the observed low polarisation of WLQs in general (\citeads{2009ApJ...699..782D}) and mild SMC-like reddening in some WLQs (\citeads{2015ApJ...805..122L}). High‑inclination (edge-on) sightlines instead pass primarily through the bulk of the dust cocoon, producing more SMC‑like extinction seen in PSS\,J0141$+$3334, SDSS\,J1012$+$0358, and SDSS\,J0155$+$2543. Mini-BAL features observed in our sample are likely produced by clumpy, dense clouds formed within the disc wind.

When the high-velocity disc wind impacts the dust cocoon, it carves a directional channel. Driven by the ram pressure of the outflow, the wind ablates the cavity walls and entrains the surrounding material, sweeping the dust grains outwards along the disc wind. The movement of gas and dust is illustrated by dashed black arrows in Fig.~\ref{fig:evolution_model}. Along the boundary of this channel, the intense velocity shear between the fast outflow and the dense cocoon triggers Kelvin-Helmholtz instabilities, which continuously ablate and strip dust from the cocoon, widening the cavity over time. This process is similar to what was described in \citetads{2013ApJ...763L..18W}, \citetads{2016MNRAS.461..967M}, and \citetads{2019MNRAS.486.4526B}. Continued feedback from the disc wind eventually clears a significant fraction of the dust cocoon. During this late blowout phase, we observe reduced SMC-like extinction as the quasar becomes nearly unveiled (e.g. SDSS\,J0155$+$2543). The re-emergence of UV emission lines likely occurs at the same time, creating the more obvious HiBELs seen in SDSS\,J1012$+$0358 and SDSS\,J0155$+$2543. Later in this period, WLQs free of dust obscuration (\citeads{2009ApJ...699..782D, 2015ApJ...805..123P}) can be observed. The ISM and circumgalactic medium of the quasar host galaxy can be ionised due to their interaction with the propagating outflows, producing the extended Ly$\alpha$ emission observed in SDSS\,J1012$+$0358.

The launching of disc wind from the accretion disc may cease, potentially driven by the episodic accretion observed in LoBAL quasars (e.g. \citeads{2024ApJ...970....9B}). As a result, the disc wind becomes truncated, and can account for the observed abnormal BAL profiles in GQ\,1237$+$1233 that display broader red wings than blue wings (see Sect.~\ref{sec:velocities}). The unusual velocity profiles imply that the highest column densities are associated with higher outflow velocities. Assuming that outflows accelerate (\citeads{2022SciA....8.3291H}; \citeads{2024ApJ...968...49W}) and undergo dilution with increasing distance from the BH (\citeads{2024MNRAS.528.6496H}), such profiles would arise when the dense, low-velocity wind near the black hole dissipates, leaving the faster, more distant BAL clouds to dominate the absorption. The observed extremely weak {\AlIII} BAL in GQ\,1237$+$1233 suggests a drop in gas column density to a level where prominent LoBAL features can no longer be formed, and the transition from LoBAL to HiBAL quasars seen in SDSS\,J0827$+$4252 (\citeads{2022ApJ...930....5Y}) can happen.

After the quasar has cast off the dust cocoon, some dusty clouds near the disc plane may be left and form the so-called torus, which is commonly present in the unified model of AGNs (\citeads{2012agn..book.....B}). The quasar is completely exposed from now on and the outflow may disappear or persist for a while (\citeads{2019MNRAS.487.2594T}). If the disc wind persists, the quasars can appear as non-BAL and BAL quasars from different inclinations.

\section{Conclusions} \label{sec:conclusions}
In this paper, we offer an in-depth analysis of a sample of $z\approx$ 2.07--3.28 peculiar LoBAL quasars characterised by very weak UV emission lines and strong absorption lines, akin to GQ\,1309$+$2904 from \citetaliasads{2020A&A...634A.111F}. We validate the highly blueshifted nature of their BALs through the observation of strong ${\mathrm{H} \alpha}$ emission in the near-IR spectra. In the case of PSS\,J0141$+$3334, we use the onset of the Ly$\alpha$ forest to constrain the redshift as ${\mathrm{H} \alpha}$ is redshifted out of the $K$ band. Through a comprehensive multi-wavelength study involving spectroscopic, photometric, and polarimetric data, our key findings are as follows:
\medskip

{\it (i)} We observe deep, high-velocity HiBAL and LoBAL troughs in these objects, covering velocities from \SI{-5000}{\km\per\s} to \SI{-47000}{\km\per\s}. These outflows are accompanied by blueshifted and remarkably faint BELs. In particular, {\CIV} and {\HeII} emission is either very weak or entirely absent, suggesting softer SEDs of our objects than normal quasars. This suppression of HiBELs appears to be physically linked to the high-speed outflows. While these findings are consistent with the trends reported by \citetaliasads{2020MNRAS.492.4553R}, the features observed here are significantly more extreme.

{\it (ii)} The quasars in our sample have $M_{\mathrm{BH}} \approx 10^{8.7}$--$10^{9.4}\,M_{\odot}$ and $L_{\mathrm{bol}} / L_{\mathrm{Edd}} \approx$ 0.14--0.34, showing their massive and active nature. However, their BELs are much more blueshifted than those of non-BAL and HiBAL quasars with similar $M_{\mathrm{BH}}$ and $L_{\mathrm{bol}} / L_{\mathrm{Edd}}$ in \citetaliasads{2020MNRAS.492.4553R} and \citetads{2023MNRAS.523..646T}.

{\it (iii)} We find that our quasars are reddened by either a SMC-like dust extinction curve or an anomalous (UV-steeper) dust extinction curve. Of the four objects exhibiting anomalous extinction, three (including GQ\,1309$+$2904) show low degrees of polarisation, while the fourth lacks polarisation data. Interpreting this low polarisation as a signature of low inclination, we suggest that our sightlines to these objects pass directly through the disc wind. These results imply that small dust grains may be entrained within the disc wind itself.

{\it (iv)} SDSS\,J1012$+$0358 is the only object in our sample that is found to have photometric variability, relatively high radio intensity, potential jets, and extended Ly$\alpha$ emission.

{\it (v)} The peculiar LoBAL quasars in our sample, characterised by highly blueshifted BALs and weak UV emission lines, represent a specific subpopulation of WLQs observed through the disc wind. These viewing angles have been systematically overlooked because standard WLQ studies intentionally exclude sources with BALs. Specifically, three of our objects (GQ\,0109$-$0719, GQ\,1353$+$2554, and GQ\,1237$+$1233) are classified as WLQs, based on their {\CIV} emission upper limits, while two (SDSS\,J1012$+$0358 and SDSS\,J0155$+$2543) appear to be bridge quasars transitioning from the WLQ phase to normal quasars using the criterion from \citetads{2025ApJ...994..213C}. We propose that this orientation allows us to witness the quasars emerging from their dust cocoons. In this scenario, powerful outflows pierce the cocoon, creating shocks that shatter large SMC-like grains into the smaller dust grains in the disc wind. Thus, quasars with extreme outflow velocities and weak UV emission lines likely signal the critical transition from early-type quasars to the unobscured non-BAL and HiBAL population.

\begin{acknowledgements}
We thank the anonymous referee for very constructive suggestions which substantially improved this paper.
We thank Amy L. Rankine, Karen M. Leighly, Fuyan Bian, Clara Giménez-Arteaga, Stanislav G. Djorgovski, David Turnshek, and Patrick B. Hall for helpful and instructive discussions.

Based on observations made with the Gran Telescopio Canarias (GTC) and with the Nordic Optical Telescope (NOT), installed in the Spanish Observatorio del Roque de los Muchachos of the Instituto de Astrofísica de Canarias.

The Cosmic Dawn Center (DAWN) is funded by the Danish National Research Foundation under grant DNRF140. GM and LC are supported by the Independent Research Fund Denmark (DFF-2032-00071). 
JPUF and SV are supported by the Independent Research Fund Denmark (DFF-4090-00079) and thank the Carlsberg Foundation for support. JPUF acknowledges support from the European Research Council (ERC) under the European Union’s research and innovation program (ERC Grant HEAVYMETAL No.~101071865). KEH acknowledges support from the Carlsberg Foundation Reintegration Fellowship Grant CF21-0103. 

This work made use of Astropy\footnote{\href{https://www.astropy.org}{https://www.astropy.org}}: a community-developed core Python package and an ecosystem of tools and resources for astronomy \citepads{2022ApJ...935..167A}.
\end{acknowledgements}

\bibliographystyle{aa}
\bibliography{ref}

\object{GQ\,0109$-$0719, GQ\,1353$+$2554, GQ\,1237$+$1233, PSS\,J0141$+$3334, SDSS\,J1012$+$0358, SDSS\,J0155$+$2543}

\begin{appendix}
\onecolumn
\section{Observation log} \label{sec: app_obs}
The log of observations is presented here in Table~\ref{tab:log}. 

\begin{table*}[hbt!]
\centering
\caption{Log of observations.}
\label{tab:log}
\addtolength{\tabcolsep}{-0.06cm}
\begin{tabular}{lllrccr}
\hline \hline \noalign{\smallskip}
Date (UT) & Grism & $\lambda$ range & Exp. time  & Airmass & S/N\tablefootmark{\textcolor{blue}{a}} & Resolution\tablefootmark{\textcolor{blue}{b}} \\
dd/mm/yyyy  &      & (\SI{}{\AA})  & (\SI{}{\s})   &      &   &   \\
\hline
\noalign{\smallskip}
\multicolumn{7}{c}{GQ\,1309$+$2904}\\
\hline
\noalign{\smallskip}
22/04/2021 & DiPol-UF & & 3600 &  \\
\noalign{\smallskip}
\hline
\noalign{\smallskip}
\multicolumn{7}{c}{GQ\,0109$-$0719}\\
\hline
\noalign{\smallskip}
02/08/2021 & OSIRIS/R1000B & \num{3630}--\num{7500} & 2$\times$350 & 1.34--1.35 & 38 & 1018 \\
04/08/2021 & ALFOSC/grism20 & \num{5650}--\num{10150} & 2$\times$500 & 1.34--1.37 & 51 & 770 \\
03/08/2022 & OSIRIS/R2000B & \num{3950}--\num{5700} & 2$\times$1350 & 1.29--1.35 & 52 & 2165 \\
13/08/2022 & EMIR/LR(HK) & \num{14500}--\num{24200} & 16$\times$120 & 1.23--1.43 & 15 & 987 \\
\noalign{\smallskip}
\hline
\noalign{\smallskip}
\multicolumn{7}{c}{GQ\,1353$+$2554}\\
\hline
\noalign{\smallskip}
23/05/2020 & OSIRIS/R1000B & \num{3630}--\num{7500} & 2$\times$420 & 1.06--1.07 & 45 & 1018 \\
29/05/2020 & OSIRIS/R2500U & \num{3440}--\num{4610} & 3$\times$1800 & 1.05--1.17 & 10 & 2555 \\
08/03/2021 & ALFOSC/grism20 & \num{5650}--\num{10150} & 2$\times$500 & 1.27--1.32 & 31 & 770 \\
22/04/2021 & DiPol-UF & & 3600 &  \\
18/12/2022 & EMIR/LR(HK) & \num{14500}--\num{24200} & 16$\times$120 & 1.23--1.36 & 8 & 987 \\
\noalign{\smallskip}
\hline
\noalign{\smallskip}
\multicolumn{7}{c}{GQ\,1237$+$1233}\\
\hline
\noalign{\smallskip}
05/01/2019 & OSIRIS/R1000B & \num{3630}--\num{7500} & 2$\times$750 & 1.08--1.10 & 37  & 1018 \\
22/04/2021 & DiPol-UF & & 3600 &  \\
06/01/2023 & EMIR/LR(HK) & \num{14500}--\num{24200} & 16$\times$120 & 1.06--1.11 & 8 & 987 \\
\noalign{\smallskip}
\hline
\noalign{\smallskip}
\multicolumn{7}{c}{PSS\,J0141$+$3334}\\
\hline
\noalign{\smallskip}
01/01/2022 & ALFOSC/grism19 & \num{4400}--\num{6950} & 2$\times$900 & 1.16--1.21 & 10 & 970\\
01/01/2022 & ALFOSC/grism20 & \num{5650}--\num{10150} & 2$\times$900 & 1.27--1.32 & 16 & 770\\
18/02/2022 & EMIR/LR(HK)   & \num{14500}--\num{24200} & 12$\times$120 & 1.55--1.72 & 30 & 987 \\
13/08/2022 & EMIR/LR(YJ)    & \num{8500}--\num{13500} & 16$\times$120 & 1.03--1.07 & 8 & 987 \\
04/08/2022 & OSIRIS/R2000B & \num{3950}--\num{5700} & 3$\times$1250 & 1.04--1.10 & \multirow{2}{*}{22}  & 2165 \\
19/08/2022 & OSIRIS/R2000B & \num{3950}--\num{5700} & 3$\times$1250 & 1.13--1.24 &  & 2165 \\ 
\noalign{\smallskip}
\hline
\noalign{\smallskip}
\multicolumn{7}{c}{SDSS\,J1012$+$0358}\\
\hline
\noalign{\smallskip}
24/11/2022 & OSIRIS/R2000B & \num{3950}--\num{5700} & 2$\times$1350 & 1.30--1.39 & 28 & 2165 \\
26/11/2022 & OSIRIS/R2500U & \num{3440}--\num{4610} & 3$\times$1350 & 1.35--1.61 & 10 & 2555 \\
27/11/2022 & OSIRIS/R1000R & \num{5100}--\num{10000} & 3$\times$400 & 1.14--1.16 & 65 & 1122\\
17/12/2022 & EMIR/LR(HK)   & \num{14500}--\num{24200} & 16$\times$120 & 1.22--1.32 & 12 & 987 \\
\noalign{\smallskip}
\hline
\noalign{\smallskip}
\multicolumn{7}{c}{SDSS\,J0155$+$2543}\\
\hline
\noalign{\smallskip}
24/12/2022 & EMIR/LR(HK)   & \num{14500}--\num{24200} & 16$\times$120 & 1.09--1.16 & 8 & 987 \\
23/07/2023 & OSIRIS/R2500U & \num{3440}--\num{4610} & 2$\times$1440 & 1.20--1.41 & 4 & 2555 \\
\noalign{\smallskip}
\hline
\noalign{\smallskip} 
\hline
\end{tabular}
\tablefoot{
\tablefoottext{\textcolor{blue}{a}}{Mean S/N calculated using \texttt{Specutils} by \citet{nicholas_earl_2024_11099077}.}
\tablefoottext{\textcolor{blue}{b}}{Resolution measured for $0.6^{\prime\prime}$ slit in OSIRIS and EMIR, and for $1^{\prime\prime}$ slit in ALFOSC.}
}

\end{table*}

\clearpage

\section{Normalised optical spectra and absorption line systems} \label{sec: app_nal}
\begin{figure*}[th]
    \centering
    \includegraphics[width=17cm]{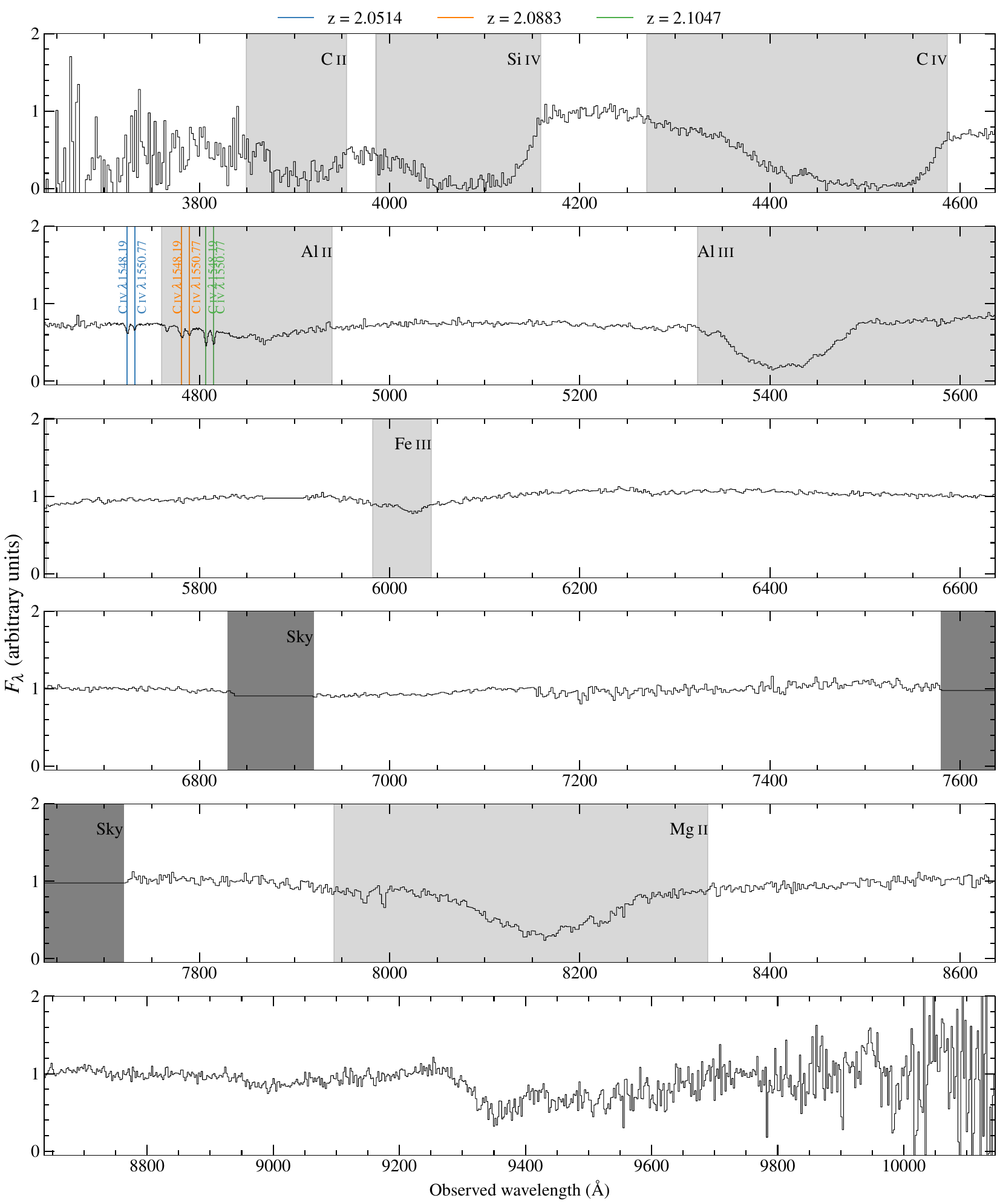}
    \captionof{figure}{Normalised optical spectrum of GQ\,0109$-$0719. The NAL systems at different redshifts ($z = 2.0514, 2.0883$, and 2.1047) are marked in different colours. We shaded all identified BALs in light grey and telluric bands in dark grey.}
    \label{fig:normalised_GQ0109}
\end{figure*}

\begin{figure*}[th]
    \centering
    \includegraphics[width=17cm]{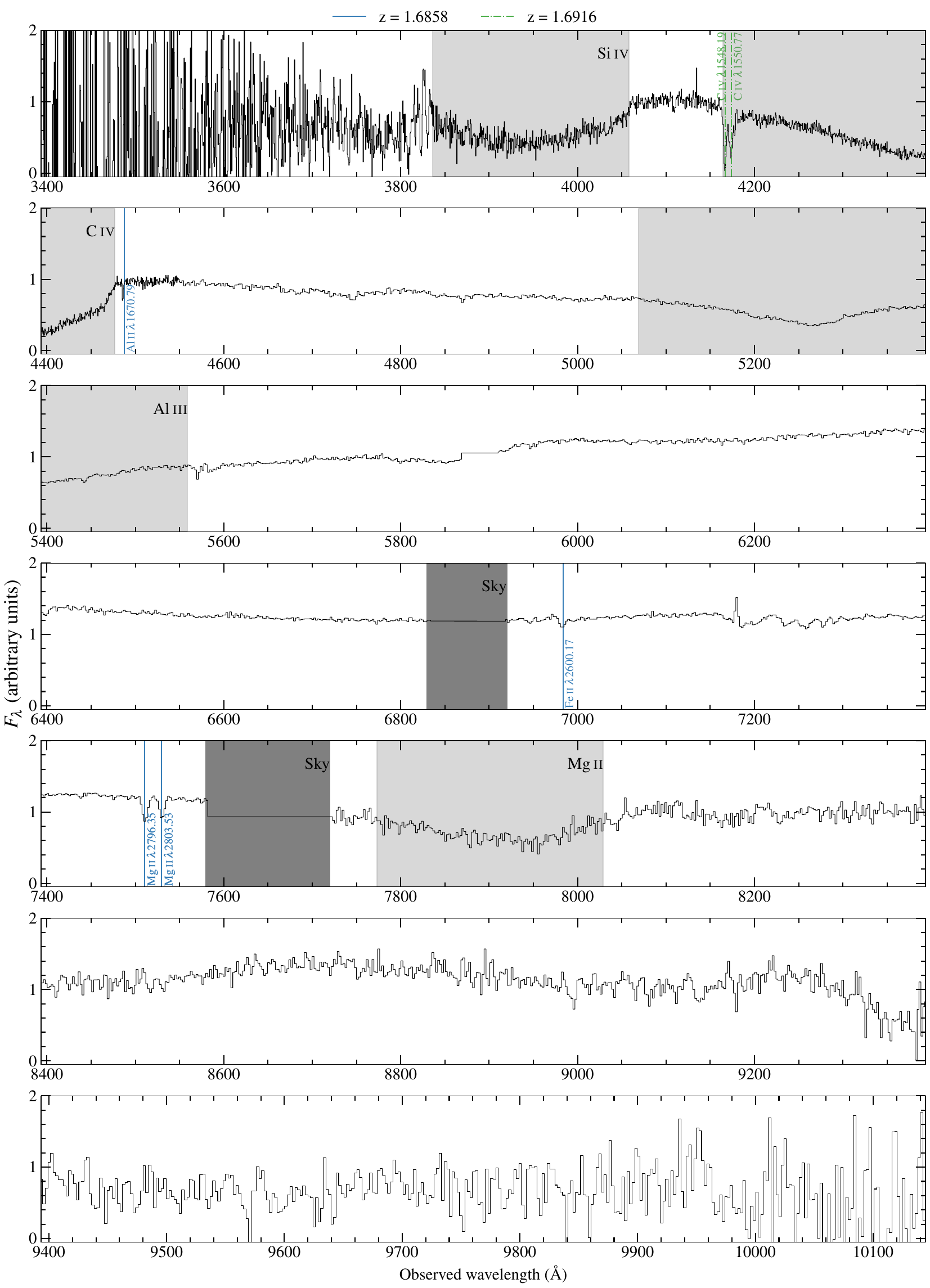}
    \captionof{figure}{Same as Fig.~\ref{fig:normalised_GQ0109} but for GQ\,1353$+$2554. A narrow line system is found at $z = 1.6858$ and a {\CIV} mini-BAL system is found at $z = 1.6916$. We show a double-Gaussian fit to this mini-BAL in Fig.~\ref{fig:GQ1353_CIV}.}
    \label{fig:normalised_GQ1353}
\end{figure*}

\begin{figure*}[th]
\centering
\includegraphics[width=17cm]{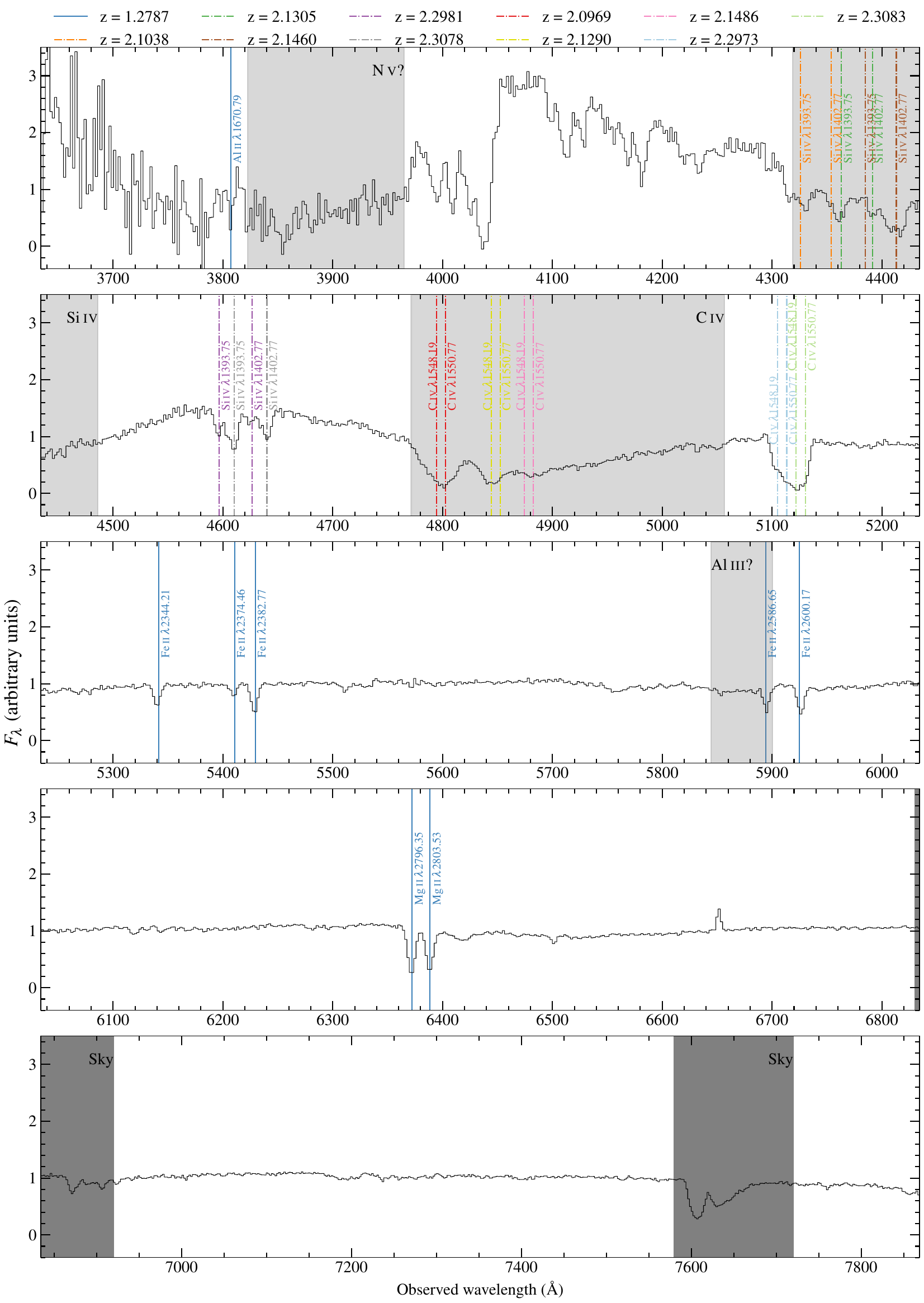}
\caption{Same as Fig.~\ref{fig:normalised_GQ0109} but for GQ\,1237$+$1233. We identified one NAL system at $z = 1.2787$, three mini-BAL systems with high velocities, and two mini-BALs close to the systemic redshift of the quasar (for more see Figs.~\ref{fig:GQ1237_SiIV_blue}--\ref{fig:GQ1237_CIV}).
}
\label{fig:normalised_GQ1237}
\end{figure*}

\begin{figure*}[th]
\centering
\includegraphics[width=17cm]{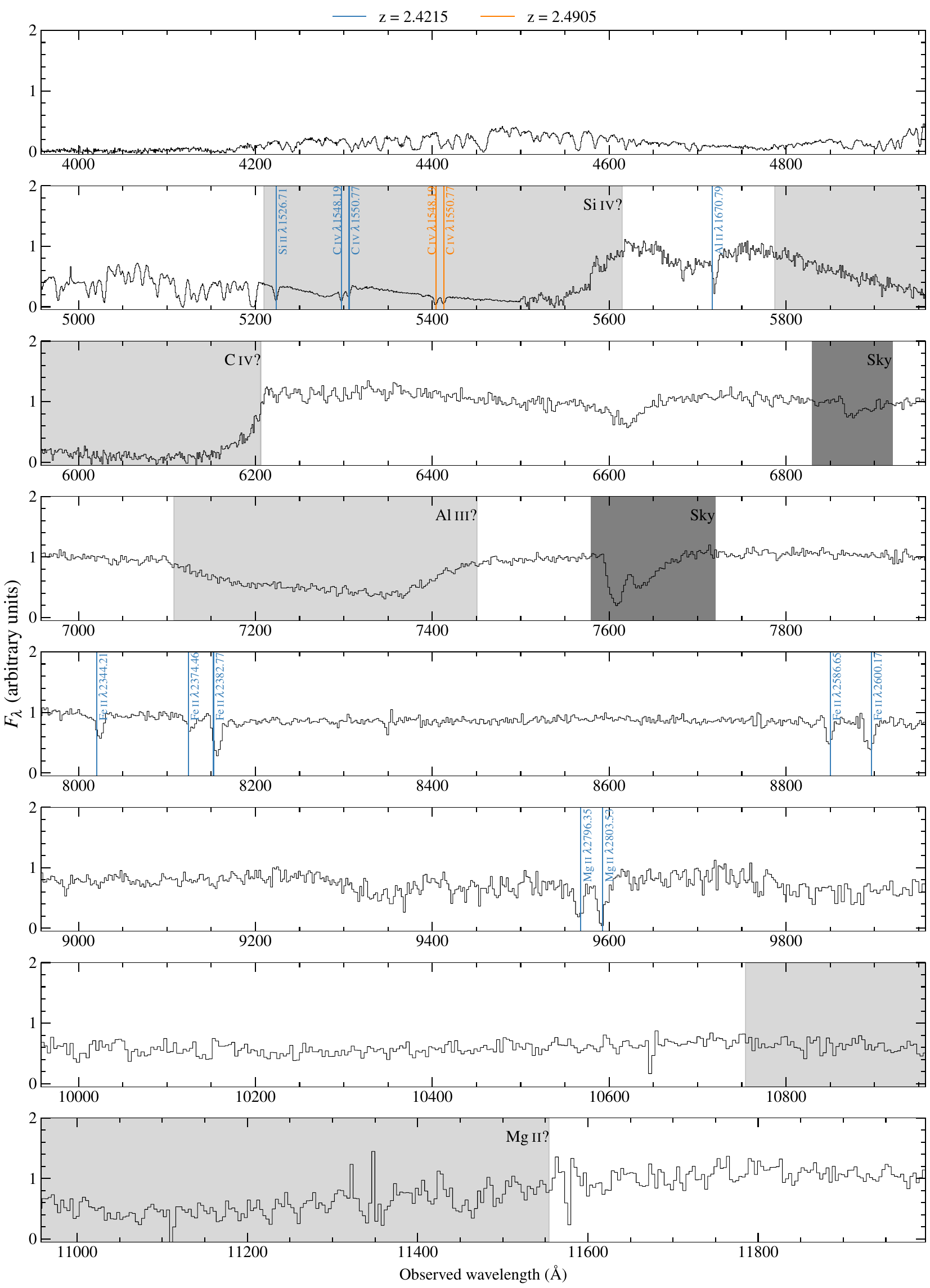}
\caption{Same as Fig.~\ref{fig:normalised_GQ0109} but for PSS\,J0141$+$3334. Two NAL systems are found at $z = 2.4215$ and 2.4905. The BALs are marked very tentatively based on the lower limit of its systemic redshift from Sect.~\ref{sec:PSSJ0141}.
}
\label{fig:normalised_PSS0141}
\end{figure*}

\begin{figure*}[th]
\centering
\includegraphics[width=17cm]{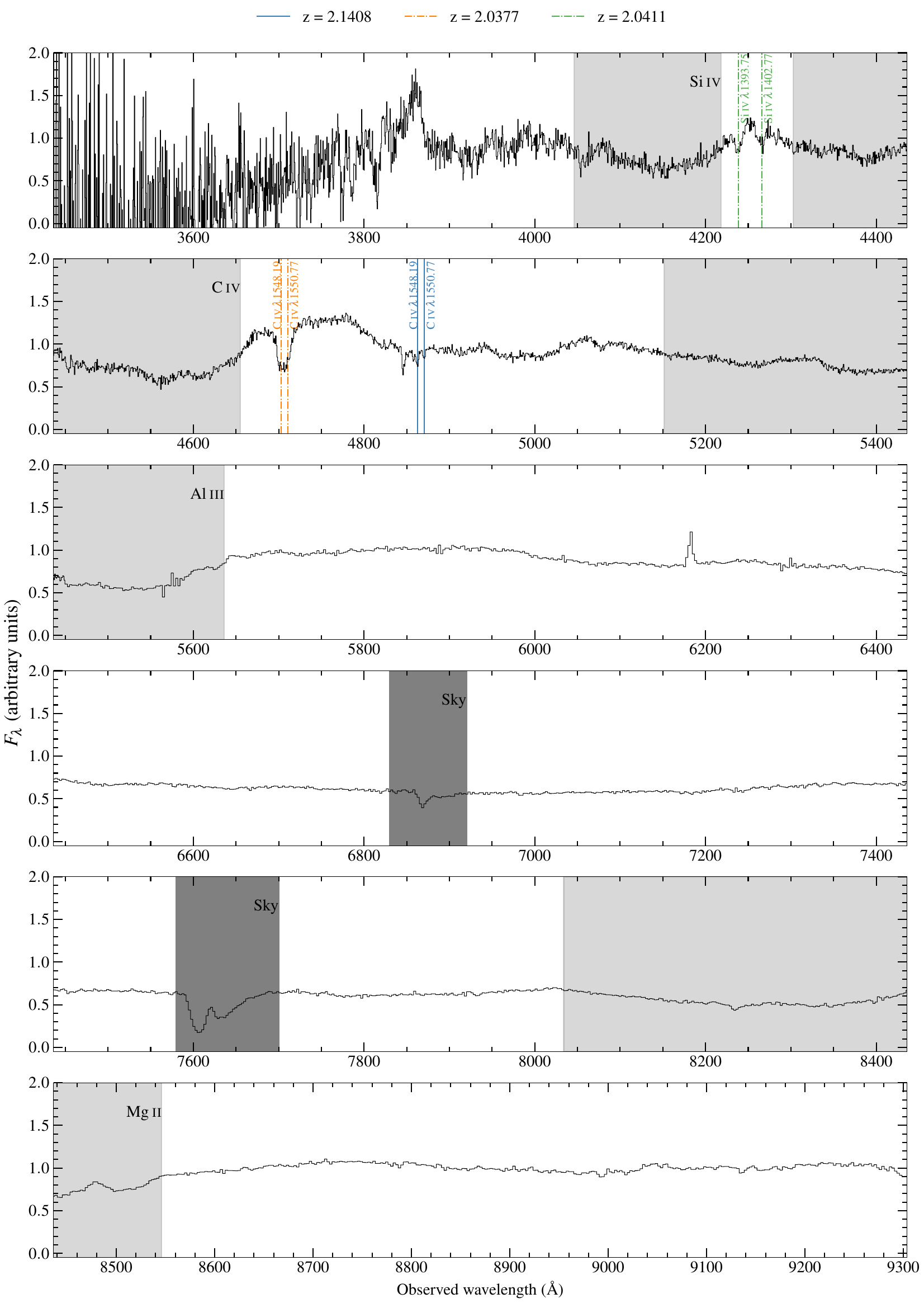}
\caption{Same as Fig.~\ref{fig:normalised_GQ0109} but for SDSS\,J1012$+$0358. One NAL system at $z = 2.1408$ is identified, along with {\CIV} and {\SiIV} mini-BALs at $z = 2.0377$ and $z = 2.0411$, respectively. These two mini-BALs potentially originate from different regions within the same outflowing cloud and detailed double-Gaussian fits can be found in Figs.~\ref{fig:SDSSJ1012_SiIV}--\ref{fig:SDSSJ1012_CIV}. 
}
\label{fig:normalised_SDSSJ1012}
\end{figure*}

\begin{figure*}[th]
    \centering
    \includegraphics[width=17cm]{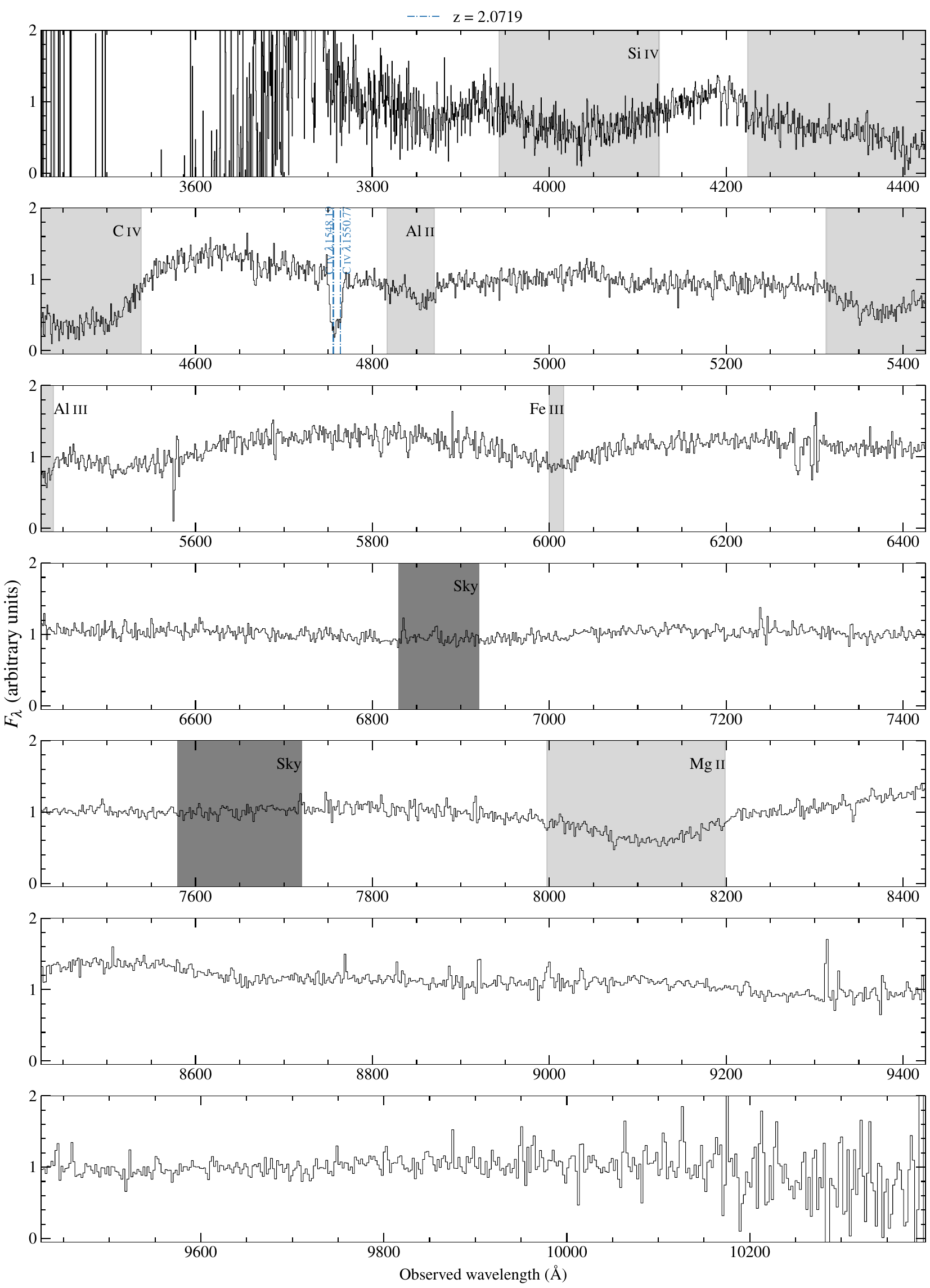}
    \captionof{figure}{Same as Fig.~\ref{fig:normalised_GQ0109} but for SDSS\,J0155$+$2543. Here we find a {\CIV} mini-BAL system at $z = 2.0719$. We also present a double-Gaussian fit of this {\CIV} mini-BAL in Fig.~\ref{fig:SDSSJ0155_CIV}.}
    \label{fig:normalised_SDSSJ0155}
\end{figure*}

We show the optical spectra, normalised by the continuum (dash-dotted blue lines in Figs.~\ref{fig:GQ0109-0719_z_full}--\ref{fig:SDSSJ0155+2543_z_full}), here in Figs.~\ref{fig:normalised_GQ0109}--\ref{fig:normalised_SDSSJ0155}. All identified absorption line systems are marked with different colours indicating their different redshifts, with confirmed NAL systems in vertical solid lines and mini-BAL systems in dash-dotted lines. The BAL troughs are shaded in light grey, while the telluric bands are shaded in dark grey.

\FloatBarrier
\twocolumn
\section{Mini-BAL systems} \label{sec: app_mini_bal}
\begin{figure}[th]
\centering
\resizebox{\hsize}{!}{\includegraphics[scale=0.4]{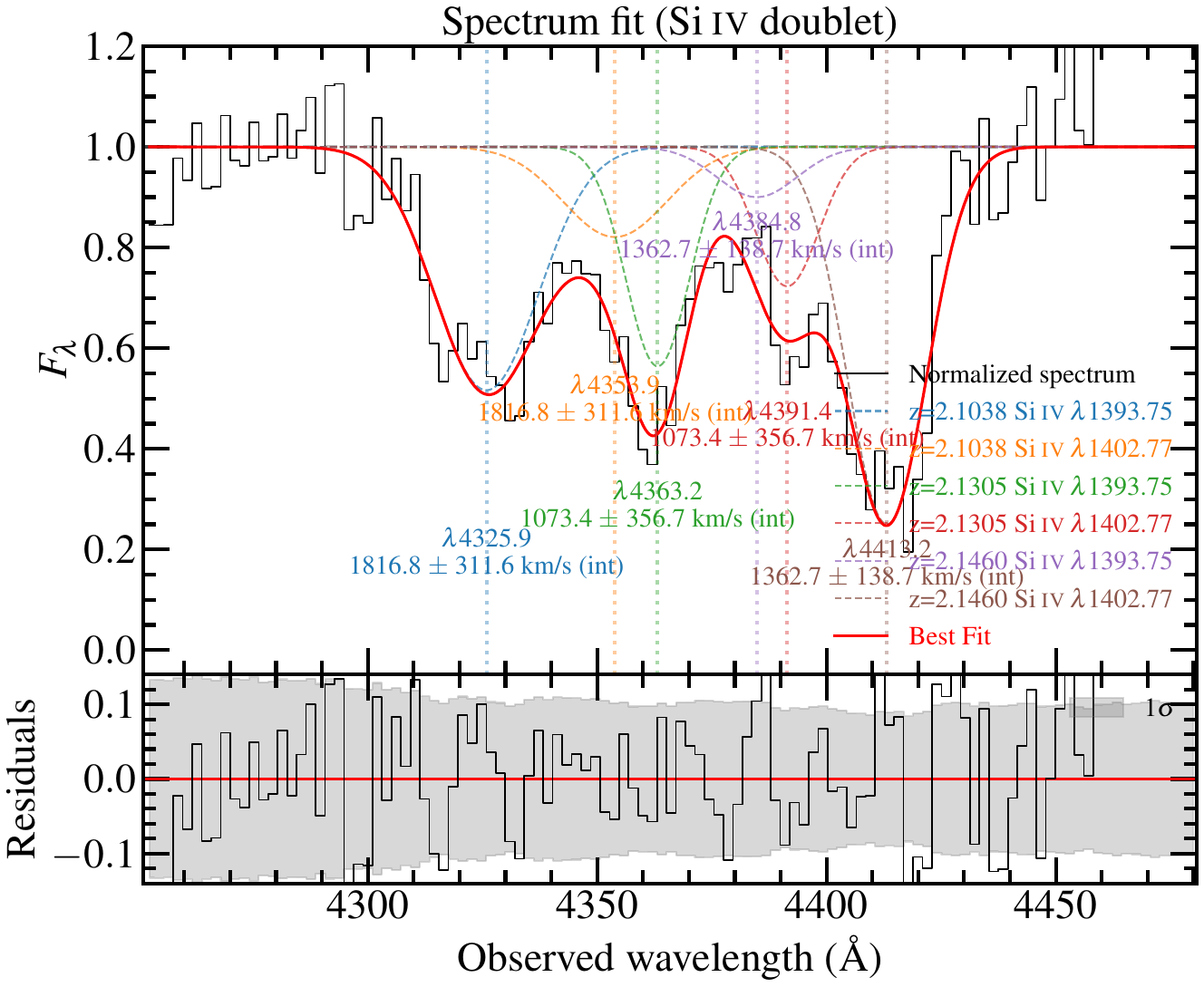}}
\caption{Double-Gaussian fit to the {\SiIV} doublet absorption for GQ\,1237$+$1233. \textit{Top}: Potential mini-BAL profile normalised by the local continuum is plotted in black, with the total best-fit profile overplotted in red. Three {\SiIV} mini-BAL systems are identified within this absorption trough at $z = 2.1038$, 2.1305, and 2.1460. The corresponding central wavelengths and intrinsic FWHMs (corrected for the instrumental resolution $R = 1018$) for each component are annotated in the plot. \textit{Bottom}: Residuals after subtracting the best-fit profile from the spectrum, with the grey shading representing the 1$\sigma$ error.
}
\label{fig:GQ1237_SiIV_blue}
\end{figure}

\begin{figure}[th]
\centering
\resizebox{\hsize}{!}{\includegraphics[scale=0.4]{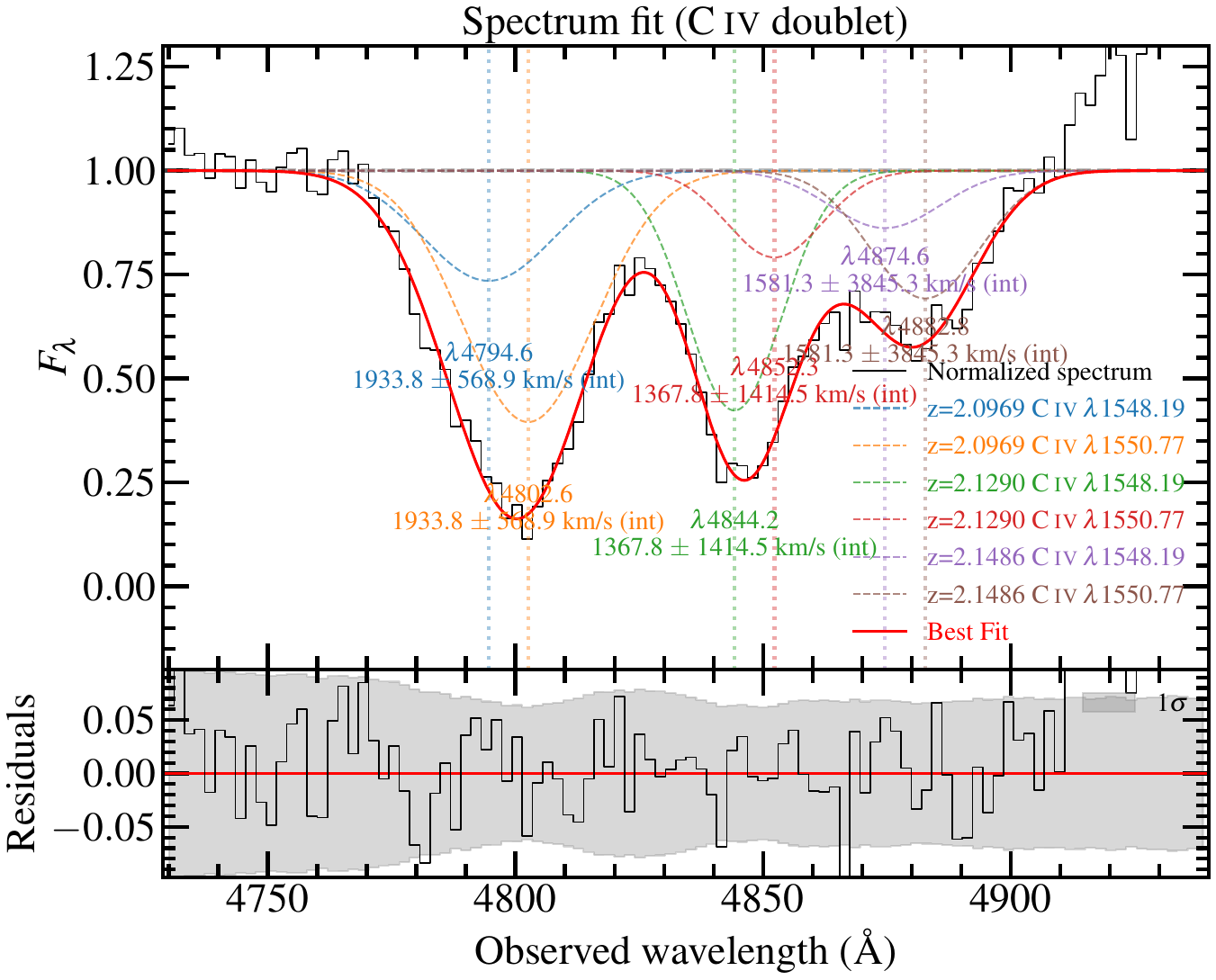}}
\caption{Same as Fig.~\ref{fig:GQ1237_SiIV_blue} but for {\CIV} doublets of GQ\,1237$+$1233. Following the identification of {\SiIV} mini-BALs in Fig.~\ref{fig:GQ1237_SiIV_blue}, three {\CIV} mini-BALs are found in this absorption profile, at $z = 2.0969$, 2.1290, and 2.1486.
}
\label{fig:GQ1237_CIV_blue}
\end{figure}

\begin{figure}[th]
\centering
\resizebox{\hsize}{!}{\includegraphics[scale=0.4]{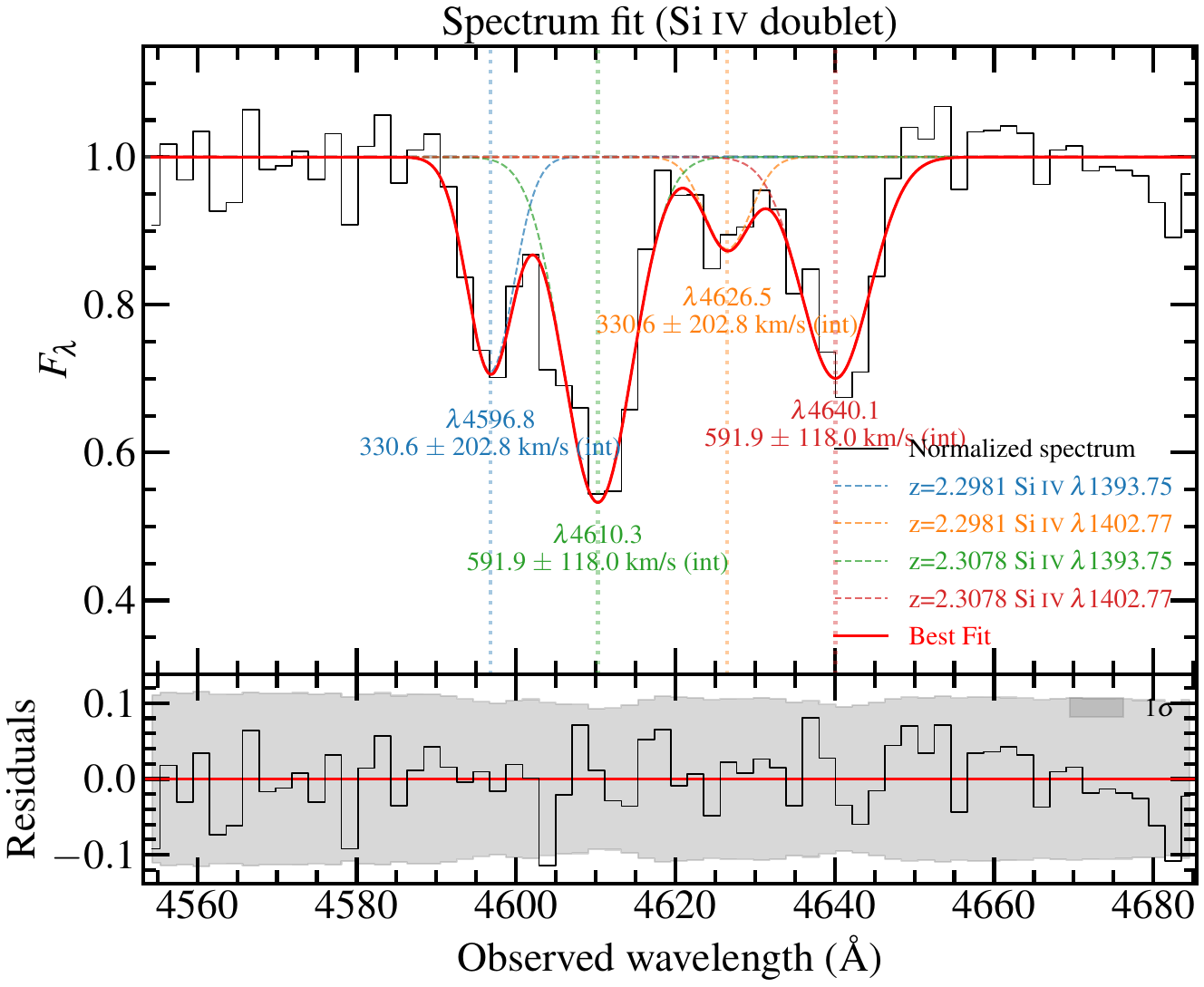}}
\caption{Same as Fig.~\ref{fig:GQ1237_SiIV_blue} but for another {\SiIV} absorption profile of GQ\,1237$+$1233 at longer wavelengths. Two {\SiIV} mini-BALs at $z = 2.2981$ and $z = 2.3078$ are found in this case.
}
\label{fig:GQ1237_SiIV}
\end{figure}

\begin{figure}[th]
\centering
\resizebox{\hsize}{!}{\includegraphics[scale=0.4]{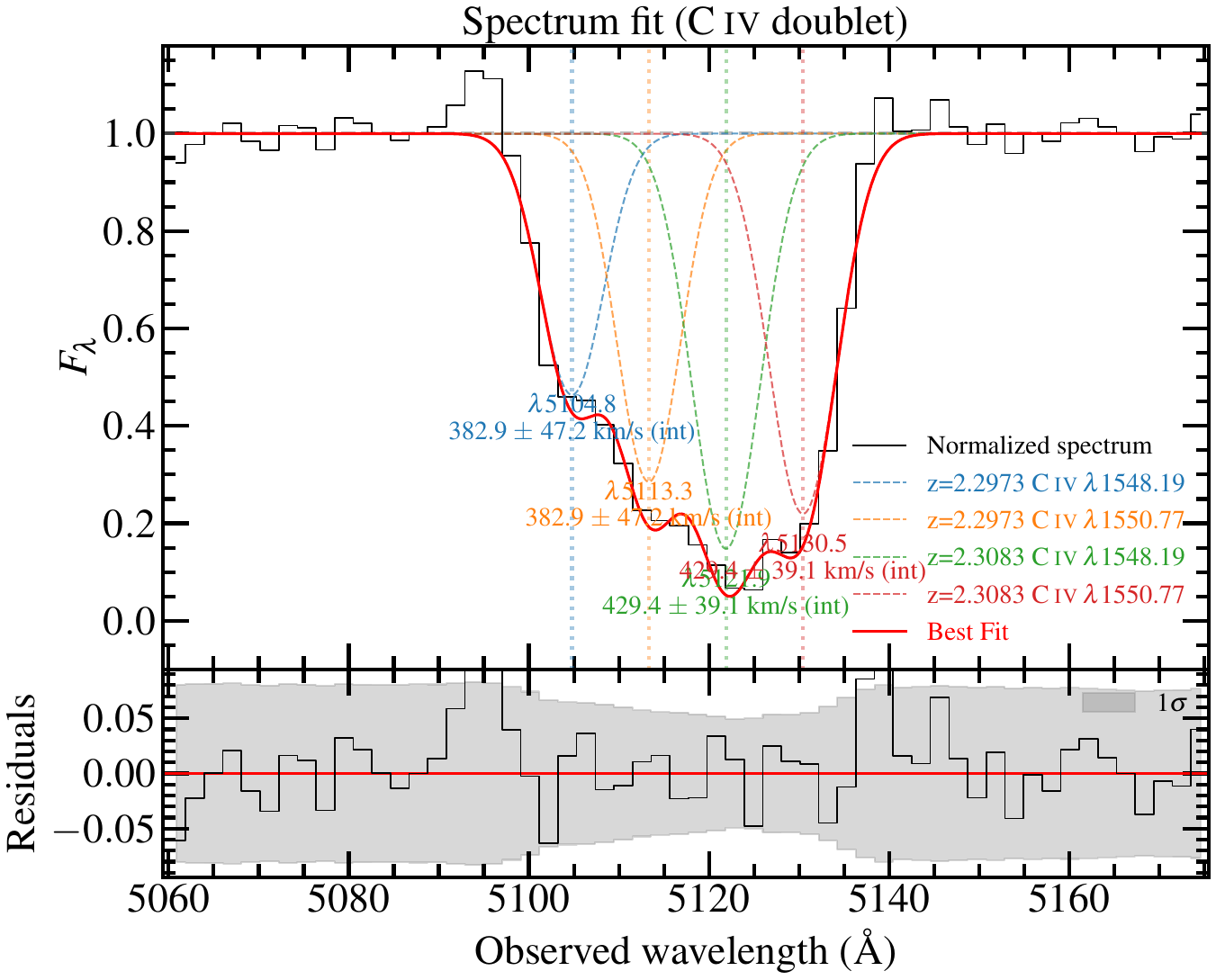}}
\caption{Same as Fig.~\ref{fig:GQ1237_SiIV_blue} but for the {\CIV} doublet absorption of GQ\,1237$+$1233 at longer wavelengths. Based on the {\SiIV} mini-BALs shown in Fig.~\ref{fig:GQ1237_SiIV}, we fitted two double-Gaussian profiles and identified two {\CIV} mini-BAL systems at $z = 2.2973$ and $z = 2.3083$.
}
\label{fig:GQ1237_CIV}
\end{figure}

\begin{figure}[th]
\centering
\resizebox{\hsize}{!}{\includegraphics[scale=0.4]{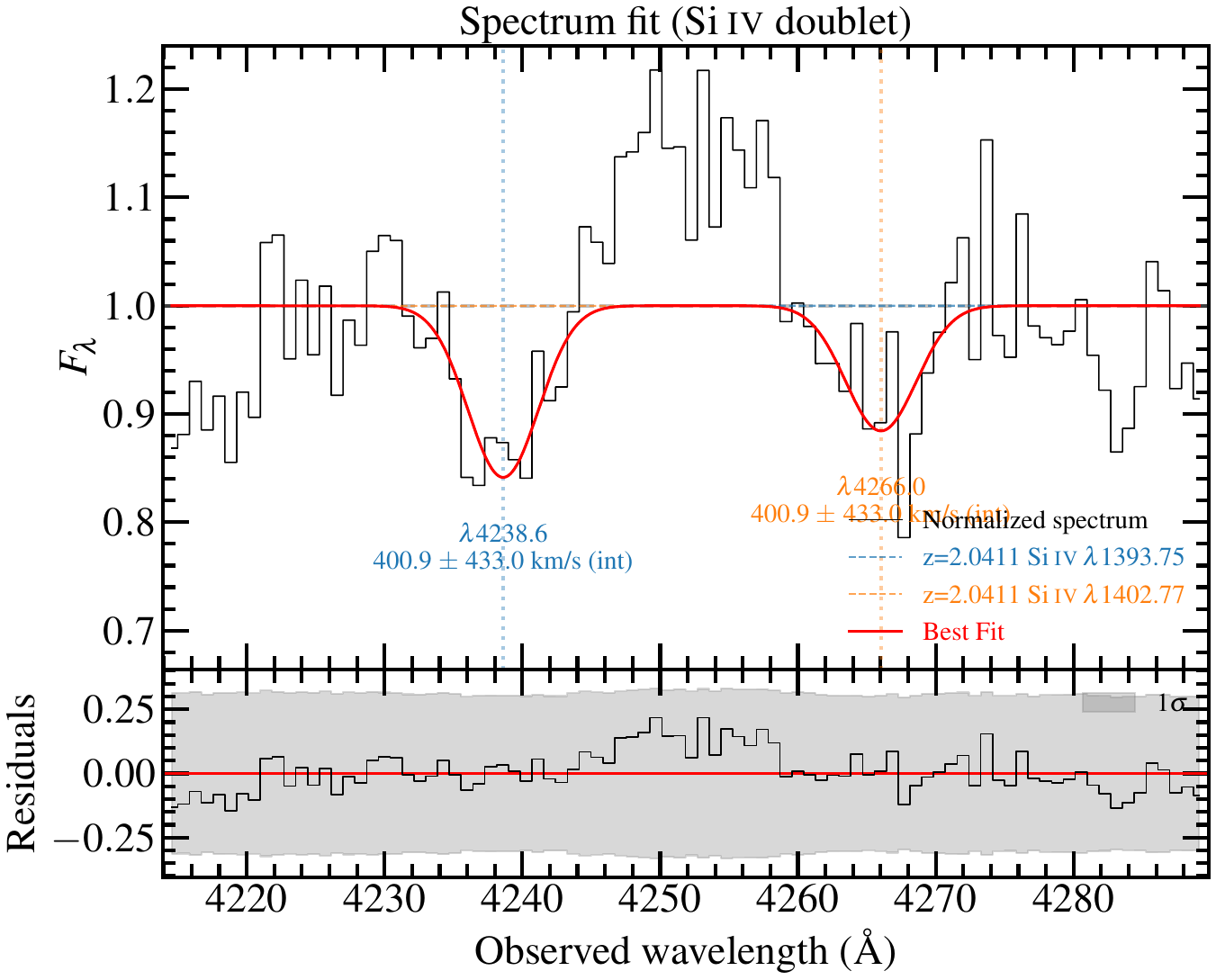}}
\caption{Same as Fig.~\ref{fig:GQ1237_SiIV_blue} but for the {\SiIV} mini-BAL of SDSS\,J1012$+$0358. Here one {\SiIV} mini-BAL is identified at $z = 2.0411$. The intrinsic FWHM has been corrected for the instrumental resolution $R = 2165$.
}
\label{fig:SDSSJ1012_SiIV}
\end{figure}

\begin{figure}[th]
\centering
\resizebox{\hsize}{!}{\includegraphics[scale=0.4]{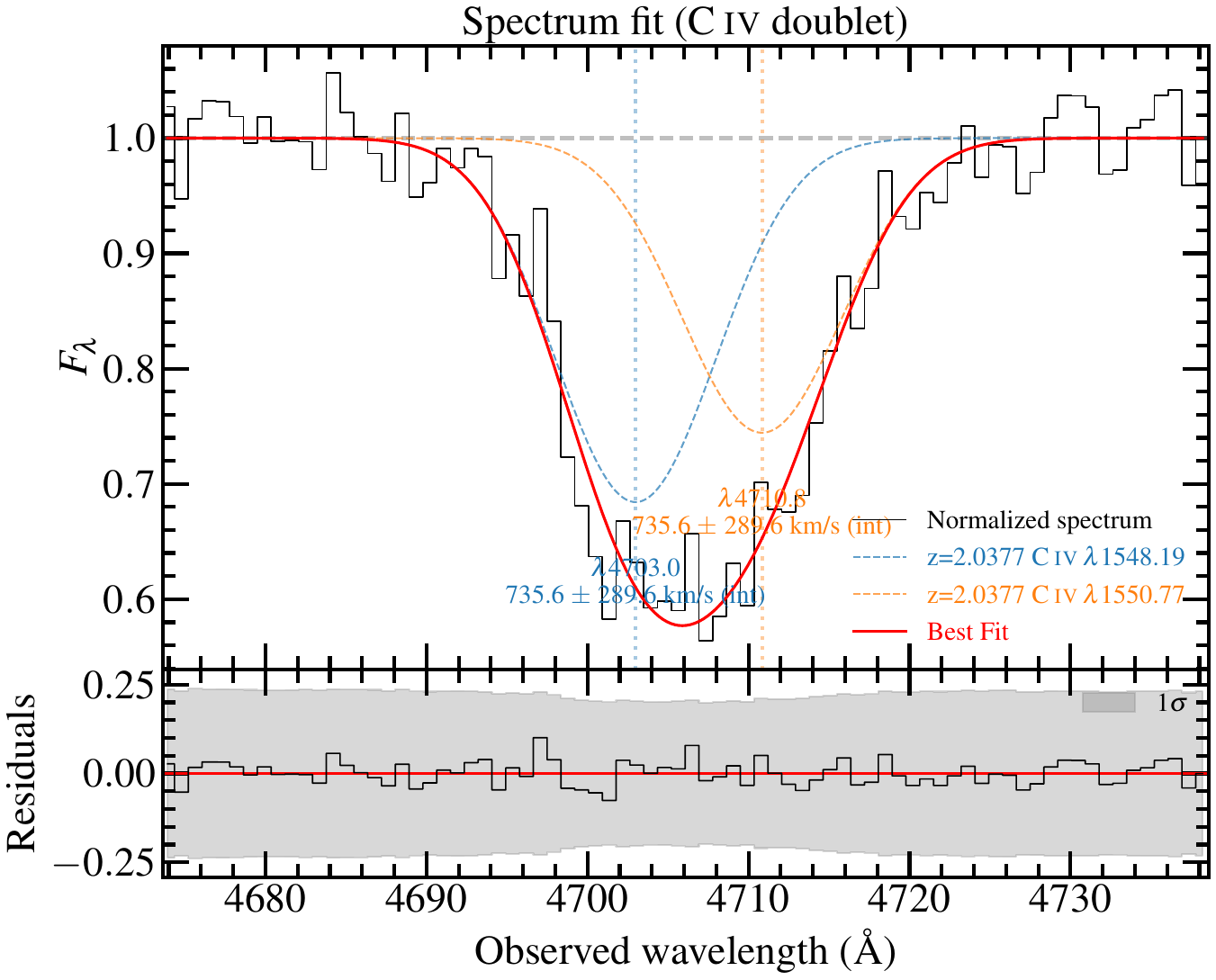}}
\caption{Same as Fig.~\ref{fig:GQ1237_SiIV_blue} but for the {\CIV} mini-BAL of SDSS\,J1012$+$0358. One {\CIV} mini-BAL is found at $z = 2.0377$.
}
\label{fig:SDSSJ1012_CIV}
\end{figure}

\begin{figure}[th]
\centering
\resizebox{\hsize}{!}{\includegraphics[scale=0.4]{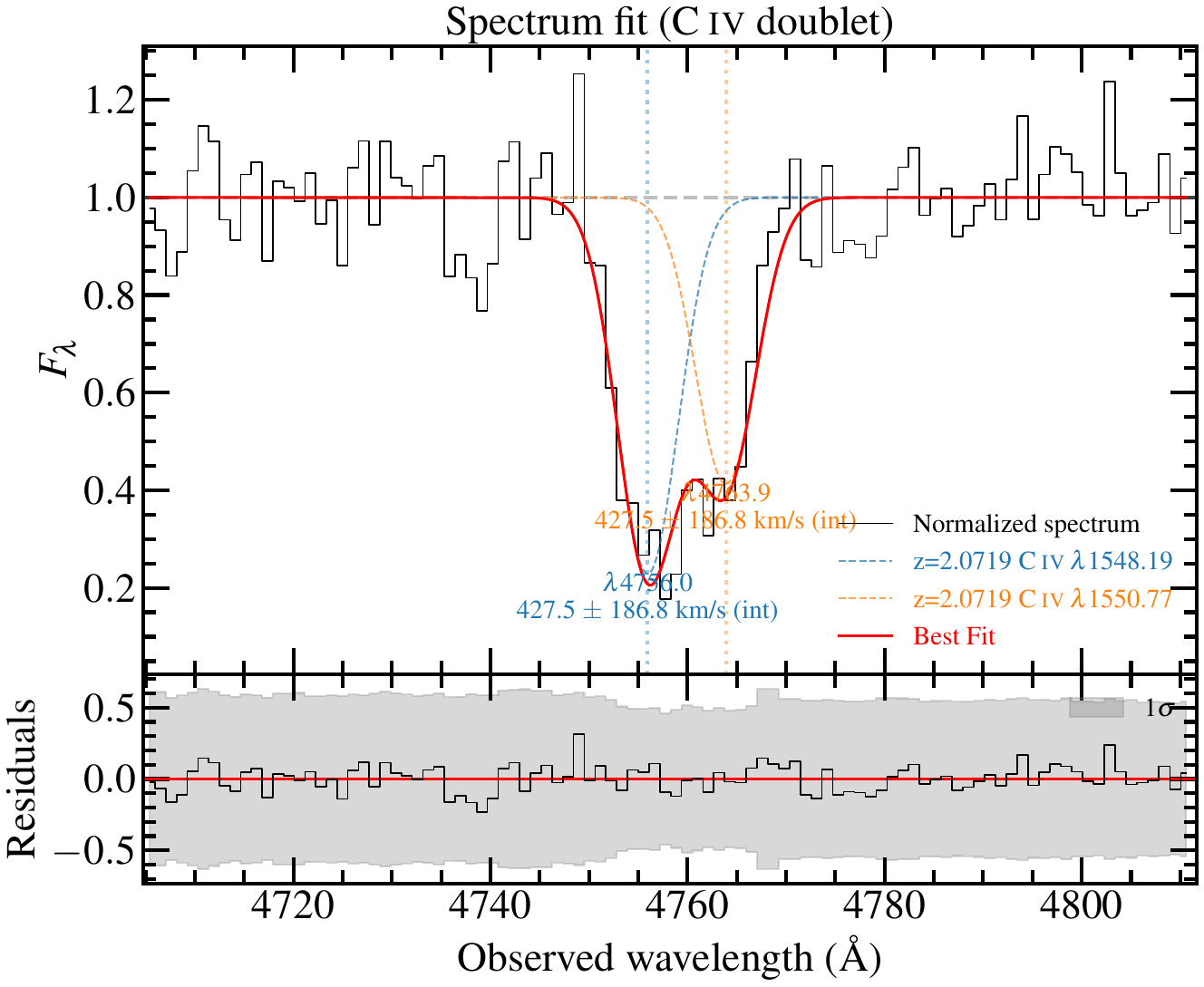}}
\caption{Same as Fig.~\ref{fig:GQ1237_SiIV_blue} but for the {\CIV} mini-BAL of SDSS\,J0155$+$2543. One {\CIV} mini-BAL system is found at $z = 2.0719$. The intrinsic FWHM has been corrected for the instrumental resolution $R = 1850$, which is adapted from the SDSS resolution at around \SI{4700}{\AA}.
}
\label{fig:SDSSJ0155_CIV}
\end{figure}

\begin{figure}[th]
\centering
\resizebox{\hsize}{!}{\includegraphics[scale=0.4]{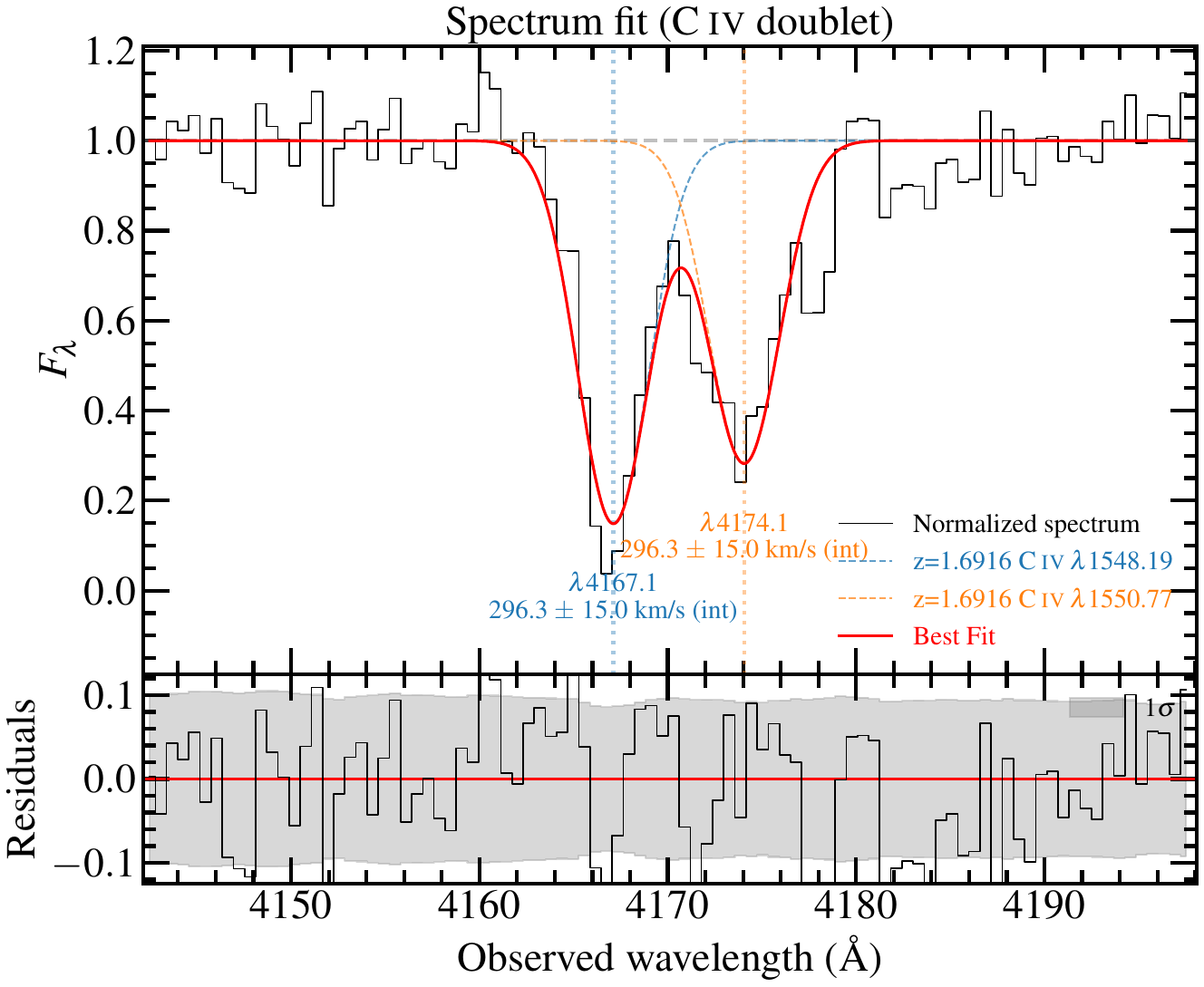}}
\caption{Same as Fig.~\ref{fig:GQ1237_SiIV_blue} but for the {\CIV} mini-BAL of GQ\,1353$+$2554. We identified one {\CIV} mini-BAL system at $z = 1.6916$. The intrinsic FWHM has been corrected for the instrumental resolution $R = 2555$.
}
\label{fig:GQ1353_CIV}
\end{figure}

We show detailed double-Gaussian fits to potential mini-BAL systems of our sample in Figs.~\ref{fig:GQ1237_SiIV_blue}--\ref{fig:GQ1353_CIV}. Fitted central wavelengths, EWs, and intrinsic FWHMs are listed in Table~\ref{tab:mini-BALs}.

\end{appendix}

\end{document}